\RequirePackage{fix-cm}
\documentclass[smallextended]{svjour3}       
\smartqed  
\usepackage{graphicx,amssymb,amsmath}

\usepackage{natbib,xcolor}
\usepackage{rotating,dcolumn}
\newcolumntype{d}[1]{D{.}{.}{#1}}

\PassOptionsToPackage{uniquelist=minyear,uniquename=allfull}{biblatex}

\newcommand{\fn}[1]{${}^{\rm #1}$}
\newcommand{\cdaweb}[1]{CDAWeb (\texttt{#1})}

\journalname{Living Reviews in Solar Physics}
\begin{document}

\title{The multi-scale nature of the solar wind}


\titlerunning{The multi-scale nature of the solar wind}        

\author{Daniel Verscharen  \and
        Kristopher G.~Klein \and
        Bennett A.~Maruca
}

\authorrunning{D.~Verscharen et al.} 

\institute{D. Verscharen \at
              Mullard Space Science Laboratory \\
              University College London\\
              Dorking, RH5 6NT\\
              United Kingdom\\
              Tel.: +44 1483-204-951\\
              \email{d.verscharen@ucl.ac.uk}    \\
              \emph{Also at:} Space Science Center\\
              University of New Hampshire\\
              Durham, NH 03824\\
              United States
           \and
           K.~G.~Klein \at
	  Lunar and Planetary Laboratory and Department of Planetary Sciences\\
	  University of Arizona\\
	  Tucson, AZ 85719 \\
	  United States
	  \and
	  B.~A.~Maruca \at
	  Bartol Research Institute, Department of Physics and Astronomy\\
	  University of Delaware\\
	  Newark, DE 19716\\
	  United States
	  }

\date{Received: 9 February 2019 / Accepted: 9 November 2019}

\maketitle

\begin{abstract}
The solar wind is a magnetized plasma and as such exhibits collective plasma behavior associated with its characteristic spatial and temporal scales. The characteristic length scales include the size of the heliosphere, the collisional mean free paths of all species, their inertial lengths, their gyration radii, and their Debye lengths. The characteristic timescales include the expansion time, the collision times, and the periods associated with gyration, waves, and oscillations. We review the past and present research into the multi-scale nature of the solar wind based on in-situ spacecraft measurements and plasma theory. We emphasize that couplings of processes across scales are important for the global dynamics and thermodynamics of the solar wind. We describe methods to measure in-situ  properties of particles and fields. We then discuss the role of expansion effects, non-equilibrium distribution functions, collisions, waves, turbulence, and kinetic microinstabilities for the multi-scale plasma evolution. 
\end{abstract}

\setcounter{tocdepth}{3}
\tableofcontents

\section{Introduction}\label{sec:intro}

The solar wind is a continuous magnetized plasma outflow that emanates from the solar corona. This extension of the Sun's outer atmosphere propagates through interplanetary space.  Its existence was first conjectured based on its interaction with planetary bodies in the solar system. Although the connection between solar activity and disturbances in the Earth's magnetic field had been established in the 19\textsuperscript{th} century \citep{sabine1851,sabine1852,hodgson1859,stewart1861}, the connection of these events with ``corpuscular radiation'' was not made until\ the early 20\textsuperscript{th} century \citep{birkeland1914,chapman1917}. The arguably first appearance of the notion of a continuous ``swarm of ions proceeding from the Sun'' in the literature dates back to a footnote by \citet{eddington1910} as an explanation for the observed shape of cometary tails. Later, \citet{hoffmeister1943} summarized multiple comet observations and suggested that some form of solar corpuscular radiation is responsible for the observed lag of comet ion tails with respect to the heliocentric radius vector \citep[for the link between solar activity and comet tails, see also][]{ahnert1943}. 
\citet{biermann1951} revisited the relation between comet tails and solar corpuscular radiation by quantifying the momentum transfer from the solar wind to cometary ions. He especially noted that the solar radiation pressure is insufficient to explain the observed structures \citep{milne1926} and that the corpuscular radiation is more variable than the electromagnetic radiation emitted by the Sun. The origin of the solar corpuscular radiation, however, remained unclear until \citet{parker1958} showed that a hot solar corona cannot maintain a hydrostatic equilibrium. Instead, the pressure-gradient force overcomes gravity and leads to a radial acceleration of the coronal plasma to supersonic velocities, which Parker called ``solar wind'' in contrast to a subsonic ``solar breeze''  \citep{chamberlain1961}, which was later found to be unstable \citep{velli1994}. Soon after this prediction, the solar wind was measured in situ by spacecraft \citep{gringauz1960,neugebauer1962}. For the last four decades, the solar wind has been monitored almost continuously in situ. 
Parker's underlying concept is the mainstream paradigm for the acceleration of the solar wind, but many questions remain unresolved. For example, we still have not identified the mechanisms that heat the solar corona to temperatures order of magnitude higher than the photospheric temperature, albeit this discovery was made some eighty years ago \citep{grotrian1939,edlen1943}. As we discuss the observed features of the solar wind in this review, we will encounter further deficiencies in our understanding that require more detailed analyses beyond Parker's model. In this process, we will find many observational facts that models of coronal heating and solar-wind acceleration must explain in order to achieve a realistic and consistent description of the physics of the solar wind. 

In the first section of this review, we lay out the various characteristic length and timescales in the solar wind and motivate our thesis that this multi-scale nature defines the evolution of the solar wind. We then introduce the observed large-scale, global features and the microphysical, kinetic features of the solar wind as well as the mathematical basis to describe the related processes.

\subsection{The characteristic scales in the solar wind}\label{sec:scales}

\begin{table}
\caption{The multiple characteristic plasma parameters (top), length scales (middle), and  timescales (bottom) in the solar wind. This table shows typical parameters in the solar wind at 1 au and in the upper solar corona ($\sim$ 100 Mm above photosphere). For each angular frequency $\omega$, the associated timescale is given by $\Pi_{\omega}\equiv 2\pi/|\omega|$.}
\label{tab_sw}
\begin{tabular}{llll}
    \noalign{\smallskip}\hline\noalign{\smallskip}    
Symbol & Solar Wind & (Upper) Corona & Definition \\ 
\noalign{\smallskip}\hline\noalign{\smallskip}
$n_{\mathrm p}$, $n_{\mathrm e}$ & $3\,\mathrm{cm}^{-3}$ & $10^6\,\mathrm{cm}^{-3}$ & proton and electron number density \\
$T_{\mathrm p}$, $T_{\mathrm e}$ & $10^5\,\mathrm{K}$ & $10^6\,\mathrm{K}$ & proton and electron temperature \\
$B$ & $3\times 10^{-5}\,\mathrm G$ &  1 G & magnetic field strength \\
\noalign{\smallskip}\hline\noalign{\smallskip}
$\lambda_{\mathrm{mfp,p}}$ & 3 au & 100 Mm & proton collisional mean free path \\ 
$L$ & 1 au & 100 Mm & characteristic size of the system\\
$d_{\mathrm p}$ & 140 km & 230 m & proton inertial length \\
$\rho_{\mathrm p}$ & 160 km & 13 m & proton gyration radius \\
$d_{\mathrm e}$ & 3 km & 5 m & electron inertial length \\
$\rho_{\mathrm e}$ & 2 km & 30 cm & electron gyration radius \\
$\lambda_{\mathrm p}$, $\lambda_{\mathrm e}$ & 12 m & 7 cm & proton and electron Debye lengths \\ 
\noalign{\smallskip}\hline\noalign{\smallskip}
$\Pi_{\nu_{\mathrm c}}$ & 120 d & 2 h & proton collision time \\
$\tau $ & 2.4 d & 10 min & expansion time \\
$\Pi_{\Omega_{\mathrm p}}$ & 26 s & $660\,\mu\mathrm s$& proton gyration period \\
$\Pi_{\omega_{\mathrm {pp}}}$ & $3\,\mathrm {ms}$ & $5\,\mu\mathrm s$ & proton plasma period\\
$\Pi_{\Omega_{\mathrm e}}$ & 14 ms & 360 ns & electron gyration period \\
$\Pi_{\omega_{\mathrm {pe}}}$ & $70\,\mu\mathrm s$ & 110 ns  & electron plasma period \\ \noalign{\smallskip}\hline\noalign{\smallskip}
\end{tabular}
\end{table}

\begin{figure}
 \includegraphics[width=\textwidth]{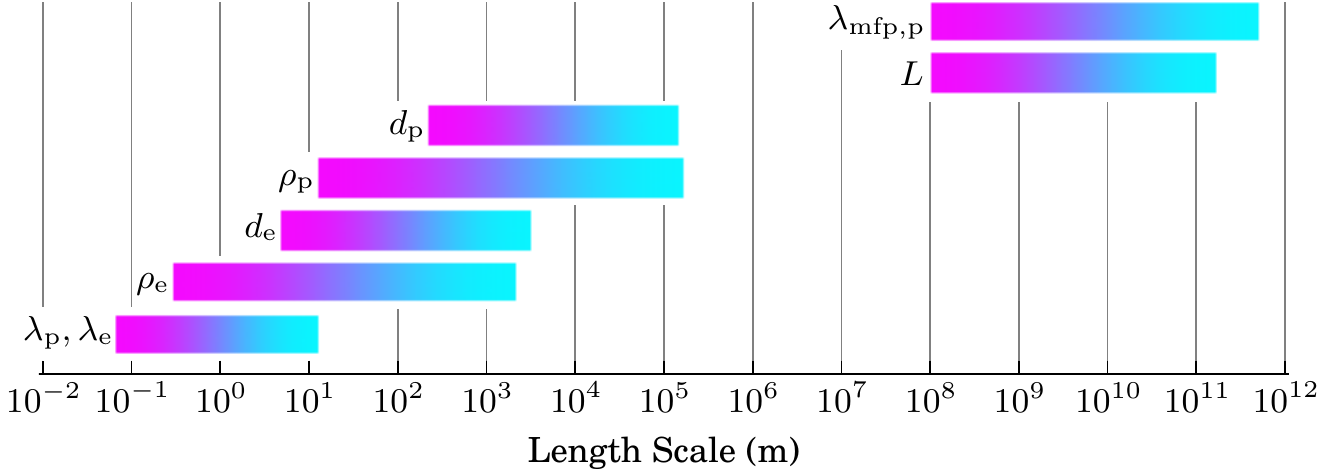}\vspace{0.5cm}
 \includegraphics[width=\textwidth]{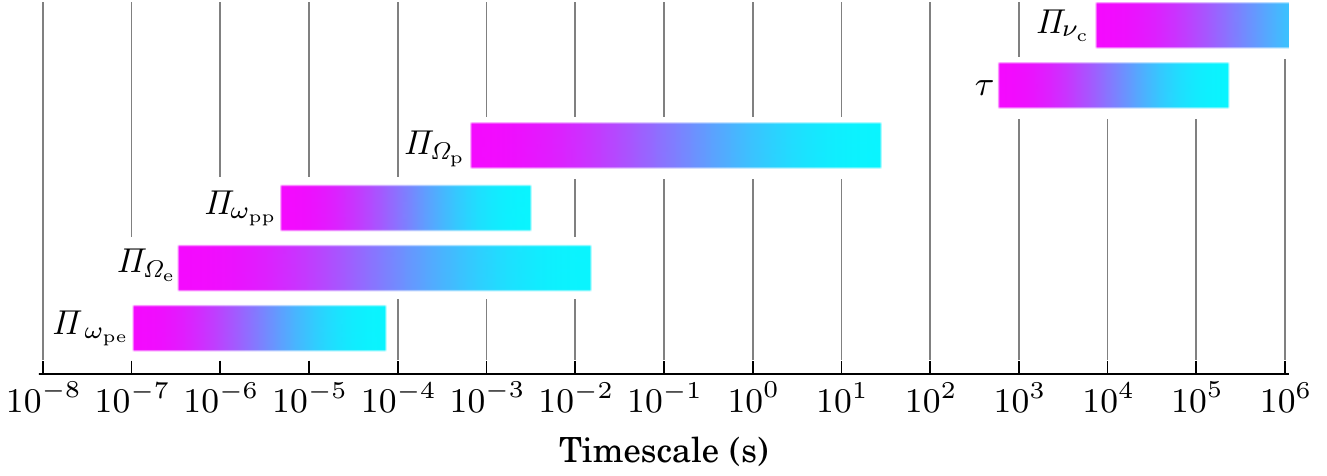}
\caption{Graphical representation of the characteristic length scales (top) and timescales (bottom) in the solar wind. The bar lengths represent the typical range for each scale given in Table~\ref{tab_sw}. The magenta end of each bar indicates the typical coronal value, and the cyan end of each bar indicates the typical value at 1~au.}
\label{fig:scales}       
\end{figure}

Table~\ref{tab_sw} lists typical values for the characteristic plasma parameters and scales in the solar wind at 1~au and in the upper solar corona that we introduce and define in this section. It is important to remember that all of these quantities vary widely in time and may differ significantly between thermal and superthermal particle populations.  We illustrate the broad range of the characteristic length scales and timescales in Fig.~\ref{fig:scales}.

The solar wind expands to a heliocentric distance of about 90 au, where it transitions to a subsonic flow by crossing the solar-wind termination shock \citep{stone2005,burlaga2008}. Although we do not expound upon the physics of the outer heliosphere and the interaction of the solar wind with the interstellar medium, this is the largest spatial scale in the supersonic solar wind. Considering the inner heliosphere (i.e., the spherical volume centered around the Sun within Earth's orbit), we identify the characteristic \emph{size of the system} as $L\sim 1\,\mathrm{au}$. For a typical radial solar-wind flow speed $U_r$ in the range of 300~km/s to 800~km/s \citep{lopez1986}, we find an \emph{expansion time} of
\begin{equation}\label{expansion_time}
\tau\sim \frac{L}{U_r}\sim 2.4\,\mathrm d
\end{equation}
for the solar wind from the Sun to 1~au. The Sun's siderial rotation period at its equator, 
\begin{equation}
\tau_{\mathrm{rot}}\sim 25\,\mathrm d,
\end{equation}
introduces another characteristic global timescale.

In addition to the outer size of the system, a plasma has multiple characteristic scales due to the interactions of its free charges with electric and magnetic fields. In a homogeneous and constant magnetic field $\vec B_0$, a plasma particle with charge $q_j$ and mass $m_j$ (where $j$ denotes the particle species) experiences a continuous deflection of its trajectory due to the Lorentz force. The frequency associated with this helical motion is given by the \emph{gyro-frequency}\footnote{Following the prevalent convention in space plasma physics, we adopt the metric system of Gaussian-cgs units. The NRL Plasma Formulary \citep{huba2016} includes a guide to converting formul\ae{} between cgs and SI units. In some figures, we plot magnetic field in nT for consistency with the published plots on which they are based.} (also called the \emph{cyclotron frequency})
\begin{equation}\label{gyrofreq}
\Omega_j\equiv \frac{q_jB_0}{m_jc},
\end{equation}
where $c$ is the speed of light in vacuum. The timescale for one closed loop around the magnetic field is then given by the \emph{gyro-period} $\Pi_{\Omega_j}\equiv 2\pi/|\Omega_j|$. In the solar wind at 1 au, $\Pi_{\Omega_{\mathrm p}}\sim 26\,\mathrm s$ and $\Pi_{\Omega_{\mathrm e}}\sim 14\,\mathrm {ms}$, where the index $\mathrm p$ represents protons and the index $\mathrm e$ represents electrons. On the other hand, in the upper corona (about 100 Mm above the photosphere), where the magnetic field is much stronger than in the solar wind, $\Pi_{\Omega_{\mathrm p}}\sim 660\,\mu\mathrm s$ and $\Pi_{\Omega_{\mathrm e}}\sim 360\,\mathrm{ns}$. Aside from protons, $\alpha$-particles (i.e., fully ionized helium atoms) are also dynamically important in the solar wind since they account for $\lesssim 20\%$ of the mass density.

We define the perpendicular thermal speed as
\begin{equation}\label{wperp}
w_{\perp j}\equiv\sqrt{\frac{2k_{\mathrm B}T_{\perp j}}{m_j}}
\end{equation} 
and the parallel thermal speed as
\begin{equation}\label{wpar}
w_{\parallel j}\equiv\sqrt{\frac{2k_{\mathrm B}T_{\parallel j}}{m_j}},
\end{equation}
where $T_{\perp j}$ ($T_{\parallel j}$) is the temperature of particle species $j$ in the direction perpendicular  (parallel) to $\vec B_0$ and $k_{\mathrm B}$ is the Boltzmann constant. We define the concept of temperatures perpendicular and parallel to $\vec B_0$ in Equations~(\ref{Tperp}) and (\ref{Tpar}). Assuming a thermal distribution of particles with a perpendicular thermal speed $w_{\perp j}$, the characteristic size of the gyration orbit is given by the \emph{gyro-radius}
\begin{equation}
\rho_j\equiv \frac{w_{\perp j}}{\left|\Omega_j\right|}.
\end{equation}
At 1~au, solar-wind gyro-radii are typically $\rho_{\mathrm p}\sim 160\,\mathrm{km}$ and $\rho_{\mathrm e}\sim 2\,\mathrm{km}$. In the upper corona, the gyro-radii are smaller: $\rho_{\mathrm p}\sim 13\,\mathrm{m}$ and $\rho_{\mathrm e}\sim 30\,\mathrm{cm}$.

The \emph{plasma frequency}
\begin{equation}\label{eq:plasmafreq}
\omega_{\mathrm pj}\equiv \sqrt{\frac{4\pi n_{0j}q_j^2}{m_j}},
\end{equation}
where $n_{0j}$ is the background number density of species $j$, corresponds to the characteristic timescale for electrostatic interactions in the plasma: $\Pi_{\omega_{\mathrm pj}}\equiv 2\pi/\omega_{\mathrm pj}$. In the solar wind at 1 au, $\Pi_{\omega_{\mathrm {pp}}}\sim 3\,\mathrm {ms}$, and $\Pi_{\omega_{\mathrm {pe}}}\sim 70\,\mu\mathrm s$. These timescales are even shorter in the corona: $\Pi_{\omega_{\mathrm {pp}}}\sim 5\,\mu\mathrm s$ and $\Pi_{\omega_{\mathrm {pe}}}\sim 110\,\mathrm {ns}$. A reduction of the local electron number density (e.g., through a spatial displacement of a number of electrons with respect to the ions) leads to an oscillation of the electrons with respect to the ions, in which the electrostatic force due to the displaced charge serves as the restoring force. This \emph{plasma oscillation} occurs with a frequency $\sim\omega_{\mathrm {pe}}$. In addition, light waves cannot propagate at frequencies $\lesssim \omega_{\mathrm {pe}}$ in a plasma as the free plasma charges shield the wave's electromagnetic fields so that the wave amplitude drops off exponentially with distance when the wave frequency is $\lesssim \omega_{\mathrm {pe}}$.  The exponential decay length associated with this shielding is given by the skin-depth $d_{\mathrm e}\equiv c/\omega_{\mathrm {pe}}$.

  More generally, we define the \emph{skin-depth} (also called the \emph{inertial length}) of species $j$ as
\begin{equation}
d_j\equiv\frac{c}{\omega_{\mathrm pj}}=\frac{v_{\mathrm Aj}}{|\Omega_j|},
\end{equation}
where 
\begin{equation}\label{Alfspeed}
v_{\mathrm Aj}\equiv \frac{B_0}{\sqrt{4\pi n_{0j} m_j}}
\end{equation}
is the \emph{Alfv\'en speed} of species $j$. In the solar wind at 1 au, $d_{\mathrm p}\sim 140 \,\mathrm{km}$, and $d_{\mathrm e}\sim 3\,\mathrm{km}$. In the upper corona, on the other hand, $d_{\mathrm p}\sim 230 \,\mathrm{m}$, and $d_{\mathrm e}\sim 5\,\mathrm{m}$. In processes that occur on length scales greater than $d_{\mathrm p}$ and timescales greater than $\Pi_{\Omega_{\mathrm p}}$, protons exhibit a \emph{magnetized behavior}, which means that their trajectory is closely tied to the magnetic field lines, following a quasi-helical gyration pattern with the frequency given in Equation~(\ref{gyrofreq}). Likewise, electrons exhibit magnetized behavior in processes that occur on length scales greater than $d_{\mathrm e}$ and timescales greater than $\Pi_{\Omega_{\mathrm e}}$. 

An important length scale associated with electrostatic effects is the \emph{Debye length}
\begin{equation}\label{intro:debye}
\lambda_j\equiv\sqrt{\frac{k_{\mathrm B}T_j}{4\pi n_{0j} q_j^2}},
\end{equation}
where $T_j$ is the (scalar, isotropic) temperature of species $j$. We note that $\lambda_{\mathrm p}\sim \lambda_{\mathrm e}$ through much of the heliosphere, which makes the Debye length unique among the scales we discuss. The total Debye length 
\begin{equation}\label{intro:debye_total}
\lambda_{\mathrm D}\equiv \left(\sum\limits_j\frac{1}{\lambda_j}\right)^{-1}
\end{equation}
is the characteristic exponential decay length for a time-independent global electrostatic potential in a plasma. In the solar wind at 1 au, $\lambda_{\mathrm p}\sim\lambda_{\mathrm e}\sim 12\,\mathrm m$, while the plasma in the upper corona exhibits $\lambda_{\mathrm p}\sim\lambda_{\mathrm e}\sim 7\,\mathrm{cm}$.  Collective plasma processes (i.e., particles behaving as if they only interact with a smooth macroscopic electromagnetic field rather than with individual moving charges) become important if the number of particles within a sphere of radius $\lambda_{\mathrm D}$ is large, 
\begin{equation}\label{plasmacond}
n_{0\mathrm e}\lambda_{\mathrm D}^3\gg 1,
\end{equation}
 and if 
 \begin{equation}\label{plasmacond2}
 \lambda_{\mathrm D}\ll L.
 \end{equation}
 Equations~(\ref{plasmacond}) and (\ref{plasmacond2}) guarantee that electrostatic single-particle effects are shielded by neighboring charges from the surrounding plasma (known as \emph{Debye shielding}). If either of these conditions is not fulfilled, common plasma-physics methods do not apply and a material is merely an ionized gas rather than a plasma. The solar wind, however, satisfies both of these conditions and, therefore, is a plasma. 

In addition to these collective plasma length scales and timescales, collisional effects are associated with with their own characteristic scales, which depend on the type of collisional interaction under consideration (e.g., temperature equilibration or isotropization) and on different combinations of plasma parameters. We discuss these effects and the associated timescales in Sect.~\ref{sec:col}.  

Comparing the coronal electron Debye length as the smallest plasma length scale of the solar wind with the size of the system reveals that the solar wind covers over twelve orders of magnitude in its characteristic length scales (neglecting length scales associated with collisions, which can be even greater than $L$). Similarly, comparing the corona's electron plasma period with the solar wind's expansion time reveals that the solar wind also covers over twelve orders of magnitude in its characteristic timescales (again neglecting timescales associated with collisions, which can be even greater than $\tau$). These ratios demonstrate the \emph{intrinsically multi-scale nature of the solar wind}. The broad range of scales also illustrates the difficulty in treating the solar wind and all related physics processes numerically since complete numerical simulations would need to resolve this entire range of scales.

This review describes plasma processes that depend upon or modify the multi-scale nature of the solar wind. As a truly Living Review, its first edition is limited to small-scale processes that affect the large-scale evolution of the plasma. In a later major update, we will describe how large-scale processes affect the small-scale structure of the plasma such as expansion effects on particle properties, wave reflection and the creation of turbulence, streaming interactions, mixing from different solar sources in co-rotating interaction regions, and magnetic focusing effects, as well as the impact of these processes on global solar-wind modeling. Although every plasma process is conceivably a multi-scale process, we, by practical necessity, only address the physics processes we consider most relevant to the multi-scale evolution of the solar wind. The most prominent processes \emph{not} covered in this review include detailed discussions of reconnection \citep{pontin2011,gosling2012,paschmann2013}, shock waves \citep{balogh1995,chashei1997,lepping2000,rice2003}, the physics of the outer heliosphere \citep[pick-up ions, energetic neutral atoms, etc.,][]{zank1995,gloeckler1998a,zank1999,richardson2004,mccomas2012,zank2018}, interplanetary dust \citep{krueger2007,mann2010}, interactions with planetary bodies \citep{grard1991,kivelson2007,gardini2011,bagenal2013}, eruptive events such as coronal mass ejections \citep{zurbuchen2006,howard2009,webb2012}, solar energetic particles \citep{ryan2000,mikic2006,klein2017}, and (anomalous) cosmic rays \citep{heber2006,potgieter2008,giacalone2012,potgieter2013}. We also limit our discussion of minor-ion physics.

\subsection{Global structure of the solar wind}\label{sec:global}

\begin{figure}
  \includegraphics[width=\textwidth]{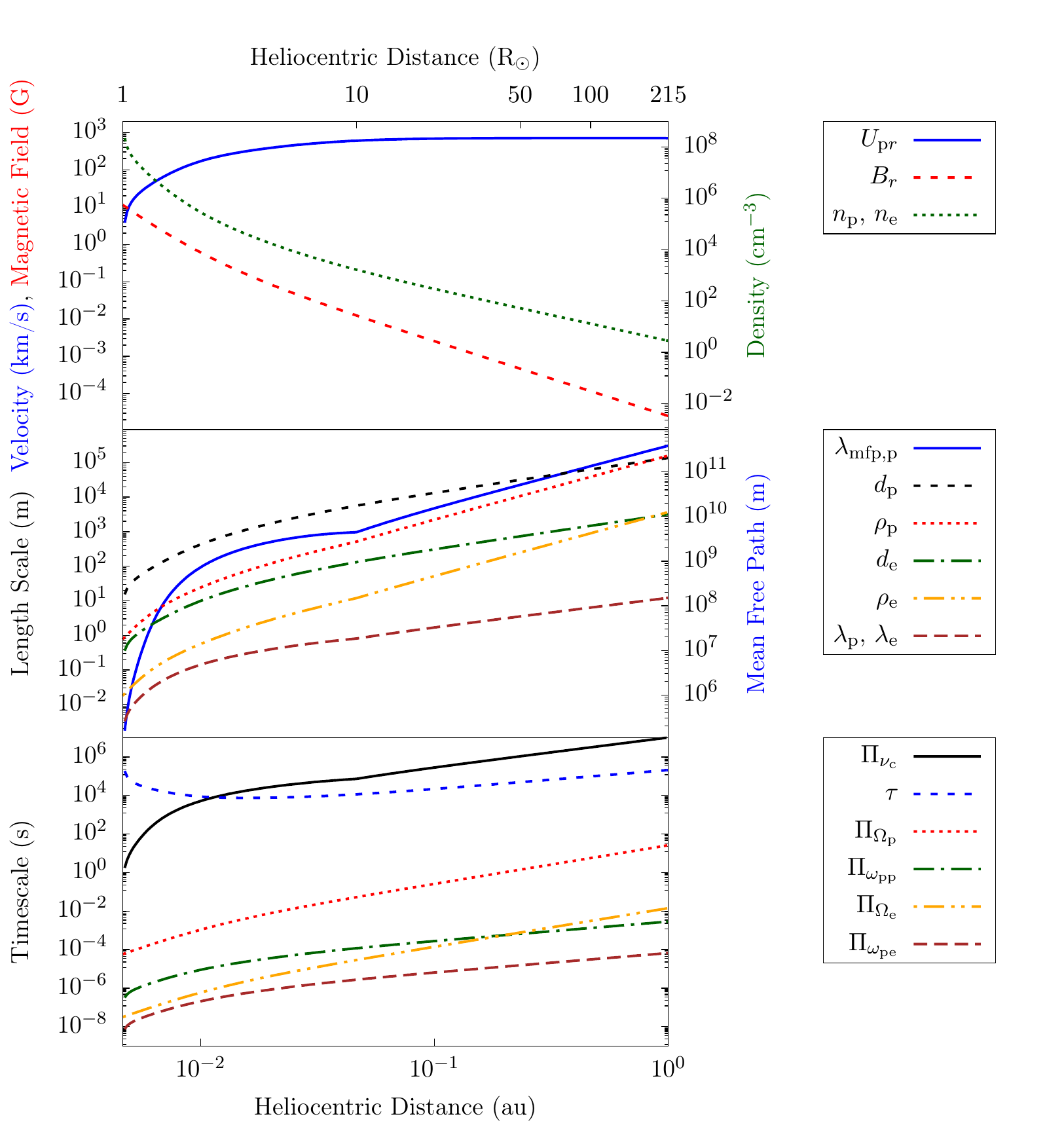}
\caption{Characteristic average quantities, length scales, and timescales as functions of distance from the Sun in the inner heliosphere for typical fast-solar-wind conditions. We calculate these scales based on typical radial profiles of the solar-wind magnetic-field strength, density, and velocity (shown in the top panel). The profiles for the magnetic field and the density are taken from \citet{smith2012} for a radial polar flux tube. The radial velocity profile then follows from flux conservation, $n_{j}U_{jr}/B_r=\mathrm{constant}$. The electron temperature is taken from a fit to measurements at $r< 10R_{\odot}$ \citep{cranmer1999} and then connected to a power-law with a power index corresponding to the radial temperature profiles observed with Helios in the fast solar wind \citep{stverak2015}. We take $T_{\mathrm p}\approx T_{\mathrm e}$ for simplicity.}
\label{fig:radial_profiles}       
\end{figure}

At heliocentric distances greater than a few solar radii $R_{\odot}$, the solar wind's expansion is, to first order, radial, which creates large-scale radial gradients in most of the plasma parameters. For this discussion of the global structure, we concentrate only on long-term averages of the plasma quantities and neglect their frequent -- and, as we will see later, sometimes comparable to order unity -- variations. Figure~\ref{fig:radial_profiles} illustrates these average quantities as functions of distance in the inner heliosphere and demonstrates the resulting profiles for the characteristic length scales and timescales. Beyond a distance of about $10\,R_{\odot}$, the average radial velocity stays approximately constant. Continuity under steady-state conditions requires that 
\begin{equation}
\nabla\cdot \left(n_{j}\vec U_{j}\right) = 0,
\end{equation}
where $\vec U_j$ is the bulk velocity of species $j$. In spherical coordinates and under the assumption that $\vec U_j\approx U_{jr} \hat{\vec e}_r\approx \mathrm{constant}$, the average density then decreases $\propto r^{-2}$. In the acceleration region and in regions of super-radial expansion connected to coronal holes, continuity requires steeper gradients closer to the Sun as confirmed by white-light polarization measurements \citep{cranmer2005}. In addition, the deceleration of streaming $\alpha$-particles leads to a small deviation from the $r^{-2}$ density profile \citep{verscharen2015}.

To first order, the average magnetic field follows the Parker spiral in the plane of the ecliptic \citep{parker1958,levy1976,behannon1978,mariani1978,mariani1979}
as a result of the frozen-in condition of ideal magnetohydrodynamics (MHD; see Sect.~\ref{sec:MHD}) and the rotation of the Sun. We define
\begin{equation}
\beta_{j}\equiv\frac{8\pi n_{j} k_{\mathrm B}T_j}{B^2},
\end{equation}
where $B$ is the magnetic field, as the ratio between the thermal pressure of species $j$ and the magnetic pressure. In the solar corona, $\beta_j\ll 1$, so that 
the magnetic field constraints the plasma to co-rotate with the Sun. However, the magnetic field's torque on the plasma decreases with distance from the Sun until the plasma outflow dominates the evolution of the magnetic field and convects the field into interplanetary space \citep{weber1967}. 
In the Parker model, the \emph{Parker angle} $|\phi_{Br}|$ between the direction of the magnetic field and the radial direction increases with distance $r$ from the Sun, 
\begin{equation}\label{tantheta}
\tan \,\phi_{Br}=\frac{B_{\phi}}{B_r}=\frac{\Omega_{\odot}\sin \theta}{U_{\mathrm p r}}\left(r_{\mathrm{eff}}-r\right),
\end{equation}
where $B_{\phi}$ and $B_{r}$ are the azimuthal and radial components of the magnetic field, $\Omega_{\odot}$ is the angular speed of the Sun's rotation, $\theta$ is the polar angle, and $r_{\mathrm{eff}}$ is the effective co-rotation radius. In our sign and coordinate convention, $\phi_{Br}\le 0$ if $B_r>0$ since the Sun rotates in the $+\hat{\vec e}_{\phi}$-direction, which differs from Parker's (\citeyear{parker1958}) original choice. The radius $r_{\mathrm{eff}}$ is an auxiliary quantity to describe the heliospheric distance beyond which the solar wind behaves as if it were co-rotating for $r\le r_{\mathrm{eff}}$ \citep{hollweg1989}. Observations indicate that $r_{\mathrm{eff}}\sim 10R_{\odot}$ in the fast wind and $r_{\mathrm{eff}}\sim 20R_{\odot}$ in the slow wind \citep{bruno1997}. The Parker angle $|\phi_{Br}|$ increases from $0^{\circ}$ at $r_{\mathrm{eff}}$ to about $45^{\circ}$ at $r=1\,\mathrm{au}$. This trend continues into the outer heliosphere as shown by observations \citep{thomas1980,forsyth2002}. The magnitude of the \emph{Parker field} decreases with distance as
\begin{equation}\label{Bnull}
B_0\propto \frac{\sqrt{1+\tan^2\,\phi_{Br}}}{r^2},
\end{equation}
which is $\propto r^{-2}$ in the limit $\tan^2\phi_{Br}\ll 1$ at small $r$ and $\propto r^{-1}$  in the limit  $\tan^2\phi_{Br}\gg 1$ at large $r$.
We note that the original Parker model is not completely torque-free, although a torque-free treatment leads to only minor modifications \citep{verscharen2015}. Further details about the heliospheric magnetic field can be found in the review by \citet{owens2013}.

\begin{figure}
  \includegraphics[width=\textwidth]{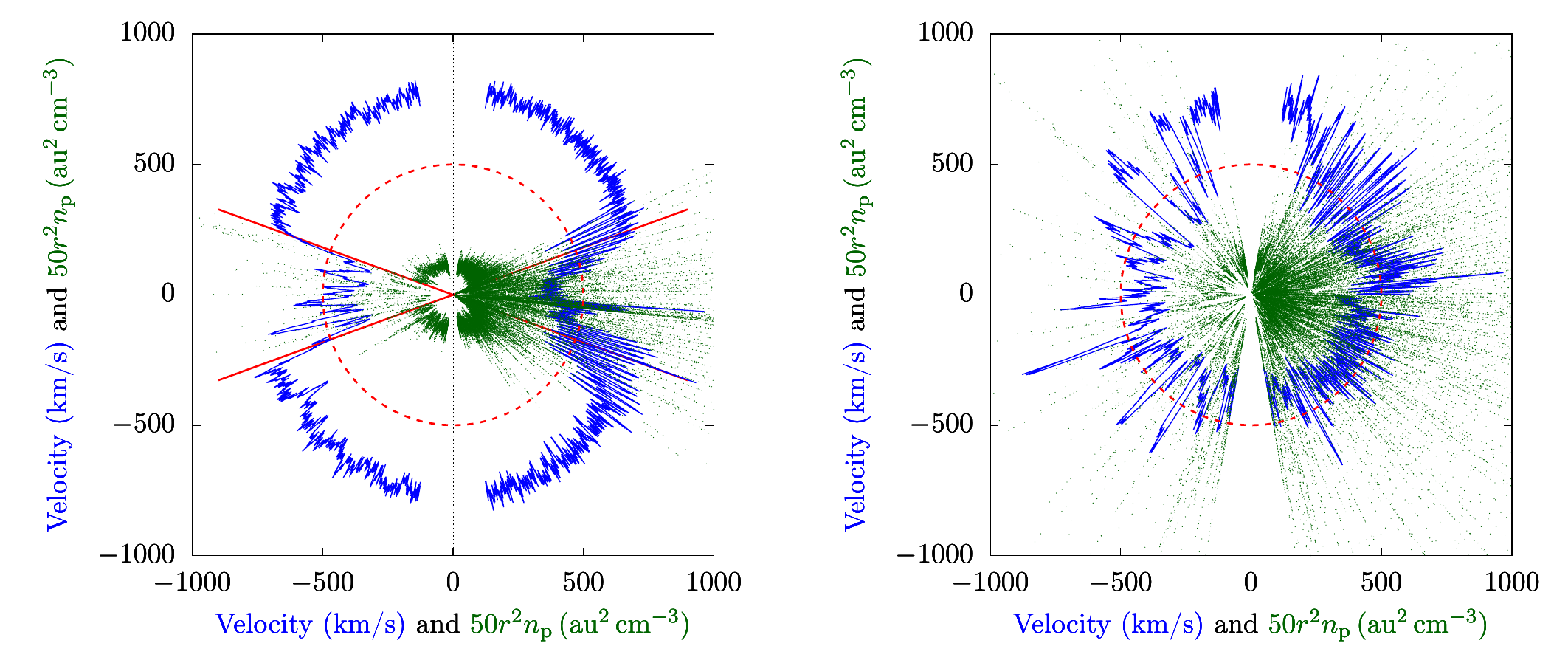}
\caption{Ulysses/SWOOP observations of the solar-wind proton radial velocity and density at different heliographic latitudes. The distance from the center in each of these polar plots indicates the velocity (blue) and density (green). The polar angle represents the heliographic latitude. Since these measurements were taken at varying distances from the Sun, we compensate for the density's radial decrease by multiplying $n_{\mathrm p}$ with $r^2$. The red circle represents $U_{\mathrm pr}=500\,\mathrm{km/s}$ and $r^2n_{\mathrm p}=10\,\mathrm{au}^2\,\mathrm{cm}^{-3}$. The straight red lines  indicate the sector boundaries at $\pm 20^{\circ}$ latitude. Left panel: Ulysses' first polar orbit during solar minimum (1990-12-20 through 1997-12-15). Right panel: Ulysses' second polar orbit during solar maximum (1997-12-15 through  2004-02-22). After \citet{mccomas2000} and \citet{mccomas2008}. }
\label{fig:polar_plots}       
\end{figure}

\subsection{Categorization of solar wind }

Traditionally, the solar wind has been categorized into three groups  \citep{srivastava2000}: 
\begin{enumerate}
\item \emph{fast wind} with bulk velocities between about 500 km/s and 800 km/s, 
\item \emph{slow wind} with bulk velocities between about 300 km/s and 500 km/s, and \item variable/eruptive events such as \emph{coronal mass ejections} with speeds from a few hundreds up to 2000 km/s.
\end{enumerate}
Measurements from the Ulysses spacecraft during solar minimum dramatically demonstrate that the fast wind emerges predominantly from polar coronal holes and the slow wind from the streamer belt at the solar equator \citep{phillips1995,mccomas1998,mccomas2000,mccomas2003,ebert2009}.  The left-hand panel in Fig.~\ref{fig:polar_plots} illustrates the clear sector boundary between fast and slow wind during solar minimum. During solar maximum, however, fast and slow wind emerge from neighboring patches everywhere in the corona. The right-hand panel  in Fig.~\ref{fig:polar_plots} shows that the occurrence of fast and slow wind streams does not strongly correlate with heliographic latitude during solar maximum. On average, fast polar wind exhibits both a lower density and less variation in density than slow wind. The association of different wind streams with different source regions suggests that the magnetic-field configuration in the corona plays a crucial role in determining the properties of the wind streams. In addition to the differences in speed and density, fast and slow wind exhibit further distinguishing marks. Fast wind, relative to slow wind, generally is more steady, is more Alfv\'enic \citep[i.e., it exhibits a higher correlation or anti-correlation between fluctuations in vector velocity and vector magnetic field; see Sect.~\ref{sec:waves} and][]{tu1995}, and has a higher proton temperature \citep{neugebauer1976,wilson2018}. Importantly for its multi-scale evolution, fast wind is also less collisional (both in terms of the local collisional relaxation times and the cumulative time for collisions to act) than slow wind  \citep{marsch1982,marsch1983,livi1986,kasper2008,bourouaine2011,durovcova2017}, which allows for more kinetic non-equilibrium features to survive the thermalizing action of Coulomb collisions. Fast wind, therefore, exhibits more non-Maxwellian structure in its distribution functions \citep{marsch2006,marsch2018} as we discuss in the next section. 

The elemental composition and the heavy-ion charge states also differ between fast and slow wind \citep{bame1975,ogilvie1995,vonsteiger1995,bochsler2000,vonsteiger2000,aellig2001,zurbuchen2002,kasper2007,kasper2012,lepri2013}. 
Elements with a low \emph{first ionization potential (FIP)} such as magnesium, silicon, and iron exhibit enhanced abundances in the solar corona and in the solar wind with respect to their photospheric abundances \citep{gloeckler1989,raymond1999,laming2015}. Conversely, elements with a high FIP such as oxygen, neon, and helium have much lower enhancements or even depletions with respect to their photospheric abundances. This FIP fractionation bias also varies with wind speed and is generally smaller in fast wind than in slow wind \citep{zurbuchen1999,bochsler2007}. Since the elemental composition of a plasma parcel does not change as it propagates through the heliosphere unless it mixes with neighboring parcels, composition measurements are a reliable method to distinguish solar-wind source regions. 
Moreover, studies of heavy ions constrain proposed models of solar-wind acceleration and heating. For instance, proposed acceleration and heating scenarios must explain the observed preferential heating of minor ions. In the solar wind, most heavy ion species $i$ exhibit $T_i/T_{\mathrm p}\approx 1.35 m_i/m_{\mathrm p}$ \citep{tracy2015,heidrich2016,tracy2016}. 

Lately, the traditional classification of wind streams by speed has experienced some major criticism \citep[e.g.,][]{maruca2013,xu2015,camporeale2017}. Speed alone does not fully classify the properties of the wind, and there is a smooth transition in the distribution of wind speeds. At times, fast solar wind shows properties traditionally associated with slow wind and vice versa, such as collisionality, Alfv\'enicity, FIP-bias, anisotropy, beam structures, etc. Although these atypical behaviors suggest a false dichotomy between fast and slow wind, we retain the traditional nomenclature, albeit defining ``fast wind'' as wind with the typical fast-wind properties and ``slow wind'' as wind with the typical slow-wind properties under consideration instead of relying on the flow speeds alone. Nevertheless, we expressly caution the reader against assuming wind speed alone as a reasonable indication of wind  type.

\subsection{Kinetic properties of the solar wind}\label{sec:kinetic}

Kinetic plasma physics describes the statistical properties of a plasma by means of the \emph{particle velocity distribution functions} $f_j(\vec x,\vec v,t)$ for each plasma species $j$. We define and normalize the distribution function so that
\begin{equation}
f_j(\vec x,\vec v,t)\,\mathrm d^3\vec x\,\mathrm d^3\vec v
\end{equation}
represents the number of particles of species $j$ in the phase-space volume $\mathrm d^3\vec x\,\mathrm d^3\vec v$ centered on the phase-space coordinates $(\vec x,\vec v)$ at time $t$. The distribution function relates to the bulk properties (i.e., density, bulk velocity, temperature, \dots) through its velocity moments as described in Sect.~\ref{sec:moments}.  A continuous definition of $f_j$ is appropriate when Equation~(\ref{plasmacond}) is fulfilled. 

The central equation in kinetic physics is the \emph{Boltzmann equation}, 
\begin{equation}\label{boltzmann}
\frac{\partial f_j}{\partial t}+\vec v\cdot \frac{\partial f_j}{\partial \vec x}+\vec a\cdot \frac{\partial f_j}{\partial \vec v}=\left(\frac{\delta f_j}{\delta t}\right)_{\mathrm c},
\end{equation}
where $\vec a$ is the acceleration of a $j$-particle due to macroscopic forces, and the right-hand side describes the temporal change in $f_j$ due to particle collisions,  which are mediated by microscopic electric forces among individual particles \citep[see also Sect.~\ref{sec:col:theory} of this review;][]{lifshitz1981}. We use the term \emph{macroscopic fields} to indicate that these are locally averaged to remove the rapidly fluctuating Coulomb electric fields due to individual charges, which are responsible for Coulomb collisions. The applicability of this mean-field approach is a key quality of a plasma and distinguishes it from other types of ionized gases, in which Equation~(\ref{plasmacond}) is not fulfilled. Without the collision term, the Boltzmann equation represents a fluid continuity equation for the density in phase space. It is thus related to \emph{Liouville's theorem} and describes the conservation of the phase-space density along trajectories in the absence of collisions.\footnote{We refrain from discussing the multiple ways of deriving the Boltzmann equation such as the closure of the BBGKY hierarchy \citep{bogoliubov1946} or the Klimontovich--Dupree formalism \citep{dupree1961,klimontovich1967}. Instead, we express the Boltzmann equation in terms of Liouville's theorem and subsume all higher-order particle interactions in the collision term on the right-hand side of Equation~(\ref{boltzmann}). For more details, see also Sect.~\ref{sec:col:theory}.} In this case, and when using only macroscopic electromagnetic forces in the acceleration term, we obtain the \emph{Vlasov equation},
\begin{equation}\label{vlasov}
\frac{\partial f_j}{\partial t}+\vec v\cdot \frac{\partial f_j}{\partial \vec x}+\frac{q_j}{m_j}\left(\vec E+\frac{1}{c}\vec v\times \vec B\right)\cdot \frac{\partial f_j}{\partial \vec v}=0,
\end{equation} 
which is the fundamental equation of collisionless kinetic plasma physics. 
These macroscopic electric and magnetic fields obey \emph{Maxwell's equations},
\begin{align}
\nabla\cdot \vec E&=4\pi \rho_{\mathrm c}, \\ \label{Maxw1full}
\nabla \cdot \vec B&=0,\\
\nabla \times \vec E&=-\frac{1}{c}\frac{\partial \vec B}{\partial t}, \label{Maxw4full}
\end{align}
and
\begin{equation}
\nabla \times \vec B=\frac{4\pi }{c}\vec j+\frac{1}{c}\frac{\partial \vec E}{\partial t},\label{Maxw5full}
\end{equation}
where the charge density $\rho_{\mathrm c}$ and the current density $\vec j$ are given by integrals over the distribution functions as
\begin{equation}\label{chargedens}
\rho_{\mathrm c}=\sum \limits_j q_j \int f_j\,\mathrm d^3\vec v
\end{equation}
and
\begin{equation}\label{currentdens}
\vec j=\sum \limits_j q_j \int \vec v f_j\,\mathrm d^3\vec v.
\end{equation}
Equations~(\ref{vlasov}) through (\ref{currentdens}) form a closed set of integro-differential equations in six-dimensional phase space and time that fully describe the evolution of collisionless plasma.

\subsubsection{Fluid moments and fluid equations}\label{sec:moments}

Although the distribution functions $f_j$ contain all of the microphysical properties of the plasma, it is often sufficient to rely on a reduced set of macrophysical parameters that only depend on time and three-dimensional configuration space (versus time and six-dimensional phase space). These parameters are called \emph{bulk parameters} and correspond to the velocity moments as integrals over the full velocity space of the distribution function. Certain velocity moments represent named fluid bulk parameters. For instance, the zeroth velocity moment corresponds to the \emph{number density}
\begin{equation}
n_j=\int f_j\,\mathrm d^3\vec v.
\end{equation}
Using $n_j$, the first velocity moment corresponds to the \emph{bulk velocity}
\begin{equation}\label{bulkvel}
\vec U_j=\frac{1}{n_j}\int \vec v f_j\,\mathrm d^3\vec v,
\end{equation}
while the second moment represents the \emph{pressure tensor}
\begin{equation}\label{pressuretensor}
\mathsf P_j=m_j\int \left(\vec v-\vec U_j\right)\left(\vec v-\vec U_j\right) f_j\,\mathrm d^3\vec v.
\end{equation}
The third moment corresponds to the \emph{heat-flux tensor}
\begin{equation}\label{heatflux}
\mathsf Q_j=m_j\int \left(\vec v-\vec U_j\right) \left(\vec v-\vec U_j\right) \left(\vec v-\vec U_j\right)  f_j\,\mathrm d^3\vec v.
\end{equation}

For many applications in magnetized-plasma physics, it is useful to choose the coordinate system to be aligned with the direction  $\hat{\vec b}\equiv \vec B/|\vec B|$ of the magnetic field and to define the pressure components with respect to the direction of the magnetic field. In this coordinate system, Equation~({\ref{heatflux}}) reduces through contraction to the perpendicular heat-flux vector
\begin{equation}
\vec q_{\perp j}=\frac{1}{2}\mathsf Q_j : \left(\mathsf I_3-\hat{\vec b}\hat{\vec b}\right)
\end{equation}
and the parallel heat-flux vector
\begin{equation}
\vec q_{\parallel j}=\mathsf Q_j : \left(\hat{\vec b}\hat{\vec b}\right),
\end{equation}
where $\mathsf I_3$ is the three-dimensional unit matrix. We define the double-dot and triple-dot products in a similar way to the usual dot product as
\begin{equation}
\mathsf A:\mathsf B=\sum\limits_{i,j}\mathsf A_{ij}\mathsf B_{ji}\qquad \text{and} \qquad \mathsf A \dot{:} \mathsf B=\sum\limits_{i,j,k}\mathsf A_{ijk}\mathsf B_{kji}.
\end{equation}
Although higher moments do not give rise to named bulk parameters like these four, the \emph{moment hierarchy} can be continued to infinity by multiplying the integrand with further powers of velocity. 

Taking velocity moments of the full Vlasov equation and exploiting the definitions of the lowest moments above leads to the \emph{multi-fluid plasma equations} \citep{barakat1982,marsch2006}. The zeroth and first moments of the Vlasov equation are the \emph{continuity equation},
\begin{equation}
\frac{\partial n_j}{\partial t}+\nabla\cdot \left(n_j\vec U_j\right) = 0,
\end{equation}
and the \emph{momentum equation},
\begin{equation}\label{fluidmomentum}
n_jm_j\left(\frac{\partial}{\partial t}+\vec U_j\cdot \nabla\right)\vec U_j=-\nabla \cdot \mathsf P_j+n_j q_j\left(\vec E+\frac{1}{c}\vec U_j\times \vec B\right).
\end{equation}
We define the perpendicular pressure and the parallel pressure as 
\begin{equation}
p_{\perp j}\equiv\mathsf P_j:\frac{\mathsf I_3-\hat{\vec b}\hat{\vec b}}{2}
\end{equation}
 and 
 \begin{equation}
 p_{\parallel j}\equiv\mathsf P_j:\left(\hat{\vec b}\hat{\vec b}\right),
 \end{equation}
  respectively, which are related to the temperatures in the directions perpendicular and parallel to $\vec B$ through 
  \begin{equation}\label{Tperp}
    T_{\perp j}=\frac{p_{\perp j}}{n_j k_{\mathrm B}}  
  \end{equation}
  and
  \begin{equation}\label{Tpar}
    T_{\parallel j}=\frac{p_{\parallel j}}{n_j k_{\mathrm B}}  
  \end{equation}
  We  write the \emph{perpendicular energy equation} as
\begin{multline}
\left(\frac{\partial}{\partial t}+\vec U_j\cdot \nabla\right)p_{\perp j}+p_{\perp j}\left(\nabla\cdot \vec U_j+\nabla_{\perp}\cdot  \vec U_j\right)=\left(\hat{\vec b}\hat{\vec b}-\mathsf I_3\right):\left(\vec{\tau}_j\cdot \nabla \vec U_j\right)\\
-\nabla\cdot \vec q_{\perp j}-\frac{1}{2}\vec{\tau}_j:\left(\frac{\partial}{\partial t}+\vec U_j\cdot \nabla\right)\left(\hat{\vec b}\hat{\vec b}\right)-\frac{1}{2}\mathsf Q_j \dot{:} \nabla\left(\hat{\vec b}\hat{\vec b}\right)
\end{multline}
and the \emph{parallel energy equation} as
\begin{multline}
\left(\frac{\partial}{\partial t}+\vec U_j\cdot \nabla\right)p_{\parallel j}+p_{\parallel j}\left(\nabla\cdot \vec U_j+2\nabla_{\parallel}\cdot \vec U_j\right)=-2\hat{\vec b}\hat{\vec b}:\left(\vec{\tau}_j\cdot \nabla\vec U_j\right)\\
-\nabla\cdot \vec q_{\parallel j}+\vec{\tau}_j:\left(\frac{\partial}{\partial t}+\vec U_j\cdot \nabla\right)\left(\hat{\vec b}\hat{\vec b}\right)+\mathsf Q_j \dot{:}  \nabla\left(\hat{\vec b}\hat{\vec b}\right),
\end{multline}
where 
\begin{equation}
\vec \tau_j\equiv \mathsf P_j-p_{\perp j}\mathsf I_3-\left(p_{\parallel j}-p_{\perp j}\right)\hat{\vec b}\hat{\vec b}
\end{equation}
is the stress tensor, 
\begin{equation}
\nabla_{\perp}\equiv\left(\mathsf I_3-\hat{\vec b}\hat{\vec b}\right)\nabla,\qquad \text{and}\qquad  \nabla_{\parallel}\equiv\left(\hat{\vec b}\hat{\vec b}\right)\nabla. 
\end{equation}
The hierarchy of moments of the Vlasov equation continues to infinity, and similar fluid equations exist for the stress tensor, the heat-flux tensor, and all higher-order moments. However, this gives rise to a closure problem since the $n$\textsuperscript{th} moment of the Vlasov equation always includes the $(n+1)$\textsuperscript{st} moment of the distribution function. For example, the continuity equation, which is the zeroth moment of the Vlasov equation, includes the bulk velocity, which corresponds to the first moment of $f_j$. The $(n+1)$\textsuperscript{st} moment of the distribution function, in turn, requires the $(n+1)$\textsuperscript{st} moment of the Vlasov equation as a description of its dynamical evolution. Every fluid model is, therefore, fundamentally susceptible to a closure problem since the solution of an infinite chain of non-degenerate equations is formally impossible. For most practical purposes, the moment hierarchy is thus truncated by expressing a higher-order moment of $f_j$ through lower moments of $f_j$ only. Closing the moment hierarchy introduces limitations on the physics of the problem at hand and deviations in the solutions to  the multi-fluid system of equations from the solutions to the full Vlasov equation. For example, a typical closure of the moment hierarchy is the assumption of an isotropic and adiabatic pressure, i.e., $\mathsf P_j=p_j\,\mathsf I_3$ and $p_j\propto n_j^{\kappa}$, where $\kappa$ is the adiabatic exponent. This closure of the momentum equation neglects heat flux and small velocity-space structure in $f_j$. Therefore, any finite closure is only applicable if the physics of the problem at hand justifies the neglect of higher-order velocity moments of $f_j$. We note, for example, that collisions are such a process that can produce conditions under which higher-order moments are negligible (see Sect.~\ref{sec:col}).

Assuming only slow changes of the magnetic field compared to $\Pi_{\Omega_j}$ and that $\vec{\tau}_j=0$, the second velocity moment of the Vlasov equation (\ref{vlasov}) leads to the useful  \emph{double-adiabatic energy equations} \citep{chew1956,whang1971,sharma2006,chandran2011},
\begin{equation}\label{cgl1}
n_j B\left(\frac{\partial}{\partial t}+\vec U_j\cdot \nabla\right)\left(\frac{p_{\perp j}}{n_jB}\right)=-\nabla \cdot \vec q_{\perp j}-q_{\perp j}\nabla\cdot \hat{\vec b}
\end{equation}
and
\begin{equation}\label{cgl2}
\frac{n_j^3}{B^2}\left(\frac{\partial}{\partial t}+\vec U_j\cdot \nabla\right)\left(\frac{B^2p_{\parallel j}}{n_j^3}\right)=-\nabla\cdot \vec q_{\parallel j}+2q_{\perp j}\nabla\cdot \hat{\vec b}.
\end{equation}
If we neglect heat flux by setting the right-hand sides of Equations~(\ref{cgl1}) and (\ref{cgl2}) to zero, we obtain the conservation laws for the \emph{double-adiabatic invariants}, which are also referred to as the \emph{Chew--Goldberger--Low (CGL) invariants} \citep{chew1956}
\begin{equation}
\frac{p_{\perp j}}{n_jB}\approx\mathrm{constant}\qquad \text{and} \qquad \frac{B^2p_{\parallel j}}{n_j^3}\approx\mathrm{constant.}
\end{equation}

\subsubsection{Magnetohydrodynamics}\label{sec:MHD}

Magnetohydrodynamics (MHD) is a single-fluid description that results from summing the fluid equations of all species and defining the moments of the single \emph{magnetofluid} as the mass density
\begin{equation}
\rho\equiv \sum\limits_jm_jn_j,
\end{equation}
the bulk velocity
\begin{equation}
\vec U\equiv\frac{1}{\rho}\sum \limits_j m_jn_j\vec U_j,
\end{equation}
and the total scalar pressure
\begin{equation}
P\equiv \frac{1}{3} \sum\limits_j \mathsf P_j : \mathsf I_3
\end{equation}
under the assumption that $\mathsf P_j$ is isotropic and diagonal. This procedure leads to the \emph{MHD continuity equation},
\begin{equation}
\frac{\partial \rho}{\partial t}+\nabla\cdot\left(\rho\vec U\right)=0,
\end{equation}
and the \emph{MHD momentum equation},
\begin{equation}\label{MHDmoment}
\rho \left(\frac{\partial }{\partial t}+\vec U\cdot \nabla \right)\vec U=-\nabla P+\frac{1}{c}\left(\vec j\times \vec B\right).
\end{equation}
The electric-field term from Equation~(\ref{fluidmomentum}) vanishes under the quasi-neutrality assumption that $\rho_{\mathrm c}$ from Equation~(\ref{chargedens}) is negligible, which is justified on scales $\gg \lambda_{\mathrm D}$. Faraday's law describes the evolution of the magnetic field as
\begin{equation}\label{MHDfaraday}
\frac{\partial \vec B}{\partial t}=-c\nabla\times \vec E.
\end{equation}
The electric field follows from the electron momentum equation (\ref{fluidmomentum}) as the generalized Ohm's law,
\begin{equation}\label{ohm_general}
\vec E=\frac{m_{\mathrm e}}{q_{\mathrm e}}\left(\frac{\partial}{\partial t}+\vec U_{\mathrm e}\cdot \nabla\right)\vec U_{\mathrm e}+\frac{1}{n_{\mathrm e}q_{\mathrm e}}\nabla\cdot \mathsf P_{\mathrm e}-\frac{1}{n_{\mathrm e}q_{\mathrm e}c}\vec j\times \vec B+\frac{1}{n_{\mathrm e}q_{\mathrm e}c}\vec j_{\mathrm i}\times \vec B,
\end{equation}
where 
\begin{equation}
\vec j_{\mathrm i}\equiv \vec j-n_{\mathrm e}q_{\mathrm e}\vec U_{\mathrm e}
\end{equation}
  is the ion contribution to the current density. The terms on the right-hand side of Equation~(\ref{ohm_general}) represent the contributions from electron inertia, the electron pressure gradient (i.e., the ambipolar electric field), the Hall term, and the ion convection term, respectively. Under the assumptions of quasi-neutrality in a proton--electron plasma and the negligibility of terms of order $m_{\mathrm e}/m_{\mathrm p}$, we find
  \begin{equation}\label{ohm_next}
 \vec E=\frac{1}{n_{\mathrm e}q_{\mathrm e}}\nabla\cdot \mathsf P_{\mathrm e}-\frac{1}{n_{\mathrm e}q_{\mathrm e}c}\vec j\times \vec B-\frac{1}{c}\vec U\times \vec B.
  \end{equation}
 If we furthermore assume small or moderate $\beta_{\mathrm e}$ and consider processes occurring on scales $\gg d_{\mathrm p}$ \citep{chiuderi2015}, we can neglect the contributions of the electron pressure gradient  and the Hall term to $\vec E$. We then 
  find the common expression for Ohm's law in MHD:
\begin{equation}\label{MHDOhm}
\vec E=-\frac{1}{c}\vec U\times \vec B.
\end{equation}
Equations~(\ref{MHDfaraday}) and (\ref{MHDOhm}) describe \emph{Alfv\'en's frozen-in theorem}, stating that magnetofluid bulk motion across field lines is forbidden, since otherwise the infinite resistivity of the magnetofluid would lead to infinite eddy currents. Instead, the magnetic flux through a co-moving surface is conserved.\footnote{Interestingly, the inclusion of the pressure-gradient term from Equation~(\ref{ohm_next}) in Equation~(\ref{MHDOhm}) does not affect the frozen-in condition since it cancels when taking the curl in Equation~(\ref{MHDfaraday}).}
The assumptions leading to Equation~(\ref{MHDOhm}) are fulfilled for processes on time scales much greater than $\Pi_{\Omega_{j}}$ and $\Pi_{\omega_{\mathrm pj}}$ as well as on spatial scales much greater than $d_{j}$ and $\rho_{j}$. In this limit, the displacement current in Amp\`ere's law is also negligible, which allows us to write the current density in Equation~(\ref{MHDmoment}) in terms of the magnetic field:
\begin{equation}
\vec j=\frac{c}{4\pi}\nabla\times \vec B.
\end{equation}
The MHD equations are often closed with the adiabatic closure relation,
\begin{equation}
\left(\frac{\partial }{\partial t}+\vec U\cdot \nabla \right)\left(\frac{P}{\rho^{\kappa}}\right)=0,
\end{equation}
where $\kappa$ is the adiabatic exponent. The MHD equations are intrinsically scale-free and, therefore, only valid for processes that do not occur on any of the characteristic plasma scales of the system introduced in Sect.~\ref{sec:scales}. Thus, MHD only applies to large-scale phenomena that occur
\begin{enumerate}
\item on length scales $\lesssim L$,
\item on length scales $\gg \max(d_{j},\rho_j)$, and
\item on timescales $\gg \max(\Pi_{\Omega_j}, \Pi_{\omega_{\mathrm pj}})$
\end{enumerate}
 for all $j$.

\subsubsection{Standard distributions in solar-wind physics}\label{sec:distributions}

Although solar-wind measurements often reveal irregular plasma distribution functions \citep[see Sects.~\ref{sec:ion-prop} and \ref{sec:elec-prop}, as well as][]{marsch2012}, it is sometimes helpful to invoke closed analytical expressions for the distribution functions in a plasma. In the following description, we use the cylindrical coordinate system in velocity space introduced in Sect.~\ref{sec:moments} with its symmetry axis to be parallel to $\hat{\vec b}$. 

A gas in thermodynamic equilibrium has a \emph{Maxwellian} velocity distribution,
\begin{equation}\label{maxwellian}
f_{\mathrm M}(\vec v)=\frac{n_j}{\pi^{3/2}w_j^3}\exp\left(-\frac{\left(\vec v-\vec U_j\right)^2}{w_j^2}\right),
\end{equation}
where 
\begin{equation}
w_j\equiv \sqrt{\frac{2k_{\mathrm B}T_j}{m_j}}
\end{equation}
 is the (isotropic) thermal speed of species $j$. Equation~(\ref{maxwellian}) has a thermodynamic justification in equilibrium statistical mechanics based on the Gibbs distribution \citep{landau1969}. An empirically motivated extension of the Maxwellian distribution is the so-called \emph{bi-Maxwellian} distribution, which introduces temperature anisotropies with respect to the background magnetic field yet follows the Maxwellian behavior on any one-dimensional cut at constant $v_{\perp}$ or constant $v_{\parallel}$ in velocity space:
\begin{equation}\label{biMax}
f_{\mathrm {bM}}(\vec v)=\frac{n_j}{\pi^{3/2}w_{\perp j}^2w_{\parallel j}}\exp\left(-\frac{v_{\perp}^2}{w_{\perp j}^2}-\frac{\left(v_{\parallel}-U_{\parallel j}\right)^2}{w_{\parallel j}^2}\right),
\end{equation}
where $w_{\perp j}$ and $w_{\parallel j}$ are the thermal speeds defined in Equations~(\ref{wperp}) and (\ref{wpar}).
Advanced methods in thermodynamics such as non-extensive statistical mechanics lead to the \emph{$\kappa$-distribution} \citep{tsallis1988,livadiotis2013,livadiotis2017},
\begin{equation}\label{kappadist}
f_{\kappa}(\vec v)=\frac{n_j}{w_j^3}\left[\frac{2}{\pi(2\kappa-3)}\right]^{3/2}\frac{\Gamma(\kappa+1)}{\Gamma(\kappa-1/2)}\left[1+\frac{2}{2\kappa-3}\frac{\left(\vec v-\vec U_j\right)^2}{w_j^2}\right]^{-\kappa-1},
\end{equation}
where $\Gamma(x)$ is the $\Gamma$-function \citep{abramowitz1972} and $\kappa>3/2$. We note that $f_{\kappa}\rightarrow f_{\mathrm M}$ for $\kappa\rightarrow \infty$. The $\kappa$-distribution is characterized by having tails that are more pronounced for smaller $\kappa$ (i.e., the kurtosis of the distribution increases as $\kappa$ decreases). Analogous to the bi-Maxwellian is the \emph{bi-$\kappa$-distribution},
\begin{multline}\label{bikappadist}
f_{\mathrm b\kappa}(\vec v)=\frac{n_j}{w_{\perp j}^2w_{\parallel j}}\left[\frac{2}{\pi(2\kappa-3)}\right]^{3/2}\frac{\Gamma(\kappa+1)}{\Gamma(\kappa-1/2)}\\
\times \left\{1+\frac{2}{2\kappa-3}\left[\frac{v_{\perp}^2}{w_{\perp j}^2}+\frac{\left(v_{\parallel}-U_{\parallel j}\right)^2}{w_{\parallel j}^2}\right]\right\}^{-\kappa-1}.
\end{multline}
In the following sections, we will encounter observed distribution functions and recognize some of the uses and limitations of these analytical expressions.

\subsubsection{Ion properties}\label{sec:ion-prop}

\begin{figure}
  \includegraphics[width=\textwidth]{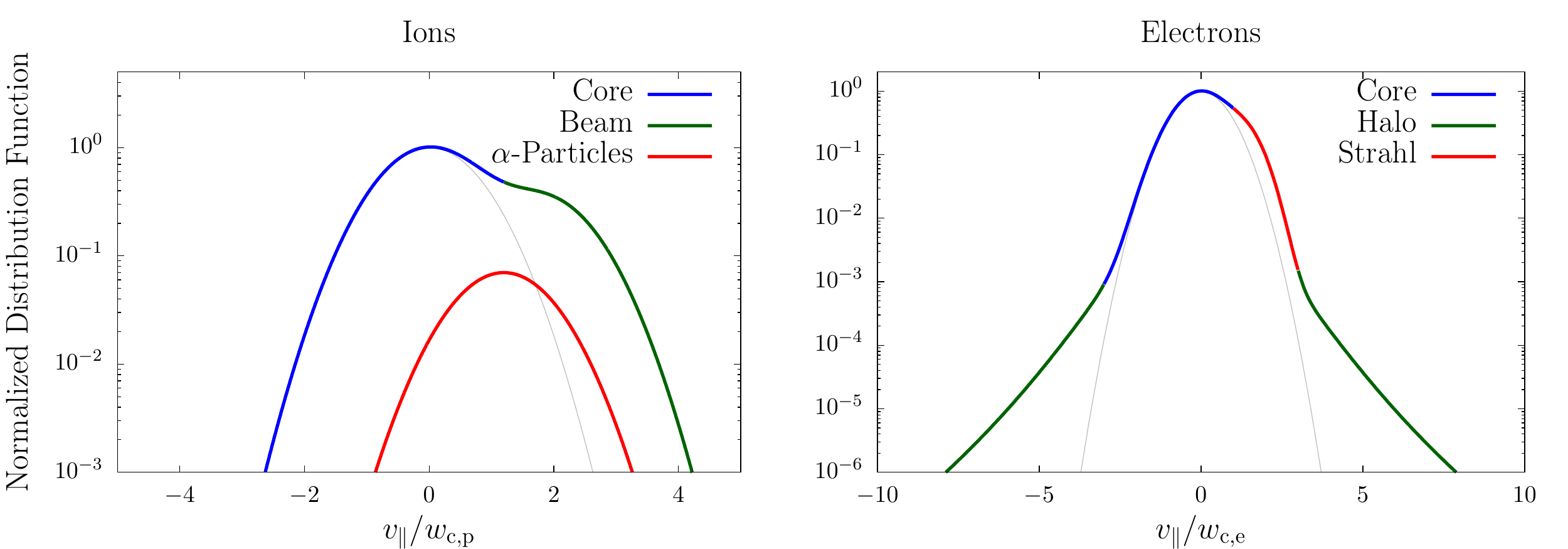}
\caption{Illustration of ion (left) and electron (right) kinetic features in the solar wind. We show cuts through the distribution function along the direction of the magnetic field. We normalize the distribution functions to the maxima of the proton and electron distribution functions, respectively. We normalize the parallel velocity to the thermal speed of the proton and electron core components, $w_{\mathrm {c,p}}$ and $w_{\mathrm {c,e}}$, respectively. We note that $w_{\mathrm {c,p}}\ll w_{\mathrm {c,e}}$. The gray curves show the underlying core distribution alone. The distributions are shown in the reference frames in which the core distribution is at rest.}
\label{fig:distr_schematics}       
\end{figure}

\begin{figure}
  \includegraphics[width=\textwidth]{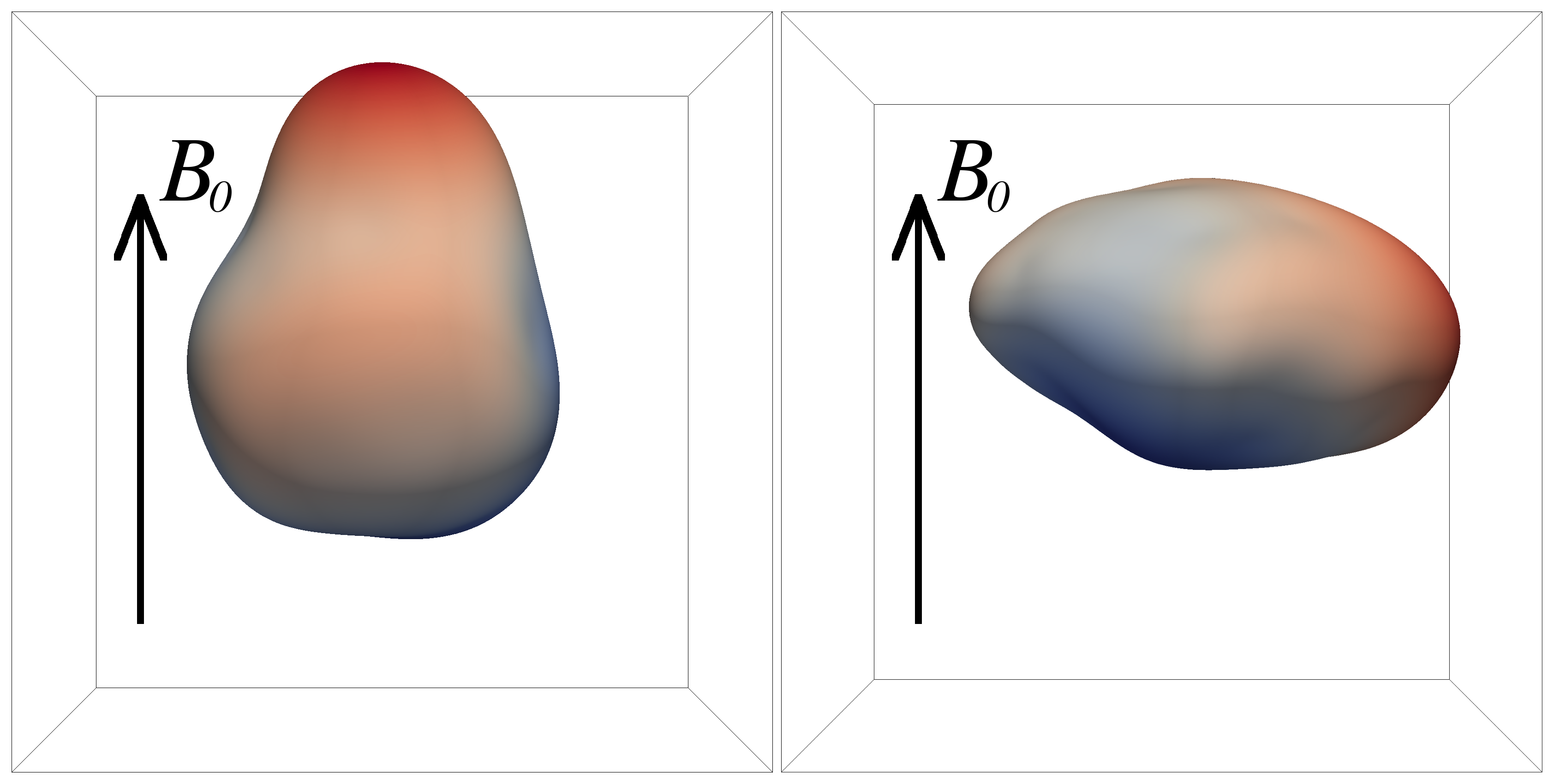}
\caption{Interpolated isosurfaces in velocity space of two proton distribution functions measured by Helios 2. The arrow $\vec B_0$ indicates the direction of the local magnetic field. The color-coding represents the distance of the isosurface from the center-of-mass velocity.  Left: measurement from 1976-02-04 at 10:21:43 UTC.  The center-of-mass velocity is 478 km/s. The elongation along the magnetic-field direction represents the proton beam. Right: measurement from 1976-04-16 at 07:50:54 UTC. The center-of-mass velocity is 768 km/s. The oblate structure of the distribution function represents a temperature anisotropy with $T_{\perp\mathrm p}>T_{\parallel \mathrm p}$.  These distribution functions are available as animations online.}
\label{fig:distributions}       
\end{figure}

In-situ spacecraft instrumentation has been measuring ion and electron velocity distributions for decades (see Sect.~\ref{sec:meas:partinst}). Figure~\ref{fig:distr_schematics} summarizes some of the observed features in ion and electron distribution functions schematically.
These observations show that proton distributions often deviate from the Maxwellian equilibrium distribution given by Equation~(\ref{maxwellian}). For instance, proton distributions often display a field-aligned \emph{beam}: a second proton component streaming faster than the proton \emph{core} component along the direction of the magnetic field with a relative speed $\gtrsim v_{\mathrm {Ap}}$ \citep{asbridge1974,feldman1974a,marsch1982,goldstein2000,tu2004,alterman2018}.  In Fig.~\ref{fig:distr_schematics} (left), the proton beam is shown in green as an extension of the distribution function toward greater $v_{\parallel}$.
Protons also show \emph{temperature anisotropies} with respect to the magnetic field \citep{hundhausen1967,hundhausen1976a,marsch1981,kasper2002,marsch2004,hellinger2006,bale2009,maruca2012}, which manifest in unequal diagonal elements of $\mathsf P_j$ in Equation~(\ref{pressuretensor}). 
Figure~\ref{fig:distributions} shows isosurfaces of $f_{\mathrm p}$ based on measurements from the Helios spacecraft. The background magnetic field is vertically aligned, and the color-coding represents the distance of the isosurfaces from the center-of-mass velocity. A standard Maxwellian distribution would be a monochromatic sphere in these diagrams. Instead, we see that the proton distribution is anisotropic. 
The example on the left-hand side  shows an extension of the isosurface along the magnetic-field direction, which indicates the proton-beam component. Almost always, the proton beam is directed away from the Sun and along the magnetic-field axis.\footnote{The proton beam may be directed toward the Sun or be bi-directional if the local radial component of the magnetic field changed its sign during the passage of the plasma parcel from the Sun to the location of the measurement.} This observation suggests that the beam represents a preferentially accelerated proton component. The existence of this beam thus puts a major observational constraint on potential mechanisms for solar-wind heating and acceleration, which must generate this almost ubiquitous feature in $f_{\mathrm p}$. 
In the example on the right-hand side of Fig.~\ref{fig:distributions}, the isosurface is spread out in the directions perpendicular to the magnetic field, which indicates that $T_{\perp\mathrm p}>T_{\parallel\mathrm p}$.  Although the plasma also exhibits periods with $T_{\perp\mathrm p}<T_{\parallel \mathrm p}$, the predominance of cases with $T_{\perp\mathrm p}>T_{\parallel\mathrm p}$ in the fast wind in the inner heliosphere \citep{matteini2007} suggests an ongoing heating mechanism in the solar wind that counter-acts the double-adiabatic expansion quantified in Equations~(\ref{cgl1}) and (\ref{cgl2}). The double-adiabatic expansion alone would create $T_{\perp\mathrm p}\ll T_{\parallel\mathrm p}$ in the inner heliosphere when we neglect the action of heat flux and collisions on protons. Therefore, only heating mechanisms that explain the observed anisotropies with $T_{\perp\mathrm p}>T_{\parallel\mathrm p}$ in the solar wind \citep[and possibly also in the corona; see][]{kohl2006} are successful candidates for a complete description of the physics of the solar wind.

The colors on the isosurfaces in Fig.~\ref{fig:distributions} illustrate that the bulk velocity of the proton distribution function differs significantly from the center-of-mass velocity. This is mostly due to the $\alpha$-particles in the solar wind \citep{ogilvie1975,asbridge1976,marsch1982a,neugebauer1994,neugebauer1996,steinberg1996,reisenfeld2001,berger2011,gershman2012,bourouaine2013}. 
Although their number density is small ($n_{\alpha}\lesssim 0.05n_{\mathrm p}$), their mass density corresponds to about 20\% of the proton mass density. We often observe the $\alpha$-particles, like the proton beam, to drift with respect to the proton core along the magnetic-field direction and away from the Sun with a typical drift speed $\lesssim v_{\mathrm {Ap}}$. In  Fig.~\ref{fig:distr_schematics} (left), the $\alpha$-particles are shown as a separate shifted distribution in red, centered around the $\alpha$-particle drift speed.

The solar wind also exhibits \emph{anisothermal behavior}; i.e., not all plasma species have equal temperatures  \citep{formisano1970,feldman1974,bochsler1985,cohen1996,vonsteiger2002,vonsteiger2006}. The $\alpha$-particles often show $T_{\parallel\alpha}\gtrsim 4 T_{\parallel \mathrm p}$ \citep{kasper2007,kasper2008,kasper2012}. Electrons are typically colder than protons in the fast solar wind but hotter than protons in the slow solar wind \citep{montgomery1968,hundhausen1970,newbury1998}. As stated in Sect.~\ref{sec:global}, heavy-ion-to-proton temperature ratios are typically greater than the corresponding heavy-ion-to-proton mass ratios for almost all observable ions in the solar wind. 
Like the other kinetic features, solar-wind heating and acceleration models are only fully successful if they explain the observed anisothermal behavior.

All of these non-equilibrium features (temperature anisotropies, beams, drifts, and anisothermal behavior) are less pronounced in the slow solar wind than in the fast wind, which is typically attributed to the greater collisional relaxation rates and the longer expansion times in the slow wind (see Sect.~\ref{sec:col:relax}). These non-equilibrium features reflect the multi-scale nature of the solar wind, since they are driven by a combination of large-scale expansion effects, local kinetic processes, and the feedback of small-scale processes on the large-scale evolution.

\subsubsection{Electron properties}\label{sec:elec-prop}

Although the mass of an electron is much less than the mass of a proton ($m_{\mathrm e}/m_{\mathrm p}\approx 1/1836$), and the electrons' contribution to the total solar-wind momentum flux is insignificant, electrons do affect the large-scale evolution of the solar wind \citep{montgomery1972,salem2003}.  As the most abundant particle species, they guarantee quasi-neutrality: $\rho_{\mathrm c}\approx 0$ and $j_{\parallel} \approx 0$ at length scales $\gg \lambda_{\mathrm e}$ and timescales $\gg \Pi_{\omega_{\mathrm {pe}}}$. Due to their small mass, they are highly mobile and have a much greater thermal speed than the protons, leading to their subsonic behavior (i.e., $U_{\mathrm e}\ll w_{\mathrm e}$). Their momentum balance in Equation~(\ref{fluidmomentum}) is dominated by their pressure gradient and  electromagnetic forces. Through these contributions, the electrons create an \emph{ambipolar electrostatic field} in the expanding solar wind. This field is the central underlying acceleration mechanism of exospheric models \citep[see Sect.~\ref{sec:col:dim};][]{lemaire1973,maksimovic2001}.  Parker's (\citeyear{parker1958}) solar-wind model does not explicitly invoke an ambipolar electrostatic field. Nevertheless, the electron contribution to the pressure gradient in Parker's MHD equation of motion is equivalent to the ambipolar electric field that follows from Equation~(\ref{fluidmomentum}) for electrons in the limit $m_{\mathrm e}\rightarrow 0$ \citep{velli1994,velli2001}. 

Although electrons typically have greater collisional relaxation rates than ions, they exhibit a number of characteristic kinetic non-equilibrium features, which, as for the ions, are more pronounced in the fast solar wind. Most notably, the electron distribution often consists of three distinct components  \citep{feldman1975,pilipp1987,pilipp1987a,hammond1996,maksimovic1997,fitzenreiter1998}: 
\begin{itemize}
\item a thermal \emph{core}, which mostly follows a Maxwellian distribution and has a  thermal energy of $\sim 10\,\mathrm{eV}$ -- blue in Fig.~\ref{fig:distr_schematics} (right); 
\item a non-thermal \emph{halo}, which mostly follows a $\kappa$-distribution, manifests as enhanced high-energy tails in the electron distribution, and has a thermal energy of $\lesssim 80\,\mathrm{eV}$  -- green in Fig.~\ref{fig:distr_schematics} (right); and 
\item a \emph{strahl},\footnote{From \emph{strahl} -- the German word for ``beam''.}  which is  a field-aligned beam of electrons and usually travels in the anti-Sunward direction with a bulk energy $\lesssim 100\,\mathrm{eV}$  -- red in Fig.~\ref{fig:distr_schematics} (right).
\end{itemize}
The core typically includes $\sim 95\%$ of the electrons. It sometimes displays a temperature anisotropy \citep{serbu1972,phillips1989,stverak2008} and a relative drift with respect to the center-of-mass frame \citep{bale2013}. A recent study suggests that a bi-self-similar distribution, which forms through inelastic particle scattering, potentially describes the core distribution better than a bi-Maxwellian distribution \citep{wilson2019}.

\begin{figure}
  \includegraphics[width=\textwidth]{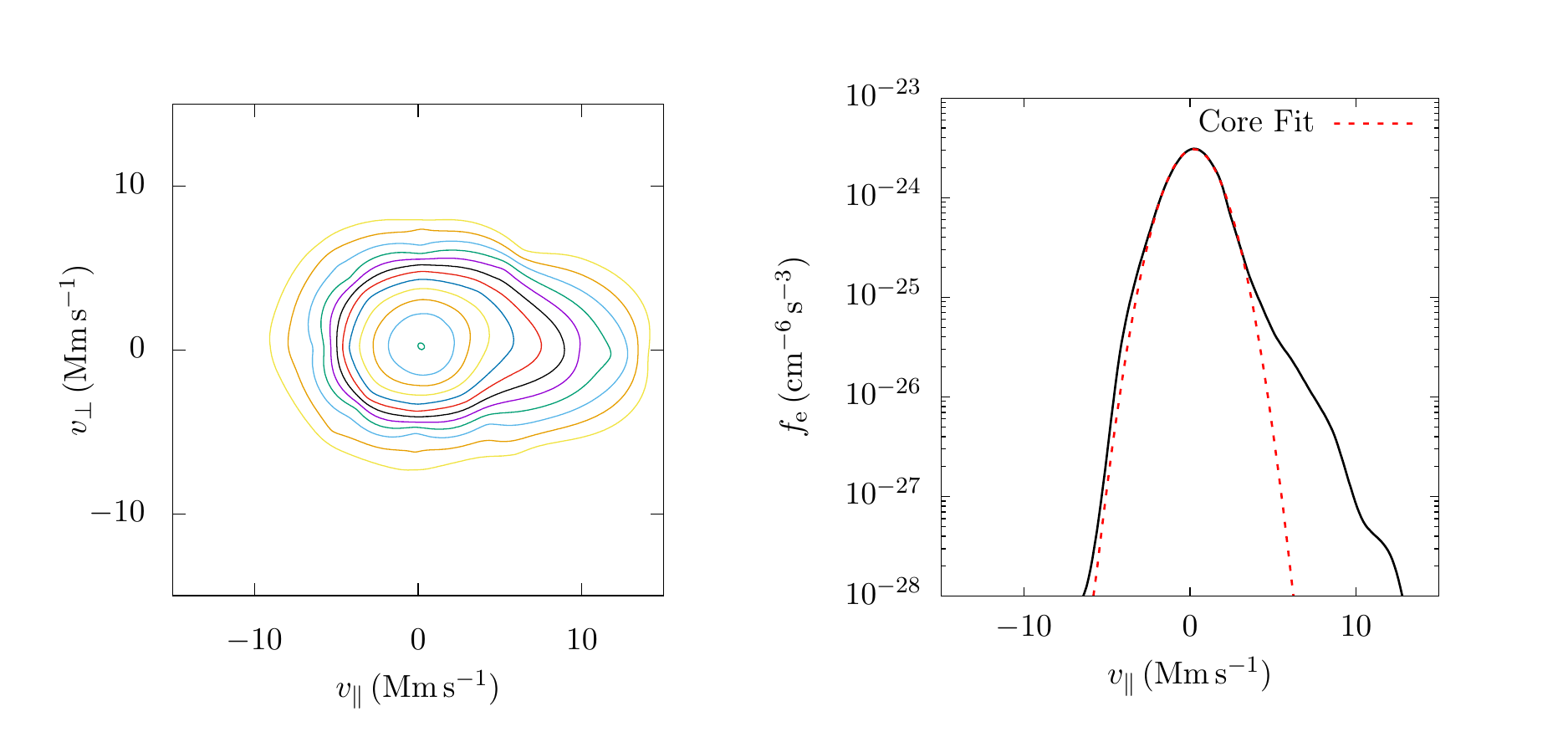}
\caption{Electron velocity distribution function measured by Helios 2 in the fast solar wind at a heliocentric distance of $0.29\,\mathrm{au}$ on 1976-04-18 at 23:38:35 UTC. Left: isocontours of the distribution in a field-aligned coordinate system. Right: a cut through the distribution function along the magnetic-field direction. The red dashed curve shows a Maxwellian fit to the core of the distribution function. The strahl is clearly visible as an enhancement in the distribution function at $v_{\parallel}>0$.}
\label{fig:electron_distribution}       
\end{figure}

The strahl probably results from a more isotropic distribution of superthermal electrons in the corona that has been focused by the mirror force in the nascent solar wind \citep{owens2008}, explaining the anti-Sunward bulk velocity of the strahl in the solar-wind rest frame. As with the ion beams, a Sunward or bi-directional electron strahl can occur when the magnetic-field configuration changes during the plasma's passage from the Sun \citep{gosling1987,owens2017}. Figure~\ref{fig:electron_distribution} shows an example of an electron velocity distribution function measured in the solar wind. This distribution exhibits a significant strahl at $v_{\parallel}>0$ but shows no clear halo component.  We reiterate our paradigm that all successful solar-wind acceleration and heating scenarios must account for the observed kinetic structure of the solar wind, including these features in the electron distributions. 
At highest energies $\gtrsim 2\,\mathrm{keV}$, a nearly isotropic \emph{superhalo} of electrons exists; however, its number density is very small compared to the densities of the other electron species ($\lesssim 10^{-5}\,\mathrm{cm}^{-3}$ at 1 au), and its origin remains poorly understood \citep{lin1998,wang2012,yang2015,tao2016}.

Observations of the superthermal electrons (i.e., strahl and halo) reveal that $(n_{\mathrm s}+n_{\mathrm h})/n_{\mathrm e}$ remains largely constant with heliocentric distance, where $n_{\mathrm s}$ is the strahl density and $n_{\mathrm h}$ is the halo density. Conversely, $n_{\mathrm s}/n_{\mathrm e}$ decreases with distance from the Sun while $n_{\mathrm h}/n_{\mathrm e}$ increases \citep{maksimovic2005,stverak2009,graham2017}. Various processes have been proposed to explain this phenomenon, most of which involve the scattering of strahl electrons into the halo \citep{vocks2005,gary2007,pagel2007,saito2007,owens2008,anderson2012,gurgiolo2012,landi2012,verscharen2019a}.

Locally, electrons often show isothermal behavior (i.e., having a polytropic index of one) due to their large field-parallel mobility. Globally, their non-thermal distribution functions carry a large heat flux according to Equation~(\ref{heatflux}) into the heliosphere \citep{feldman1976,scime1995}.
Observations of large-scale electron temperature profiles suggest that the electron heat flux, rather than local heating, dominates their temperature evolution  \citep{pilipp1990,stverak2015}. These energetic considerations also reveal that a combination of processes regulate the heat flux of the distribution.  Collisions and collective kinetic processes such as microinstabilities are the prime candidates for explaining  electron heat-flux regulation  \citep[see Sects.~\ref{sec:col:elec} and \ref{sec:inst:beam};][]{scime1994,scime1999,scime2001,bale2013,lacombe2014}.

\subsubsection{Open questions and problems}\label{open_questions}

The major outstanding science questions in solar-wind physics require a detailed understanding of the interplay between the multi-scale nature and the observed kinetic features of the solar wind. This theme applies to the coronal and solar-wind heating problem as well as the overall energetics of the inner heliosphere. We remind ourselves that any answer to the heating problem must be consistent with multiple detailed observational constraints as we have seen in the previous sections.

The observed temperature profiles and overall particle energetics of ions and electrons are consequences of the complex interactions of global heat flux, Coulomb collisions (Sect.~\ref{sec:col}), local wave action (Sect.~\ref{sec:waves}), turbulent heating  (Sect.~\ref{sec:turbulence}), microinstabilities (Sect.~\ref{sec:inst}), and double-adiabatic expansion  \citep{mihalov1978,feldman1979,gazis1982,marsch1983a,marsch1989,pilipp1990,mccomas1992,gazis1994,issautier1998,maksimovic2000,matteini2007,cranmer2009,hellinger2011,lechat2011,hellinger2013,stverak2015}. 
We still lack a detailed physics-based understanding of the majority of these processes, and the quantification of these processes and their role for the overall energetics of the solar wind remains one of the most outstanding science problems in space research.
 \begin{figure}
  \includegraphics[width=0.5\textwidth]{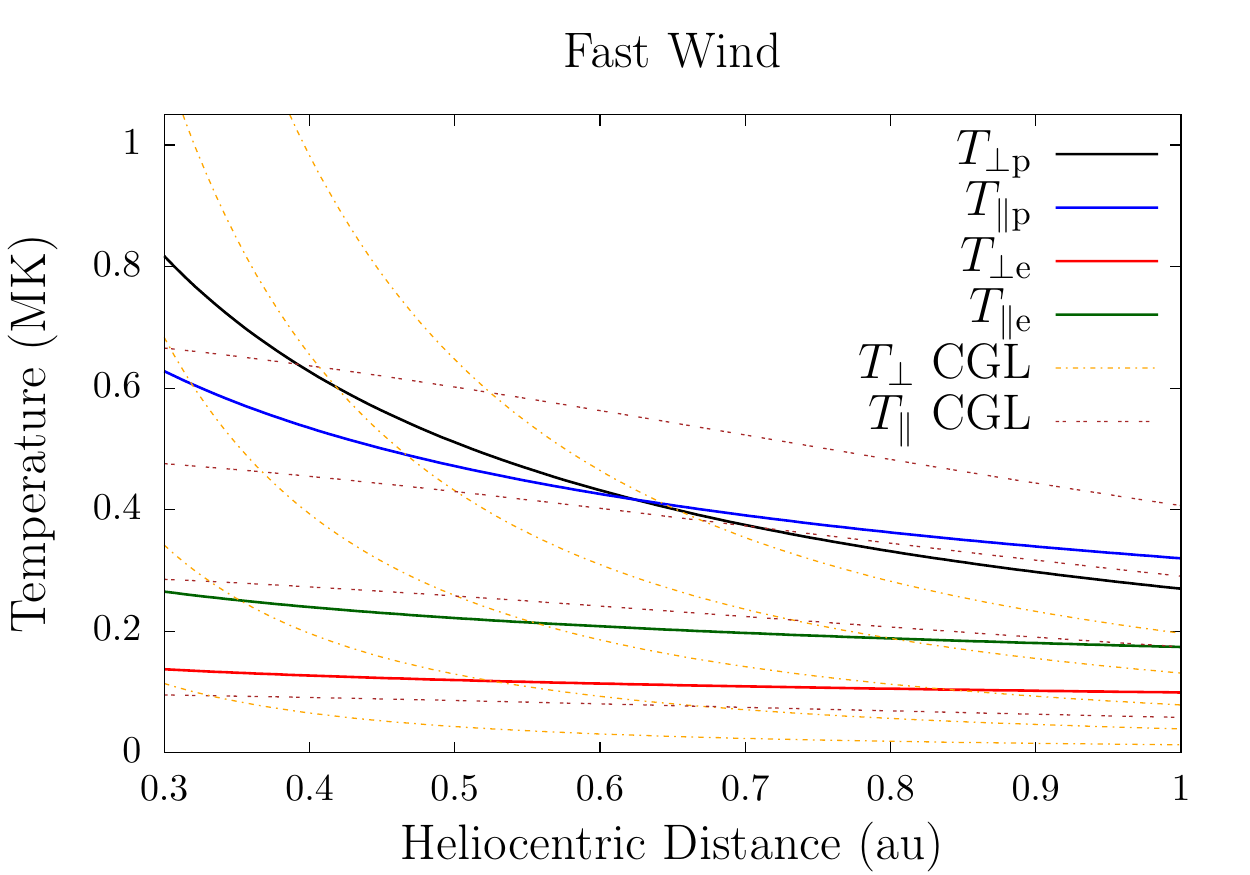}
    \includegraphics[width=0.5\textwidth]{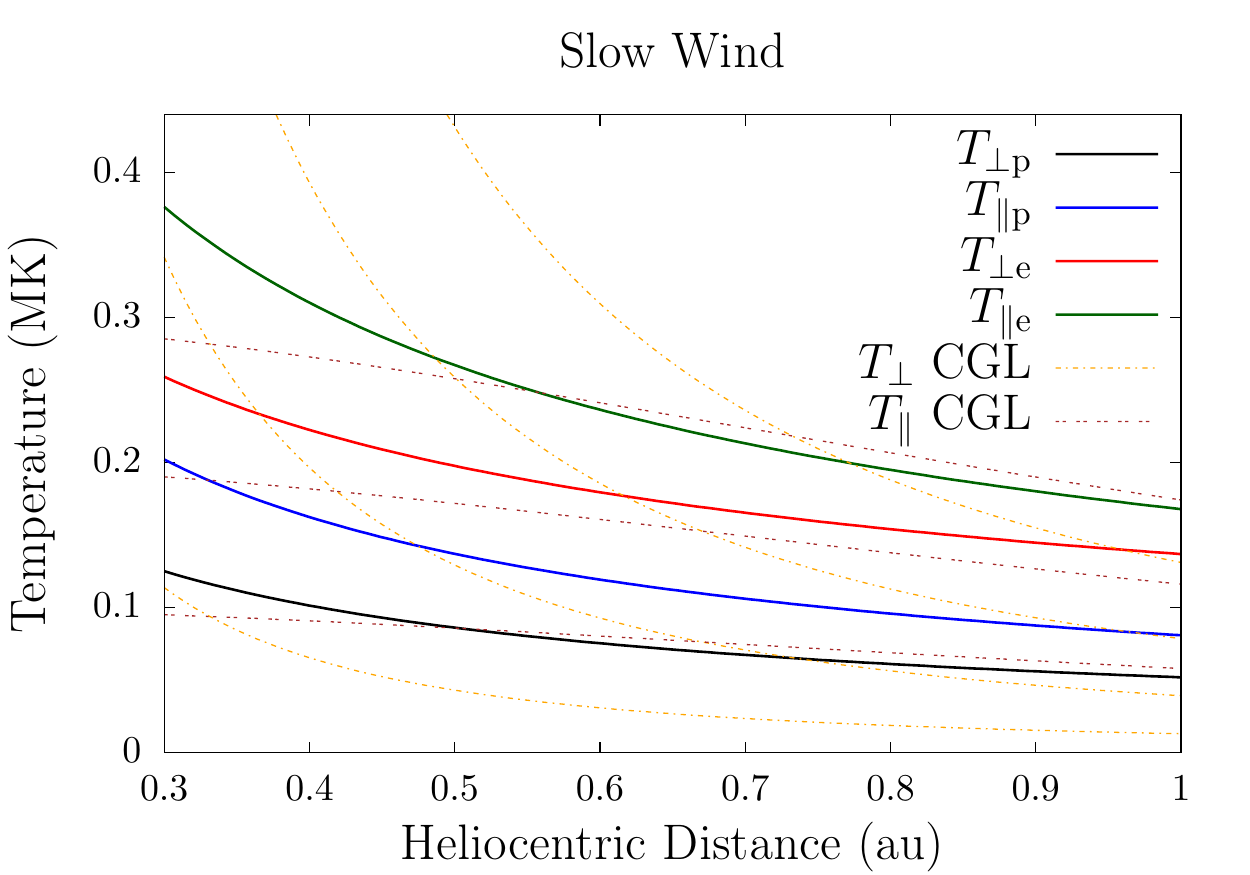}
\caption{Temperature profiles in the inner heliosphere for fast (left) and slow (right) wind. We show radial power-law fits to proton-temperature measurements separated by fast ($700\,\mathrm{km/s}\leq U_{\mathrm pr}\leq 800\,\mathrm{km/s}$)  and slow ($300\,\mathrm{km/s}\leq U_{\mathrm pr}\leq 400\,\mathrm{km/s}$) solar-wind conditions from \citet{hellinger2013}. Likewise, we show radial power-law fits to electron-temperature measurements separated by fast ($U_{\mathrm pr}\geq 600\,\mathrm{km/s}$) and slow ($U_{\mathrm pr}\leq 500\,\mathrm{km/s}$) solar-wind conditions from \citet{stverak2015}. The thin-dashed lines indicate the CGL temperature profiles according to Equations~(\ref{cgl1}) and (\ref{cgl2}), where we set the right-hand sides of both equations to zero and determine the magnetic field through Equations~(\ref{tantheta}) and (\ref{Bnull}) using $n_{j}\propto 1/r^2$, $\theta=90^{\circ}$, $r_{\mathrm{eff}}=10R_{\odot}$,  and $U_{\mathrm pr}=500\,\mathrm{km/s}$. }
\label{fig:temperature_profiles}       
\end{figure}

Observed temperature profiles (including anisotropies) are some of the central messengers about the overall solar-wind energetics, apart from velocity profiles.
Figure~\ref{fig:temperature_profiles} illustrates the radial evolution of the proton and electron temperatures in the directions perpendicular and parallel to the magnetic field and separated by fast and slow wind. We also show the expected temperature profiles under the assumption that the evolution follows the double-adiabatic (CGL) expansion according to Equations~(\ref{cgl1}) and (\ref{cgl2}) only. All of the measured temperature profiles deviate from the CGL profiles to some degree, and this trend continues at greater heliocentric distances \citep{cranmer2009}. Explaining these deviations lies at the heart of the challenge to explain coronal and solar-wind heating and acceleration. 

We intend this review to give an overview over the relevant multi-scale processes in the solar wind. In the near future, data from the Parker Solar Probe \citep{fox2016} and Solar Orbiter \citep{mueller2013} spacecraft will provide us with detailed observations of the local and global properties of the solar wind at different distances from the Sun. These groundbreaking observations will help us to quantify the roles of the multi-scale processes described in this review. 

Section~\ref{sec:meas} describes the methods to measure solar-wind particles and fields in situ. In Sect.~\ref{sec:col}, we discuss the effects of collisions on the multi-scale evolution of the solar wind. Section~\ref{sec:waves} introduces waves, and Sect.~\ref{sec:turbulence} introduces turbulence as mechanisms that affect the local and global plasma behavior. We describe the role of kinetic microinstabilities and parametric instabilities in Sect.~\ref{sec:inst}. In Sect.~\ref{sec:conclusions}, we summarize this review and consider future developments in the study of the multi-scale evolution of the solar wind.

\section{In-situ observations of space plasmas} \label{sec:meas}

Observations of space plasmas can be roughly divided into two categories: \emph{remote} and \emph{in-situ}.  Remote observations include both measurements of the plasma's own emissions (e.g., radio waves, visible light, and X-ray photons) as well as measurements of the effects that the plasma has on emissions from other sources (e.g., Faraday rotation and absorption lines).  In this way, regions such as the chromosphere that are inaccessible to spacecraft can still be studied.  Additionally, imaging instruments such as coronagraphs provide information on the global structure of space plasma.  Nevertheless, due to limited spectral and angular resolution, these instruments cannot provide information on all of the small-scale processes at work within the plasma.  Remote observations also only offer limited information on three-dimensional phenomena.  If the observed plasma is optically thick (e.g., the photosphere in visible light), its interior cannot be probed; if it is optically thin (e.g., the corona in EUV), remote observations suffer from the effects of line-of-sight integration.

In contrast, in-situ observations provide detailed information on microkinetic processes in space plasmas.  Spacecraft carry in-situ instruments into the plasma to directly detect its particles and fields and thereby to provide small-scale observations of localized phenomena.  Although an in-situ instrument only detects the plasma in its immediate vicinity, statistical studies of ensembles of measurements have provided remarkable insights into how small-scale processes affect the plasma's large-scale evolution.

This section briefly overviews both the capabilities and the limitations of instruments used to observe the solar wind in situ.  Although a full treatment of the subject is  beyond the scope of this review, a basic understanding of these instruments is essential for the proper scientific analysis of their measurements.  Section \ref{sec:meas:missions} highlights some significant heliospheric missions.  Two sections are dedicated to in-situ observations of thermal ions and electrons: Sect. \ref{sec:meas:partinst} overviews the instrumentation, and Sect. \ref{sec:meas:partanls} addresses the analysis of particle data.  Sections \ref{sec:meas:mag} and \ref{sec:meas:elec} respectively discuss the in-situ observation of the solar wind's magnetic and electric fields.  Section \ref{sec:meas:multi} presents a short description of multi-spacecraft techniques.

\subsection{Overview of in-situ solar-wind missions} \label{sec:meas:missions}

\begin{sidewaystable}
\vspace{0.58\textwidth}
\caption{\label{tab:missions} Select heliospheric missions: completed, active, and future}
\begin{tabular}{llll}
\noalign{\smallskip}\hline\noalign{\smallskip}
Mission & Years Active\fn{a} & Radial Coverage\fn{b} (au) & Source \\
\noalign{\smallskip}\hline\noalign{\smallskip}
Luna 1, 2, \& 3 & 1959 -- 1959 & $\approx$ 1.0\fn{c}       & NSSDC; \citet{johnson1979} \\
Mariner 2       & 1962 -- 1962 &  0.866  --   1.003        & COHOWeb \\
Pioneer 6       & 1965 -- 1971 &  0.814  --   0.984        & COHOWeb \\
Pioneer 7       & 1966 -- 1968 &  1.010  --   1.126        & COHOWeb \\
Pioneer 10      & 1972 -- 1995 &  0.99   --  63.04         & \cdaweb{PIONEER10\_COHO1HR\_MERGED\_MAG\_PLASMA} \\
Pioneer 11      & 1973 -- 1992 &  1.00   --  36.26         & \cdaweb{PIONEER11\_COHO1HR\_MERGED\_MAG\_PLASMA} \\
Pioneer Venus   & 1978 -- 1992 &  0.72   --   0.73         & \cdaweb{PIONEERVENUS\_COHO1HR\_MERGED\_MAG\_PLASMA} \\
ISEE-3 (ICE)    & 1978 -- 1990 &  0.93   --   1.03         &\cdaweb{ISEE-3\_MAG\_1MIN\_MAGNETIC\_FIELD} \\
Helios 1        & 1974 -- 1981 &  0.31   --   0.98         & \cdaweb{HELIOS1\_COHO1HR\_MERGED\_MAG\_PLASMA} \\
Helios 2        & 1976 -- 1980 &  0.29   --   0.98         & \cdaweb{HELIOS2\_COHO1HR\_MERGED\_MAG\_PLASMA} \\
Ulysses         & 1990 -- 2009 &  1.02   --   5.41         & \cdaweb{UY\_COHO1HR\_MERGED\_MAG\_PLASMA} \\
Cassini         & 1997 -- 2017 &  0.67   --  10.07         & COHOWeb; OMNIWeb Plus (\texttt{helio1day}) \\
STEREO B        & 2006 -- 2014 &  1.00   --   1.09         & \cdaweb{STB\_COHO1HR\_MERGED\_MAG\_PLASMA} \\
\noalign{\smallskip}\hline\noalign{\smallskip}
Voyager 1       & 1977 --      &  1.01   -- 140.71\fn{d}   & \cdaweb{VOYAGER1\_COHO1HR\_MERGED\_MAG\_PLASMA} \\
Voyager 2       & 1977 --      &  1.00   -- 118.91\fn{d}   & \cdaweb{VOYAGER2\_COHO1HR\_MERGED\_MAG\_PLASMA} \\
Wind            & 1994 --      &  0.972  --   1.017        & \cdaweb{WI\_OR\_PRE} \\
SOHO            & 1995 --      &  0.972  --   1.011        & \cdaweb{SO\_OR\_PRE} \\
ACE             & 1997 --      &  0.973  --   1.010        & \cdaweb{AC\_OR\_SSC} \\
New Horizons    & 2006 --      & 11.268  --  42.775\fn{d}  & \cdaweb{NEW\_HORIZONS\_SWAP\_VALIDSUM} \\
STEREO A        & 2006 --      &  0.96   --   0.97         & \cdaweb{STA\_COHO1HR\_MERGED\_MAG\_PLASMA} \\
DSCOVR          & 2015 --      &  0.973  --   1.007        & \cdaweb{DSCOVR\_ORBIT\_PRE} \\
PSP             & 2018 --      &  0.0459 --   0.25\fn{e,f} & \citet{fox2016} \\
\noalign{\smallskip}\hline\noalign{\smallskip}
Solar Orbiter   & 2020\fn{g,h} &  0.28   --   1.2\fn{e}    & \citet{mueller2013} \\ 
IMAP            & 2024\fn{g}   &  0.973  --   1.007\fn{i}  & NASA Release 18-046 \\
\noalign{\smallskip}\hline\noalign{\smallskip}
\end{tabular}

\fn{a}{\footnotesize Year of launch to final year (with non-fill data) in cited dataset} \\
\fn{b}{\footnotesize Incomplete for some missions due to data gaps} \\
\fn{c}{\footnotesize Exact range not available} \\
\fn{d}{\footnotesize Distance still increasing; values on 2018-01-01 (Voyager 1), 2018-10-26 (Voyager 2), or 2018-10-31 (New Horizons)} \\
\fn{e}{\footnotesize Anticipated radial coverage} \\
\fn{f}{\footnotesize Perihelion of first three orbits: 0.163 au} \\
\fn{g}{\footnotesize Anticipated launch date} \\
\fn{h}{\footnotesize \texttt{https://www.esa.int/Our\_Activities/Space\_Science/Solar\_Orbiter}, accessed 2019-09-10} \\
\fn{i}{\footnotesize Approximate radial coverage of the first Lagrangian point of the Earth-Sun system}
\end{sidewaystable}

\begin{figure}
\begin{center}
\includegraphics[width=0.9\textwidth]{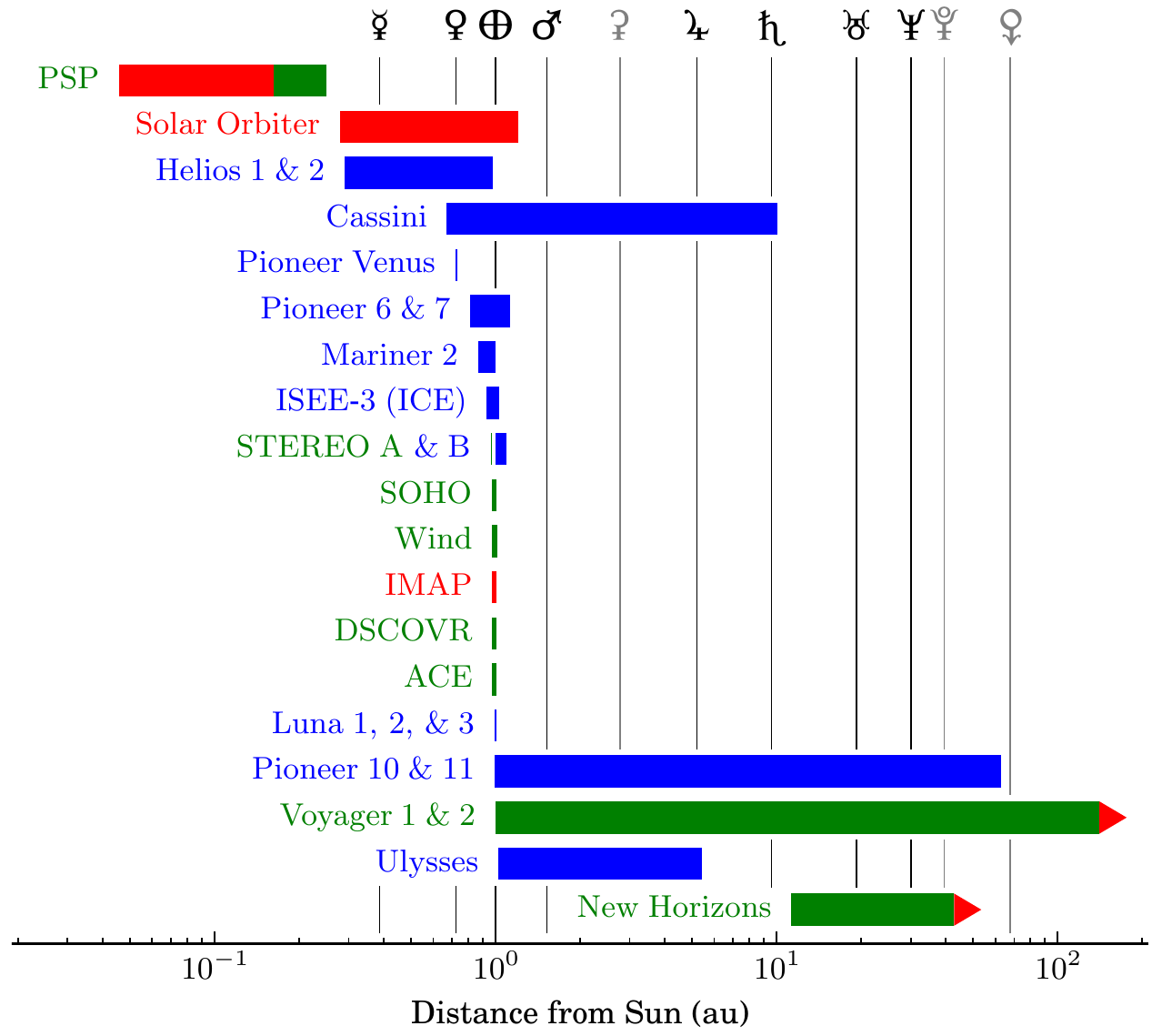}
\caption{\label{fig:missions} Radial coverage of select heliospheric missions based on Table \ref{tab:missions}.  Colors indicate the status of each mission: completed (blue), active (green), and future (red).  The colored bar for each mission does not reflect any data gaps that may be present in its dataset(s).  Mixed coloring has been used for PSP to reflect that, while the mission is active, final radial coverage has not yet been achieved.  Red arrows indicate that the radial coverages of Voyager~1 \&~2 and New Horizons are still increasing.  Vertical lines indicate the semi-major axes of the eight planets (black) and the dwarf planets Ceres, Pluto, and Eris (gray).}
\end{center}
\end{figure}

In-situ plasma instruments were among the first to be flown on spacecraft.  \citet{gringauz1960} used data from Luna 1, Luna 2, and Luna 3, which at the the time were known as the Cosmic Rockets, to report the first detection of super-sonic solar-wind ions as predicted by \citet{parker1958}.  These observations were soon confirmed by \citet{neugebauer1962}, who used in-situ measurements from Mariner 2 en route to Venus.

Since then, numerous spacecraft have carried in-situ instruments throughout the heliosphere to observe the solar wind's particles and fields.  Table \ref{tab:missions} lists a  selection of these missions grouped as completed, active, and future missions. The column ``Radial Coverage'' lists the ranges of heliocentric distance for which in-situ data are available, which are presented graphically in Fig. \ref{fig:missions}.  Currently, Voyager 1 \citep{kohlhase1977} is the most distant spacecraft from the Sun -- a superlative that it will continue to hold for the foreseeable future.  Helios~2 \citep{porsche1977} held for several decades the record for closest approach to the Sun, but, in late 2018, Parker Solar Probe \citep{fox2016} achieved a substantially closer perihelion.

\subsection{Thermal-particle instruments} \label{sec:meas:partinst}

Thermal particles constitute the most abundant but lowest-energy particles in solar-wind plasma.  Although no formal definition exists, the term commonly refers to particles whose energies are within several (``a few'') thermal widths of the plasma's bulk velocity.  We define these as protons with energies $\lesssim 10\, {\rm keV}$ and electrons with energies $\lesssim 100\, {\rm eV}$ under typical solar-wind conditions at 1~au. We note, however, that most thermal-particle instruments cover a wider range of  energies.

Although particle moments such as density, bulk velocity, and temperature are useful quantities for characterizing the plasma, these parameters generally cannot be measured directly.  Instead, thermal-particle instruments measure \emph{particle spectra}, which give the distribution of particle energies in various directions.  These spectra must then be analyzed to derive values for the particle moments (see Sect. \ref{sec:meas:partanls}).

This section focuses on the basic design and operation of three types of thermal-particle instruments: \emph{Faraday cups}, \emph{electrostatic analyzers (ESAs)}, and \emph{mass spectrometers}.  Since particle acceleration beyond thermal energies is outside of the scope of this review, we do not address instruments for measuring higher-energy particles.

Some other techniques and instruments exist for measuring thermal particles in solar-wind plasma, but we omit extensive discussion of these since they generally provide limited information about the phase-space structure of particle distributions.  For example, an electric-field instrument can be used to infer some electron properties (especially density; see Sect.~\ref{sec:meas:elec}).  Likewise \emph{Langmuir probes} provide some electron moments \citep{mottsmith1926}.  A series of bias voltages is applied to a Langmuir probe relative either to the spacecraft or to another Langmuir probe.  The electron density and temperature can then be inferred from measurements of current at each bias voltage.  The Cassini spacecraft included a spherical Langmuir probe \citep{gurnett2004} along with other plasma instruments \citep{young2004}.

\subsubsection{Faraday cups} \label{sec:meas:partinst:fc}

Faraday cups rank among the earliest instruments for studying space plasmas.  Historically noteworthy examples include the charged-particle traps on Luna 1, Luna 2, and Luna 3 \citep{gringauz1960} and the Solar Plasma Experiment on Mariner 2 \citep{neugebauer1962}, which provided the first in-situ observations of the solar wind's supersonic ions.  Since then, Faraday cups on Pioneer 6 and Pioneer 7 \citep{lazarus1966,lazarus1968}, Voyager 1 and 2 \citep{bridge1977}, Wind \citep{ogilvie1995a}, and DSCOVR \citep{aellig2001a} have continued to observe solar-wind particles.

\begin{figure}
\begin{center}
\includegraphics[width=0.80\textwidth]{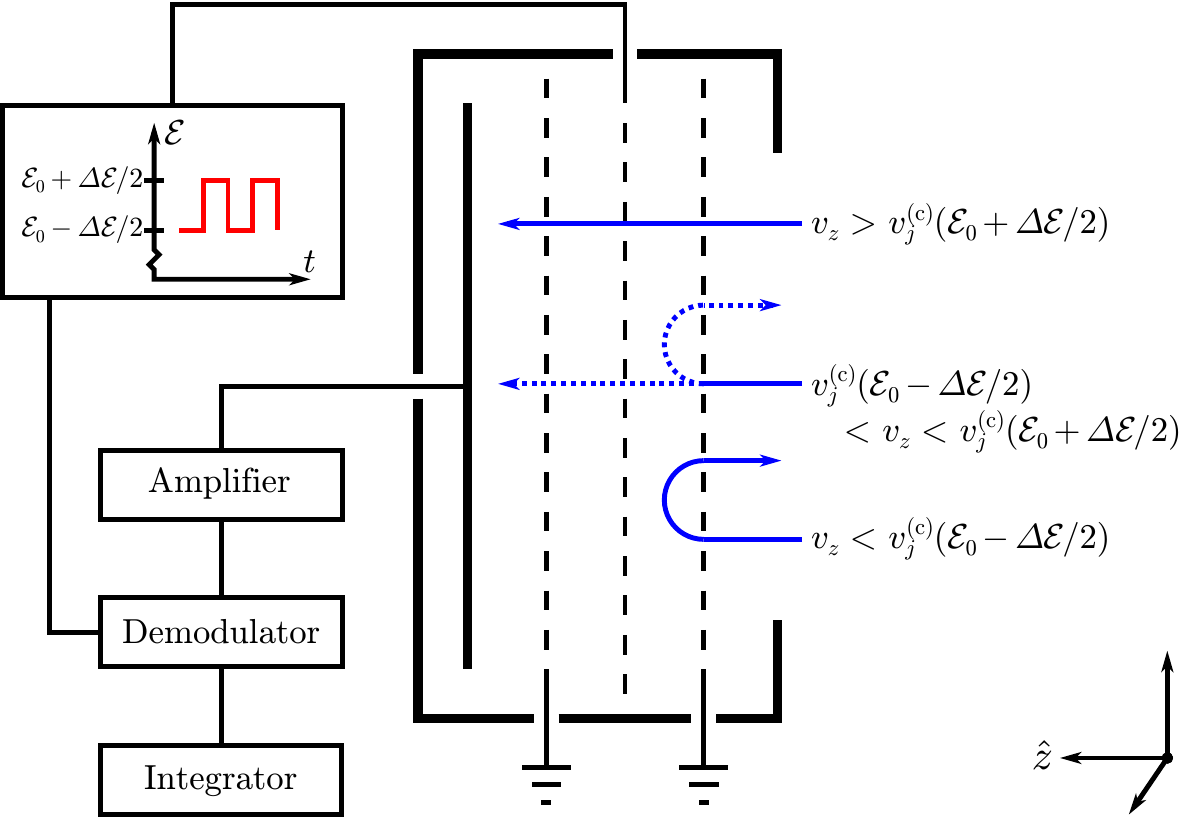}
\caption{\label{fig:fc} Simplified cross-sectional diagram of a Faraday cup for observing ions.  The cup's aperture is on the right, its \emph{collector plate} is on the left, and its three grids are indicated by dashed lines.  A square-wave voltage, $\mathcal{E} = \mathcal{E}_0 \pm \Delta\mathcal{E}/2 > 0$, is applied to the middle grid, which is known as the \emph{modulator}.  Blue arrows indicate inflowing $j$-ions.  Depending on $v_z$, the normal component of the ion's velocity, it is either always accepted by the modulator (high speed), always rejected (low speed), or only accepted when the modulator's voltage is low (intermediate speed).  The accepted ions produce a current at the collector plate, which the detection system amplifies, demodulates, and integrates to measure, in effect, the current from only the intermediate-speed ions according to Equation~(\ref{eqn:del:i}).}
\end{center}
\end{figure}

As depicted in Fig.~\ref{fig:fc}, a Faraday cup consists of a grounded metal structure with an aperture.  A typical Faraday cup has a somewhat ``squat'' geometry with a wide aperture so that it  accepts incoming particles from a wide range of directions.  For example, the full-width half-maximum field of view of each of the Wind/SWE Faraday cups is about $105^\circ$.  At the back of the cup is a metal \emph{collector} plate, which receives the current $I$ of the inflowing charged particles.

Figure~\ref{fig:fc} shows three of the fine mess grids that are placed between a Faraday cup's aperture and collector.  The inner and outer grids are electrically grounded.  A voltage $\mathcal{E}$ is applied to the middle grid, known as the \emph{modulator}, to restrict the ability of particles to reach the collector.  We define $\hat{\vec{z}}$ to indicate the direction into the Faraday cup so that $-\hat{\vec{z}}$ is the cup's \emph{look direction}.  Consider a $j$-particle of mass $m_j$ and charge $q_j$ that enters the cup with a velocity $\vec{v}$.  For a modulator voltage $\mathcal{E}$, the particle can only reach the collector if the normal component of its velocity, $v_z = \vec{v}\cdot\hat{\vec{z}}$, is greater than the \emph{cutoff speed}
\begin{equation} \label{eqn:fc:cutoff}
v^{(\mathrm c)}_j(\mathcal{E}) \equiv \begin{cases} \sqrt{\displaystyle\frac{2\,q_j\,\mathcal{E}}{m_j}} & \text{if}\; \; q_j\,\mathcal{E} > 0 \\ 0 & \text{else} \end{cases}.
\end{equation}
When $\mathcal{E}$ and $q_j$ have opposite signs, the modulator places no restriction on the particle's ability to reach the collector.

Typically, the modulator is not kept at a constant voltage but rather alternated between two voltages:
\begin{equation}\label{eqn:vwin}
\mathcal{E} = \mathcal{E}_0 \pm \frac{ \Delta \mathcal{E}}{2},
\end{equation}
where $\mathcal{E}_0$ is the offset and $\Delta\mathcal{E}$ is the peak-to-peak amplitude.  In this configuration, the detector circuit is designed to use synchronous detection to measure the difference in the collector current between the two states:
\begin{equation}\label{eqn:del:i}
\Delta I( \mathcal{E}_0, \Delta\mathcal{E}) = I\left(\mathcal{E}_0-\frac{\Delta\mathcal{E}}{2}\right) - I\left(\mathcal{E}_0+\frac{\Delta\mathcal{E}}{2}\right).
\end{equation}
Essentially, $\Delta I$ is the current from particles whose velocities are sufficient for them to reach the collector when the modulator voltage is low but not when it is high.  This method suppresses contributions to the collector current that do not vary with the modulator voltage.  These contributions include the signal from any particle species with a charge opposite that of the modulator since, per Equation~(\ref{eqn:fc:cutoff}), the modulator does not restrict the inflow of such particles.  This method also mitigates the effects of photoelectrons, which are liberated from the collector by solar UV photons and whose signal can exceed that of solar-wind particles by orders of magnitude \citep{bridge1960}.

A set of $\mathcal{E}_0$ and $\Delta\mathcal{E}$ values defines a \emph{voltage window}.  By measuring the differential current $\Delta I$ for a series of these, a Faraday cup produces an energy distribution of solar-wind particles.  The size and number of voltage windows determine the spectral resolution and range, which, for many Faraday cups, can be adjusted in flight to accommodate changing plasma conditions.  Since a Faraday cup is simply measuring  current, its detector electronics often exhibit little degradation with time.  For example, \citet{kasper2006} demonstrate that the absolute gain of each of the Wind/SWE Faraday cups \citep{ogilvie1995a} drifts $\lesssim 0.5\%$ per decade.

Various approaches exist to use Faraday cups to measure the direction of inflowing particles, which is necessary for inferring parameters such as bulk velocity and  temperature anisotropy.  The Voyager/PLS investigation \citep{bridge1977} and the BMSW solar-wind monitor on SPECTR-R \citep{safrankova2008}  include multiple Faraday cups pointed in different directions.  DSCOVR/PlasMag \citep{aellig2001a} has only a single Faraday cup but multiple collector plates: a \emph{split collector}.  Each collector is off-axis from the aperture and thus has a slightly different field of view.  Pioneer 6, Pioneer 7 \citep{lazarus1966,lazarus1968}, and Wind \citep{ogilvie1995a} are spinning spacecraft, so their Faraday cups make measurements in various directions as the spacecraft rotate.

A Faraday cup's \emph{response function} is a mathematical model for what the instrument measures under different plasma conditions: i.e., an expression for $\Delta I$ as a function of the particle distribution functions.  For simplicity, we initially consider only one particle species $j$ and assume that the distribution function $f_j$ is, during the measurement cycle, a function of $\vec v$ only.  The number density of $j$-particles in a phase-space volume $\mathrm d^3\vec v$ centered on $\vec{v}$ is 
\begin{equation}\label{eqn:dnj}
\mathrm dn_j = f_j(\vec{v})\,\mathrm d^3\vec v.
\end{equation}
The current that the Faraday cup measures from the particles in this volume is
\begin{equation}
\mathrm dI_j = q_jv_zA(\theta,\phi)\,\mathrm dn_j = q_jv_zA(\theta,\phi)f_j(\vec{v})\,\mathrm d^3\vec v,
\end{equation}
where $(v,\theta,\phi)$ are the spherical coordinates of $\vec{v}$, and $A(\theta,\phi)$ is the Faraday cup's effective collecting area as a function of particle-inflow direction.\footnote{Typically, the function $A(\theta,\phi)$ is calculated from the Faraday cup's geometry and/or is measured in ground testing.  The value of $A(\theta,\phi)$ is generally largest for $\theta = 0$, when particles flow straight into the cup, and then falls off as $\theta$ increases and less of the collector is ``illuminated'' by inflowing particles.  If a Faraday cup has an asymmetric shape and/or multiple collectors, $A(\theta,\phi)$ will also depend on $\phi$.}  If the modulator voltage spans the voltage window $\mathcal{E}_0\pm\Delta\mathcal{E}/2$, then the contribution of all $j$-particles to the measured differential current is
\begin{equation}\label{eqn:fc:dij}
\Delta I_j = \int \mathrm dI_j = q_j \int \limits _{v^{(\mathrm c)}_j(\mathcal{E}_0-\Delta\mathcal{E}/2)}^{v^{(\mathrm c)}_j(\mathcal{E}_0+\Delta\mathcal{E}/2)} \mathrm dv_z \, v_z \int\limits _{-\infty}^{\infty} \mathrm dv_y \int\limits _{-\infty}^{\infty} \mathrm dv_x \, A(\theta,\phi)  f_j(\vec{v}).
\end{equation}
Since a Faraday cup cannot distinguish current from different types of particles, the measured current is
\begin{equation}\label{eqn:fc:di}
\Delta I = \sum_j \Delta I_j,
\end{equation}
where the sum is carried out over all particle species in the plasma.

Equations (\ref{eqn:fc:dij}) and (\ref{eqn:fc:di}) provide the general form of the response function of a Faraday cup.  Section \ref{sec:meas:partanls} overviews the process of inverting the response function to determine the particle moments from a measured particle spectrum.

\subsubsection{Electrostatic analyzers} \label{sec:meas:partinst:esa}

\begin{figure}
\begin{center}
\includegraphics[width=0.70\textwidth]{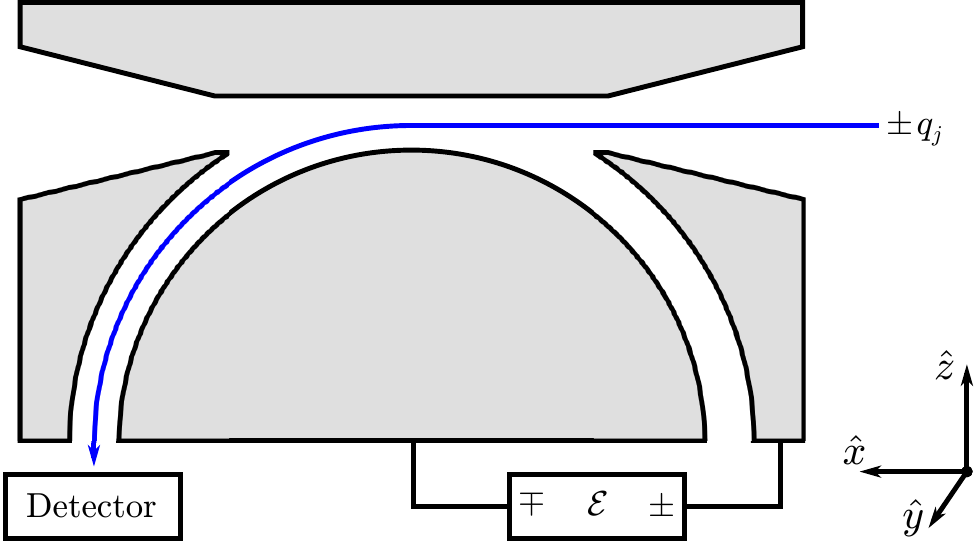}
\caption{\label{fig:esa} Simplified cross-sectional diagram of a \emph{top-hat} style  electrostatic analyzer (ESA).  The aperture is shown on the upper left and right, and can provide up to $360^\circ$ of coverage of azimuth $\phi$.  In contrast, only particles within a limited range of elevation $\theta$ are able to pass through the curved \emph{collimator} plates and reach the \emph{detector}.  A DC voltage $\mathcal{E}$ is sustained between the plates and sets the sign and value of the target \emph{energy per charge} $K/q_j$ for incoming particles. The spacing between the collimator plates defines the width of the energy windows.}
\end{center}
\end{figure}

Like Faraday cups, electrostatic analyzers (ESAs) have a long history of use in the observation of thermal particles in the solar wind.  Though ESAs are substantially more complex than Faraday cups, they enable much more direct and detailed studies of distribution functions (see Sect. \ref{sec:meas:partanls:imag}).  Additionally, they can be combined with mass spectrometers (see Sect. \ref{sec:meas:partinst:ms}) to directly probe the ion composition of the plasma.

Figure \ref{fig:esa} shows a simplified cross-section of the common \emph{top-hat} design for an ESA \citep{carlson1983}.  Such a device consists of two hemispherical shells that are nested concentrically so as to leave a narrow gap between them.  Particles enter via a hole in the top of the larger hemisphere and are then subjected to the electric field that is created by maintaining a DC voltage $\mathcal{E}$ between the two hemispheres.  The value of $\mathcal{E}$ and the curvature and spacing of the hemispheres define an \emph{energy-per-charge} range for an incoming particle to reach the detectors at the base of the hemispheres.  If an incoming particle has a kinetic energy $K$ and charge $q_j$, it can only reach the detectors if the ratio $K/q_j$ falls within that range.
To generate a particle spectrum, $\mathcal{E}$ is swept through a series of values.  The range of  particle energies is set by the range of $\mathcal{E}$ values, which, on most ESAs, can be adjusted in flight.  Nevertheless, the width of an ESA's energy window $\Delta K/K_0$ is fixed geometrically by the spacing between its collimator plates.  In contrast, the width of a Faraday cups' energy window is adjustable in flight since it is set by a voltage range according to Equation~(\ref{eqn:vwin}).

An ESA's detectors are typically arranged around the base of the hemispheres.  While Faraday cups detect incoming particles by measuring their net current, an ESA's detectors usually count particle cascades generated by the strikes from individual particles.  Such detectors would be impractical for a Faraday cup because they would be overwhelmed by solar UV photons.  On a top-hat ESA, the tight spacing of the deflectors and a low-albedo coating\footnote{For example, \emph{gold black} was used on the Wind/3DP ESAs \citep{lin1995}.} on their surfaces ensure that very few photons reach the detectors.  Each of the detectors is typically some type of electron multiplier, which uses an electrostatic potential in such a way that a strike by a single charged particle produces a cascade of electrons, which can then be registered.  \emph{Channel electron multipliers (CEMs)} were used for ACE/SWEPAM \citep{mccomas1998a}, while \emph{micro-channel plates (MCPs)} were used for Wind/3DP \citep{lin1995} and STEREO/IMPACT/SWEA \citep{sauvaud2008}.  Both CEM and MCP detectors require more complex calibration than is needed for a Faraday cup.  For example, after each particle strike, an electron multiplier experiences a \emph{dead time}, during which the electron cascade is in progress and the detector cannot respond to another particle.  Furthermore, electron multipliers (and MCPs in particular) often exhibit significant degradation in their efficiency with time.

A typical top-hat ESA has a \emph{fan-beam} field of view.  The size and number of detectors define its azimuthal resolution and coverage, and ESAs can be designed with up to $360^\circ$ of $\phi$-coverage.  In contrast, most ESAs only sample particles over a limited range of elevation $\theta$, and a number of strategies have been employed to provide $\theta$-coverage.  The ESAs in the Helios plasma investigation \citep{schwenn1975,rosenbauer1977} and in Wind/3DP \citep{lin1995} were designed to rely on spacecraft spin to sweep their fan beams.  Although the Cassini spacecraft was three-axis stabilized, its CAPS instrument suite was mounted on an actuator, which a motor rotated through about $180^\circ$ of azimuth every $3$~minutes \citep{young2004}.  The MAVEN spacecraft is likewise three-axis stabilized, but its SWIA instrument \citep{halekas2015} incorporated a second set of electrostatic deflectors to effectively steer its fan beam by adjusting the path of ions entering the top hat.  Finally, the unique design of MESSENGER/FIPS \citep{andrews2007} moved beyond the top hat to give that instrument wide $\theta$-coverage (versus a fan beam) but reduced aperture size.

For any given value of $\mathcal{E}$, each ESA detector essentially has its own effective collecting area $A_j(K,\theta,\phi)$, which depends on the energy $K = m_jv^2/2$ and direction $(\theta,\phi)$ of incoming $j$-particles.  The number of $j$-particles detected from an infinitesimal volume $\mathrm d^3\vec v$ of phase-space during a time interval $\Delta t$ is
\begin{equation}
\mathrm dN_j = \Delta t\, v A_j(K,\theta,\phi)\, \mathrm dn_j,
\end{equation}
where $\mathrm dn_j$ is the number density of $j$-particles in $\mathrm d^3\vec v$.  Substituting Equation~(\ref{eqn:dnj}) and converting to spherical coordinates gives
\begin{equation}
\mathrm dN_j = \frac{2\,\Delta t}{m_j^2}  A_j(K,\theta,\phi) f_j(K,\theta,\phi) K \sin\theta\, \mathrm dK\, \mathrm d\theta\, \mathrm d\phi,
\end{equation}
where $f_j$ has been parameterized in energy and direction rather than vector velocity.  The total number of $j$-particles detected in $\Delta t$ is
\begin{equation} \label{eqn:esa:dnj} 
\Delta N_j = \int \mathrm dN_j= \frac{2\,\Delta t}{m_j^2}  \int\limits _0^{\infty} \mathrm dK\, K \int\limits _{0}^{\pi} \mathrm d\theta\, \sin\theta \int\limits _0^{2\pi} \mathrm d\phi\, A_j(K,\theta,\phi) f_j(K,\theta,\phi).
\end{equation}
Formally, the integrals in Equation~(\ref{eqn:esa:dnj}) are carried out over all energies and directions (i.e., all of phase space) but most ESAs are designed so that a given detector is only sensitive to particles from a relatively narrow range of energies and directions.  Consequently, the detector's effective collecting area is often approximated as
\begin{equation}\label{eqn:area:approx}
A_j(K,\theta,\phi) \approx \begin{cases}\displaystyle \frac{A_0}{\sin\theta_0} & \text{if}\;\; |K-K_0|<\Delta K,\, |\theta-\theta_0|<\Delta\theta, \, |\phi-\phi_0| <\Delta\phi \\ 0 & \text{else} \end{cases},
\end{equation}
where $A_0$ is the nominal collecting area, $(\theta_0,\phi_0)$ is the look direction, $\Delta\theta$ and $\Delta\phi$ set the field of view, and $K_0$ and $\Delta K$ set the energy range of $j$-particles.  Using Equation~(\ref{eqn:area:approx}) and assuming that $\Delta K$, $\Delta\theta$, and $\Delta\phi$ are small relative to variations in $f_j(K,\theta,\phi)$, we approximate Equation~(\ref{eqn:esa:dnj}) as
\begin{equation}\label{eqn:esa:dnj:approx} 
\Delta N_j \approx \frac{2A_0K_0}{m_j^2} \Delta t\, \Delta K\, \Delta\theta\, \Delta\phi\, f_j(K_0,\theta_0,\phi_0)\approx \frac{2K_0^2}{m_j^2} G f_j(K_0,\theta_0,\phi_0),
\end{equation}
where
\begin{equation}
G \equiv A_0\, \Delta t \frac{\Delta K}{K_0} \Delta\theta\, \Delta\phi
\end{equation}
is known as the \emph{geometric factor}.  ESAs are often designed and operated in such a way that $G$ is approximately constant.

If an ESA does not have any mass-spectrometry capability (see Sect. \ref{sec:meas:partinst:ms}), then each of its detectors measures the count of all particles of any species that reach it.  Thus, the measured quantity is
\begin{equation}\label{eqn:esa:dn}
\Delta N = \sum_j \Delta N_j,
\end{equation}
where the sum is carried out over all particle species $j$.

Equations~(\ref{eqn:esa:dnj}) and (\ref{eqn:esa:dn}) specify the response function of a top-hat ESA.  A particle spectrum from such an instrument consists of a set of measured $\Delta N$-values made over various $\mathcal{E}$-values and in various directions.  Section \ref{sec:meas:partanls} describes how the response function can be used to extract information about particle distribution functions from a measured spectrum.

\subsubsection{Mass spectrometers} \label{sec:meas:partinst:ms}

As noted above, neither a Faraday cup nor an ESA can, on its own, directly distinguish among different ion species: they simply measure the current and counts, respectively, of the incoming particles.  A limited composition analysis, though, is still possible because the voltage $\mathcal{E}$ needed for either type of instrument to detect a $j$-particle of speed $v$ is proportional to $m_j/q_j$.  Though relative drift is often observed among different particle species in the solar wind, it generally remains far less than the bulk speed (see Sect.~\ref{sec:ion-prop}).  Thus, in a particle spectrum, the signals from different particle species appear shifted by their mass-to-charge ratios.  By separately analyzing these signals (see Sect. \ref{sec:meas:partanls}), values can be inferred for the moments of the various particle species.

This strategy does have significant limitations.  First, it provides no mechanism for distinguishing ions with the same mass-to-charge ratio (e.g., ${}^{12}{\rm C}^{3+}$ and ${}^{16}{\rm O}^{4+}$).  Second, even when particle species have distinct mass-to-charge ratios, ambiguity can still arise from the overlap of their spectral signal.  For example, the mass-to-charge ratios of protons and $\alpha$-particles differ enough that values for their moments can often be derived for both species from Faraday-cup \citep[e.g.,][Chapter 4]{kasper:phd} and ESA \citep[e.g.,][]{marsch1982} spectra.  Nevertheless, the $\alpha$-particle signal can suffer confusion with minor ions \citep[e.g.,][]{bame1975}, and, especially at low Mach numbers, the proton and $\alpha$-particle signals can almost completely overlap \citep[e.g.,][Sect. 3.3]{maruca:phd}.

A mass spectrometer is required to achieve the most accurate measurements of solar-wind composition \citep[see also the more complete review by][]{gloeckler1990}.  As opposed to being a separate instrument, a mass spectrometer is typically incorporated into an ESA as its detector system and is used to measure the speed of each particle.  The ESA ensures that only particles within a known, narrow range of energy per charge pass through.  As each particle enters the mass spectrometer, an electric field accelerates it by a known amount.  The particle then triggers a start signal by liberating electrons from a thin foil,\footnote{For example, a carbon foil supported by a nickel mesh was used on Ulysses/SWICS \citep{gloeckler1992}, ACE/SWICS \citep{gloeckler1998}, and STEREO/IMPACT/PLASTIC \citep{galvin2008}.} which are detected via an MCP.  Next, the particle travels a known distance  $\Delta s$  to another foil.\footnote{For example, the SWICS instruments on both Ulysses and ACE \citep{gloeckler1992,gloeckler1998} use a gold foil applied directly to the top of the detectors.}  The particle triggers a stop signal by passing through this latter foil before finally reaching the detector.  The time $\Delta t$ between the start and stop signals  is the particle's \emph{time of flight}, a measurement of which allows the particle's speed  $v=\Delta s/\Delta t$ through the mass spectrometer to be inferred.

Several different designs have been developed for mass spectrometers for heliophysics.  In a time-of-flight versus energy (TOF/E) mass spectrometer, such as Ulysses/SWICS \citep{gloeckler1992}, ACE/SWICS \citep[][Sect. 3.1]{gloeckler1998}, and STEREO/IMPACT/PLASTIC \citep{galvin2008}, \emph{solid-state detectors (SSDs)} are used to ultimately detect each ion.  Unlike an electron multiplier, an SSD is able to measure the energy of individual charged particles.  Therefore, a TOF/E instrument measures each ion's initial energy per charge, speed through the instrument, and residual energy at the detector.  Together, these quantities provide  sufficient information to determine the ion's mass, charge, and initial speed.  In contrast, a high-mass-resolution spectrometer (HMRS) such as ACE/SWIMS \citep[][Sect. 3.2]{gloeckler1998} does not need to measure the ions' residual energy and can simply use MCP detectors.  An HMRS exploits the fact that passing through the start foil tends to decrease an ion's charge state to either $0$ or $+1$.  The particle then passes through a known but non-uniform electric field, which deflects the singly ionized particle to the detectors.  The electric field causes the time of flight to be mass dependent, so each particle's mass can be inferred.

\subsection{Analyzing thermal-particle measurements} \label{sec:meas:partanls}

A particle spectrum, whether measured by a Faraday cup or an ESA, must be processed in order to extract information about the observed particles.  This involves inverting the instrument's response function -- Equations~(\ref{eqn:fc:dij}) and (\ref{eqn:fc:di}) for a Faraday cup, and Equations~(\ref{eqn:esa:dnj}) and (\ref{eqn:esa:dn}) for an ESA -- so that particle moments or phase-space densities can be derived from measured current or counts.  This section briefly describes three methods for achieving this: \emph{distribution-function imaging}, \emph{moments analysis}, and \emph{fitting of model distribution functions}.

\subsubsection{Distribution-function imaging} \label{sec:meas:partanls:imag}

Equation~(\ref{eqn:esa:dnj:approx}) suggests a very simple method for interpreting a particle spectrum from an ESA.  The number of counts $\Delta N_j$ of $j$-particles is approximately proportional to the value of the $j$-particles' distribution function $f_j$ at some point in phase space.  If only $j$-particles are considered, then the set of measured $\Delta N$-values (i.e., the particle spectrum) can be used to give a set of values for $f_j$ across phase space.
In this sense, an ESA's particle spectrum can be thought of as an image of a distribution function.  This is the method employed by \citet{marsch1982a,marsch1982} in their well-known contour-plots of proton and $\alpha$-particle distribution functions from the Helios mission (see also Figs.~\ref{fig:distributions} and~\ref{fig:electron_distribution} of this review).  Since this technique is not focused on extracting the values of particle moments, it is especially well suited to studying the three-dimensional structure of distribution functions and non-Maxwellian features.

Nevertheless, distribution-function imaging carries significant limitations.  First, in the case of ion measurements, significant confusion can arise among the various ion species in the plasma (see Sect. \ref{sec:meas:partinst:ms}).  If an ESA does not have a mass spectrometer, it simply measures the total count of particles $\Delta N$ rather than each individual $\Delta N_j$.  Second, various assumptions are made in deriving Equation~(\ref{eqn:esa:dnj:approx}).  Notably, the field of view and energy range were taken to be small relative to the scale of variations in the distribution function.  When these assumptions break down, this technique returns a distorted image of $f_j$.  Third, this technique cannot be applied to observations from a Faraday cup.  Essentially, a Faraday cup's large field of view means that each of its $\Delta I$-measurements samples a large region of phase space.  The integrals in Equation~(\ref{eqn:fc:dij}) cannot be easily simplified to give an expression like Equation~(\ref{eqn:esa:dnj:approx}).

Though ESA images of distribution functions can provide tremendous insight into phase-space structure, care must be exercised to properly account for instrumental effects.  Any ESA has finite angular and energy resolutions, which must be considered when interpreting their output.  An irregularity in a distribution function may seem significant in a contour plot but actually result from only a single datum with a low number of particle counts.  Such finite-resolution effects are often more pronounced in proton versus electron data because protons, being supersonic, are concentrated into a narrow beam of phase space.  A related effect arises in both ion and electron data from the finite period of time required for an ESA to sweep through its angular and energy ranges.  Especially during periods of high variability in the solar wind, this may result in distribution-function images that constitute ``hybrids'' of distinct plasma conditions.

\subsubsection{Moments analysis} \label{sec:meas:partanls:mom}

Moments analysis provides the most direct method for estimating particle moments from a measured particle spectrum.  Essentially, this technique relies on deriving relationships between the moments of a distribution function (see Sect.~\ref{sec:moments}) and the moments of the measured quantity: $\Delta I_j$ for a Faraday cup or $\Delta N_j$ for an ESA.  For the latter case, Equation~(\ref{eqn:esa:dnj:approx}) shows that $\Delta N_j$ is approximately proportional to $f_j$.  Thus, each moment of $f_j$ can be approximated with a discrete integral of $\Delta N_j$: a sum over all the measured $\Delta N$-values.  For a Faraday cup, the relationship between $\Delta I_j$ and $f_j$ in Equation~(\ref{eqn:fc:dij}) is more complex, but similar expressions exist to relate the moments of $f_j$ to sums of the measured $\Delta I$-values \citep[see, e.g.,][Appendix A]{kasper2006}.  In either case, the calculations are relatively simple.  For this reason, moments analyses are commonly implemented in spacecraft flight computers, which often have limited computational resources or limited down-link bandwidth for the transmission of full particle spectra.

Moments analysis carries the significant limitation that it provides no mechanism for easily distinguishing different components of a distribution function (e.g., its core and beam), or, in the case of ions, for differentiating among species (see Sect. \ref{sec:meas:partinst:ms}).  Additionally, the particle spectrum must provide excellent coverage of $f_j$ in phase space so that the discrete integrals of the measured $\Delta I$- or $\Delta N$-values can reasonably approximate the infinite integrals of $f_j$ that define its moments.

\subsubsection{Fitting model distribution functions} \label{sec:meas:partanls:fit}

In a fitting analysis of a particle spectrum, a \emph{model distribution} (such as those defined in Sect.~\ref{sec:distributions}) is chosen for each $f_j$-component and  particle species under consideration.  These model distributions are then substituted into the expression for $\Delta I$ for a Faraday cup in Equation~(\ref{eqn:fc:di}) or $\Delta N$ for an ESA in Equation~(\ref{eqn:esa:dn}).  This substitution gives an expression for the measured quantity, $\Delta I$ or $\Delta N$, in terms of the fit parameters of the model distributions: e.g., particle densities, velocities, and temperatures.  This model can then be fit to a measured spectrum to derive estimates of the particle moments.

Unlike moments analysis, fitting allows for the direct treatment of multiple $f_j$-components or ion species.  It also allows data to be weighted based on the uncertainty in each measurement and does not require that the particle spectrum cover almost all of phase space.  Indeed, \citet{kasper2006} use the microkinetic limits on temperature anisotropy to infer that fitting model distribution functions to ion measurements from the Wind/SWE Faraday cups produces temperature values that are significantly more accurate than those returned from a moments analysis.

The greatest disadvantage of fitting is the need to assume a model distribution.  If such a model does not capture all of the features of the actual distribution function, the fitting results are unreliable.  In addition, the complexity of the functions involved usually necessitates the use of non-linear fitting algorithms \citep[e.g., the Levenberg--Marquardt algorithm; see][]{marquardt1963}, which are computationally intensive and generally cannot be implemented on spacecraft computers.

\subsection{Magnetometers} \label{sec:meas:mag}

This section provides a brief overview of the three types of magnetometers most commonly used on heliophysics missions: \emph{search-coil magnetometers}, \emph{fluxgate magnetometers}, and \emph{helium magnetometers}.  The reviews by \citet{ness1970}, \citet{acuna1974,acuna2002}, and \citet{smith1976} provide much more detailed treatments of these and other types of magnetometers.

\subsubsection{Search-coil magnetometers}

Though simpler in design than fluxgate and helium magnetometers, search-coil magnetometers have been less frequently flown on space-physics missions because of their poor sensitivity to background magnetic fields and low-frequency magnetic fluctuations. The search-coil magnetometer was first used in space on Pioneer 1 \citep{sonett1960}.  Later, search coils were included in Wind/Waves \citep{bougeret1995}, Cluster/STAFF \citep{cornilleau1997}, and Themis/SCM \citep{roux2008}.

Essentially, a search-coil magnetometer is a coil of wire that wraps around a portion of a core made from a high-permeability material, which serves to amplify the magnetic field.  Let $\vec{B}_{\rm ext}$ denote the magnetic field external to the core, which is to be measured.  The magnetic field inside the core is
\begin{equation}
\vec{B}_{\rm int} = \mu_{\mathrm c} \vec{B}_{\rm ext},
\end{equation}
where $\mu_{\mathrm c}$ is the effective relative permeability of the core.  One complication is that $\mu_{\mathrm c}$ differs from $\mu_{\mathrm r}$, the relative permeability of the bulk material comprising the core.  In general,
\begin{equation}
\mu_{\mathrm c} = \frac{\mu_{\mathrm r}}{1+N_{\mathrm d}\left(\mu_{\mathrm r}-1\right)},
\end{equation}
where $N_{\mathrm d}$ is the \emph{demagnetization factor}, which reflects the core's particular geometry \citep[see, e.g.,][Sect.~2.4.3]{tumanski2011}.  For materials with relatively low permeability, $\mu_{\mathrm c} \approx \mu_{\mathrm r}$, but materials with high $\mu_{\mathrm r}$ are usually favored for search coils as they substantially boost sensitivity.

If the coil has $\mathcal{N}$ turns, then, by Faraday's law according to Equation~(\ref{Maxw4full}), the voltage induced in the coil is
\begin{equation}
\mathcal{E} = - \frac{\mathcal{N} A \mu_{\mathrm c}}{c} \, \frac{\mathrm dB_{{\rm ext},z}}{\mathrm dt},
\end{equation}
where $A$ is the core's cross-sectional area, and the core is oriented along the $z$-axis.  Thus, a measurement of $\mathcal{E}$ gives the rate of change in the axial component of $\vec{B}_{\rm ext}$.  If $B_{{\rm ext},z}(t)$ is sinusoidal,
\begin{equation}
B_{{\rm ext},z}(t) = B_{0,z} \cos\left( 2 \pi \nu t + \phi \right),
\end{equation}
the coil voltage is
\begin{equation}\label{eq:sc-volt}
\mathcal{E}(t) = \frac{2 \pi \nu \mathcal{N} A \mu_{\mathrm c} B_{0,z}}{c} \sin\left( 2 \pi \nu t + \phi \right).
\end{equation}

A single coil can only detect fluctuations in the $\vec{B}_{\rm ext}$ component parallel to the coil's axis.  Thus, search-coil magnetometers often include three orthogonal coils to enable measurements of the vector magnetic field.

The factor of $\nu$ in Equation~(\ref{eq:sc-volt}) indicates that a search coil's sensitivity scales linearly with frequency. Search-coil magnetometers are thus mostly used in the frequency range from a few Hz to several kHz. A non-accelerating search coil is completely insensitive to the background magnetic field. However, a search-coil magnetometer on a spinning spacecraft can still measure a constant field since the field is non-constant in the instrument's frame of reference.  This method was employed on Pioneer~1 to make the first measurements of the interplanetary magnetic field \citep{sonett1960,rosenthal1982}.

\subsubsection{Fluxgate magnetometers}

The fluxgate magnetometer was first invented for terrestrial use by \citet{aschenbrenner1936}, and since then, it has become the most widely used type of magnetometer in heliophysics missions.  Although the fluxgate magnetometer is more complex than the search-coil magnetometer, it is much better suited to measuring the background magnetic field and low-frequency ($\lesssim 10\,\mathrm{Hz}$) magnetic fluctuations.

\begin{figure}
\begin{center}
\includegraphics{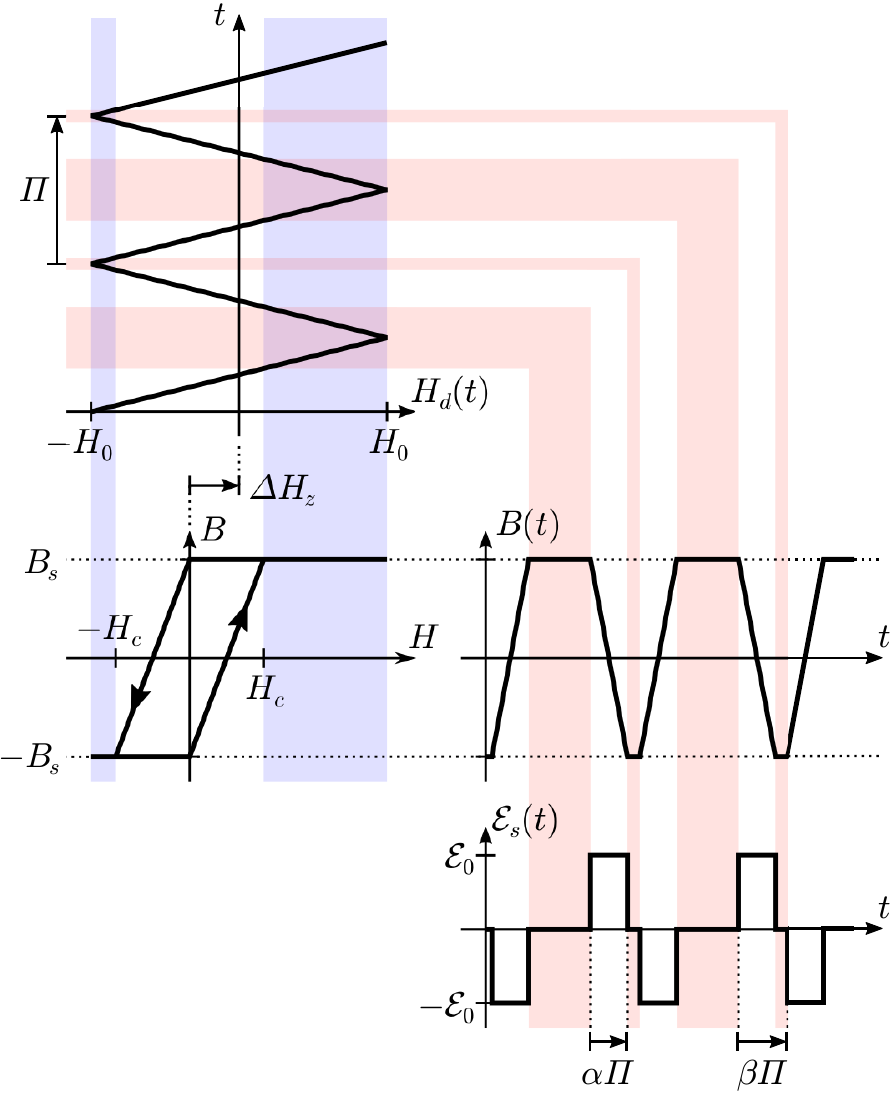}
\caption{\label{fig:fluxgate} The performance of an idealized, basic fluxgate magnetometer.  The hysteresis plot of the fluxgate's ferromagnetic core is shown in the center left and indicates the magnetic field $B$ in the core as a function of the auxiliary field $H$ applied to it.  The value of $H$ is the sum of the auxiliary field $H_{\mathrm d}$ from the fluxgate magnetometer's drive coil and the auxiliary field $\Delta H_z$ associated with the magnetic field external to the instrument.  The upper-left plot shows $H_{\mathrm d}(t)$, and $\Delta H_z$ is represented as a horizontal shift between the two left plots.  The value of $\Delta H$ has been greatly exaggerated for illustrative purposes.  The $H$-values for which the core is saturated are indicated by light-blue shading, and the times $t$ when this occurs are indicated by light-red shading.  The center-right plot shows the core's magnetic field $B(t)$, which is limited by the saturation value $B_{\mathrm s}$.  The lower-right plot shows the voltage $\mathcal{E}_{\mathrm s}(t)$ that $B(t)$ induces in the fluxgate magnetometer's sense coil. After \citet{ness1970}.}
\end{center}
\end{figure}

A fluxgate magnetometer relies on the \emph{hysteresis} of ferromagnetic materials.  The center-left plot in Fig.~\ref{fig:fluxgate} shows an idealized representation of the hysteresis curve for such a material.  The magnetic field $\vec{B}$ inside the material depends not only on the auxiliary field\footnote{Unfortunately, no widely accepted term for $\vec{H}$ exists.  Some authors \citep[e.g.,][]{jackson1975} refer to it as the ``magnetic field'' and use another term for $\vec{B}$.  Although there is some historical precedent for this naming convention, \citet{sommerfeld1952} and \citet{griffiths2013} strongly criticize it and contend that $\vec{B}$ is the more fundamental parameter.  We follow the convention used widely in modern space physics of referring to $\vec{B}$ as the ``magnetic field.''  For $\vec{H}$, we choose the term ``auxiliary field'' from \citet{griffiths2013}.} $\vec{H}$ applied to it but also on the history of the core's magnetization.  Nevertheless, there exists a critical $H$-value,  $H_{\mathrm c}$, such that the magnetic field is saturated at a strength $B_{\mathrm s}$ if $\left\lvert \vec{H} \right\rvert \geq H_{\mathrm c}$.

In a typical design, a fluxgate magnetometer consists of a ferromagnetic core wrapped by two coils of wire: a \emph{drive coil} and a \emph{sense coil}.  A triangle-wave current is applied to the drive coil to produce an auxiliary field $H_{\mathrm d}(t)$ that has an amplitude $H_0$ and period $\Pi$ (upper-left plot in Fig.~\ref{fig:fluxgate}).  The core's total auxiliary field is then
\begin{equation}
H(t) = H_{\mathrm d}(t) + \Delta H_z,
\end{equation}
where the $z$-direction corresponds to the axis of the core, and $\Delta H_z$ represents the contribution of the external magnetic field, which is to be measured.  The value of $H_0$ is chosen to be large enough that the core experiences both positive and negative saturation during each cycle of $H_{\mathrm d}(t)$.  As a result, the core's magnetic field $B(t)$ has the form of a truncated triangle wave (center-right plot in Fig.~\ref{fig:fluxgate}).  A non-zero value of $\Delta H_z$ produces a DC offset in $B(t)$, which means that the core spends different amounts of time in positive and negative saturation.  By Faraday's law according to Equation~(\ref{Maxw4full}), the voltage induced in the fluxgate magnetometer's sense coil is
\begin{equation}
\mathcal{E}_{\mathrm s} = - \frac{\mathcal{N}_{\mathrm s} A}{c} \frac{\mathrm dB}{\mathrm dt},
\end{equation}
where $\mathcal{N}_{\mathrm s}$ is the number of turns in the sense coil, and $A$ is the core's cross-sectional area.  Because of the offset and truncation in $B(t)$, $\mathcal{E}_{\mathrm s}(t)$ has the form of an irregular square wave (lower-right plot in Fig.~\ref{fig:fluxgate}).  We denote the duration of a positive or negative pulse as $\alpha \Pi$  and the time from the start of a positive pulse to the start of the next negative pulse as $\beta \Pi$.  Then,
\begin{equation}
\alpha = \frac{H_{\mathrm c}}{4 H_0} 
\end{equation}
and
\begin{equation}
\beta = \frac{1}{2}\left(1 - \frac{\Delta H}{H_0}\right).
\end{equation}
Typically, the value of $H_0$ is chosen so that it is substantially greater than $\Delta H_z$ and $H_{\mathrm c}$, in which case both $\alpha$ and $\beta$ are much less than one.
The sense-coil voltage shown in Fig.~\ref{fig:fluxgate} (lower right) has the Fourier series expansion \citep{ness1970}
\begin{equation}
\mathcal{E}_{\mathrm s}(t) = \mathcal{E}_0 \sum_{k=1}^\infty \left( 1 - e^{-i 2 \pi \beta k} \right) \frac{\sin\left(\pi \alpha k\right)}{\pi k} \cos\left( \frac{2 \pi k t}{\Pi} \right),
\end{equation}
where
\begin{equation}
\mathcal{E}_0 = - \frac{2 \mathcal{N}_{\mathrm s} A B_{\mathrm s}}{c \alpha \Pi}.
\end{equation}
In the absence of an external magnetic field, the values of $\Delta H$ and $\beta$ would both be zero, which would cause all even harmonics in the above series to vanish.  Thus, the second harmonic is typically measured in order to infer the value of $\Delta H_z$ and thereby the value of $B_z$.  

A single fluxgate sensor, like a single search-coil, is only sensitive to one component of the magnetic field.  Consequently, fluxgate magnetometers often consist of three orthogonal sensors so that the vector magnetic field can be measured.

A fluxgate magnetometer can be used to measure the background magnetic field and low-frequency magnetic fluctuations up to a few $10$'s of Hz \citep{ness1970} but it has poor sensitivity to fluctuations around or above the frequency of its drive coil.  Consequently, some missions carry not only fluxgate magnetometers but also search-coil magnetometers, which are better suited to measuring high-frequency magnetic fluctuations.  For example, the Wind spacecraft includes both the MFI fluxgate magnetometers \citep{lepping1995} and the Waves search-coil magnetometers \citep{bougeret1995}. Likewise, the four Cluster spacecraft include the FGM fluxgate magnetometers \citep{balogh1997} and the STAFF search-coil magnetometers \citep{cornilleau1997}.

More sophisticated designs for fluxgate magnetometers, which include additional coils and more complex geometries for the core, have been developed to improve sensitivity and to allow the instrument to be operated at higher frequencies.  Notably, \citet{geyger1962} introduced the use of toroidal cores, which were used, e.g., for the Pioneer 11 magnetometer \citep{acuna1974}, Voyager/MAG \citep{behannon1977}, Wind/MFI \citep{lepping1995}, and STEREO/IMPACT/MAG \citep{acuna2008}.

\subsubsection{Helium magnetometers} \label{sec:meas:mag:helium}

Helium magnetometers belong to a large class of magnetometers known as \emph{optically pumped magnetometers} \citep{ness1970,acuna2002}.  Though some optically pumped magnetometers use the vapor of an alkali metal (e.g., sodium, cesium, or rubidium) as their sensing medium, helium has been more widely used in space instruments.

The sensing element of a helium magnetometer is a \emph{cell} containing helium gas \citep{slocum1963}.  A radio-frequency oscillator is used to energize electrons in the gas, which collisionally excite helium atoms from their ground state, $1 ^{1}{\rm S}_0$, to their first excited state, $2 ^{3}{\rm S}_1$.  Since $1 ^{1}{\rm S}_0$ is a singlet state, and $2 ^{3}{\rm S}_1$ is a triplet, the transition between them via photon emission/absorption is doubly forbidden under classical selection rules.  As a result, the $2 ^{3}{\rm S}_1$ state is metastable.

Although collisional excitation produces equal populations for the three $2 ^{3}{\rm S}_1$ sub-levels, \emph{optical pumping} produces unequal populations for this triplet \citep{colegrove1960}.  A helium lamp serves a source of 1083~nm photons.  This light is then columnated into a beam, which passes through a circularly polarized filter before reaching the cell.  The 1083~nm wavelength corresponds to a helium atom's transition between the $2 ^{3}{\rm S}_1$ triplet state and the three closely-spaced $2 ^{3}{\rm P}$ states: $2 ^{3}{\rm P}_0$, $2 ^{3}{\rm P}_1$, $2 ^{3}{\rm P}_2$.  A helium atom in the $2 ^{3}{\rm S}_1$ state can transition to a $2 ^{3}{\rm P}$ state by absorbing one of these photons, after which it returns to $2 ^{3}{\rm S}_1$ via remission.  However, since the photons are circularly polarized, the atom, in the presence of a magnetic field, will preferentially return to one of the $2 ^{3}{\rm S}_1$ sub-levels over the other two.

An infrared detector is used to measure how much of the helium lamp's light is able to pass through the cell.  The transparency of helium to 1083~nm photons depends directly on the pumping efficiency, which in turn varies with the strength of the magnetic field and the field's angle with respect to the beam path.  Thus, the magnetic field can be inferred from measurements of the intensity of transmitted light.

A vector helium magnetometer typically includes three orthogonal pairs of \emph{Helmholtz coils} so that an arbitrary magnetic field can be applied to the cell in addition to the external magnetic field that is to be measured.  In the usual operating mode, a constant-magnitude magnetic field is rotated relative to the beam path at a frequency of a few 100's of Hz.  This results in a periodic variation in the intensity of transmitted light.  For a full vector measurement of the external magnetic field, the applied magnetic field is rotated through two orthogonal planes, each of which has an axis parallel to the beam path.

Vector helium magnetometers have been used on some heliophysics missions but not as many as fluxgate magnetometers.  In general, helium magnetometers are more complex and often require more mass and power than fluxgate magnetometers \citep{acuna2002}.  Nevertheless, helium magnetometers are effective for measuring strong magnetic fields, which makes them useful for planetary missions such as Pioneers~10 \&~11 \citep{smith1975}.  ISEE-3 \citep[later renamed ICE;][]{frandsen1978} also carried a vector helium magnetometer.  Some missions, including Ulysses \citep{balogh1992} and Cassini \citep{dunlop1999,dougherty2004}, carried both vector helium and fluxgate magnetometers.  The helium magnetometer on Cassini was unique in that it could be operated in either a scalar or vector mode (i.e., measure either $B$ or $\vec{B}$).  This design was developed to improve measurements of Saturn's strong magnetic field.

\subsection{Electric-field measurements} \label{sec:meas:elec}

Measurements of the vector electric field $\vec{E}$ in the solar wind are typically made over a very wide range of frequencies from a few kHz to tens of MHz.  The most common probes of $\vec{E}$ are monopole and dipole antennas, the lengths of which can vary based on scientific goals and practicalities.  For example, the length (spacecraft to tip) of each STEREO/Waves antenna is $6\,{\rm m}$ \citep{bale2008,bougeret2008}, while Wind/Waves has antennas that are $7.5\,{\rm m}$ and $50\,{\rm m}$ long \citep{bougeret1995}.

Electric-field instruments for heliophysics missions often utilize multiple receivers.  This not only helps to accommodate the wide range of frequencies but also allows for different observation modes to be implemented.  The simplest mode is \emph{waveform capture}, in which a time series of voltage measurements from each antenna is recorded.  This mode preserves the most information about $\vec{E}(t)$ but produces large amounts of data and thus is generally used only as a \emph{burst mode}.  An alternative mode is \emph{spectrum capture}, in which only the power spectral density is recorded at a predetermined set of frequencies.  This significantly lowers the data volume while preserving frequency information.  As a matter of practice, this mode is often implemented with a narrow-band receiver that is stepped through a series of discrete frequency ranges to measure the total power in each.

Electric-field instruments also have uses beyond simply measuring $\vec{E}$ for its own sake.  Although these applications are beyond the scope of this review, two merit brief mention here.  The first is the measurement of the \emph{quasi-thermal noise spectrum}, which can be used to infer the properties of electrons \citep{meyervernet1989}.  When an antenna is surrounded by a plasma, the antenna's frequency response is altered in a predictable way at frequencies near the electron plasma frequency $\omega_{\rm pe}$.  As shown in Equation~(\ref{eq:plasmafreq}), $\omega_{\rm pe}$ is proportional to $\sqrt{n_{\rm e}}$, so the determination of $\omega_{\rm pe}$ from the quasi-thermal noise spectrum is a direct measure of the electron density $n_{\rm e}$.  In addition, the temperature and some non-thermal properties of electrons can be extracted from the shape of the quasi-thermal noise spectrum.  Second, antennas can be used very effectively as dust detectors because of the large size of the antennas and the distinctive electrical signal produced by a dust grain striking an antenna \citep{couturier1981,lechat2009}.  The abundance and size-distribution of dust particles have been studied using measurements from STEREO/Waves \citep{zaslavsky2012} and Wind/Waves \citep{kellogg2016}.

\subsection{Multi-spacecraft techniques} \label{sec:meas:multi}

Most of the observational results presented in this review are based on measurements from individual spacecraft.  Nevertheless, powerful techniques have been developed to analyze simultaneous in-situ measurements from multiple spacecraft to distinguish between spatial and temporal fluctuations in the plasma.  This section offers a brief description of the key concepts.  

Spacecraft separated by relatively large distances ($\gtrsim 0.1\,{\rm au}$) offer particular benefits for observing remote or large-scale phenomena.  For example, the primary motivation of the aptly named STEREO mission \citep{kaiser2008} was to provide stereoscopic observations of the Sun and the inner heliosphere.  The in-situ particle instruments of the PLASTIC suite were designed for studies of the temporal and spatial variations of ICMEs \citep{galvin2008}.  Likewise, the Waves investigation allowed for the triangulation (\emph{radiogoniometry}) of radio-burst source regions \citep[][Sect.~3.4]{bougeret2008}, which has also been achieved using spacecraft from separate missions \citep{steinberg1984,hoang1998,reiner1998}.

Constellations of spacecraft with tighter spacings are used to observe local or small-scale plasma phenomena, especially in Earth's magnetosphere and magnetosheath.  This approach was largely pioneered with the Cluster mission \citep{escoubet1997} and later employed and expanded upon for THE\-MIS/AR\-TE\-MIS  \citep{angelopoulos2008} and MMS \citep{burch2016}.  In each of these missions, at least four spacecraft were flown in a quasi-tetrahedral formation to utilize three basic techniques \citep{dunlop1988}:
\begin{itemize}
\item In \emph{curlometry}, a four-point measurement of the magnetic field $\vec{B}$ is used to estimate $\vec{\nabla}\times\vec{B}$ and thereby the current density $\vec{j}$ \citep{robert2000}.  This technique relies on $\vec{j}$ being nearly uniform within the tetrahedron, so it is best suited to study phenomena on spatial scales of order or larger than the dimension of the constellation.
\item For the \emph{wave-telescope} technique, a Fourier analysis of $\vec{B}$-measurements from the four spacecraft is made to determine the frequency spectrum, directional distribution, and mode of plasma fluctuations \citep{neubauer1990,pincon2000,motschmann2000}.  Due to effects such as aliasing, this method is most accurate in characterizing waves comparable in scale to the spacecraft constellation \citep{sahraoui2010a}.
\item In a \emph{discontinuity analysis}, the arrival times of a magnetic discontinuity (e.g., a shock) at the spacecraft are compared so that the discontinuity's orientation and velocity can be inferred \citep{russell1983,mottez1994,dunlop2000}.  This method is most accurate for discontinuities whose boundary regions are thin relative to the spacecraft separations.
\end{itemize}

\section{Coulomb collisions}\label{sec:col}

Collisions among particles provide the fundamental mechanism through which an ionized or neutral gas increases its entropy and ultimately comes into thermal equilibrium. In a fully ionized plasma, hard scatterings rarely occur; instead, \emph{Coulomb collisions}, in which charged particles slightly deflect each other, are the primary collisional means by which particles exchange momentum and energy. The solar wind's low density ensures that the rates of particle collisions remain relatively low.  In contrast, the denser plasma of the solar corona has a much higher collision rate, and collisional processes are understood to be an important ingredient in the heating and acceleration of coronal plasma (see Sect.~\ref{sec:col:dim}). Unfortunately, this has led to the widespread misconception that, beyond the solar corona, Coulomb collisions have no impact on the evolution of solar-wind plasma. In reality, while collision rates in the solar wind can be very low, the effects of collisions on the plasma never truly vanish.

This section overviews the effects that Coulomb collisions have on the microkinetics and large-scale evolution of solar-wind plasma through interplanetary space.  Section~\ref{sec:col:dim} provides a simple dimensional analysis of Coulomb collisions, while Sect.~\ref{sec:col:theory} overviews the more complete kinetic theory of particle collisions in plasmas.  Section~\ref{sec:col:relax} describes observations of solar-wind collisional relaxation.

\subsection{Dimensional analysis of Coulomb collisions}\label{sec:col:dim}

Before addressing the detailed kinetic treatment of collisions, we use dimensional analysis to derive a very rough expression for the rate of collisions in a plasma among particles of the same species.

We consider a species whose particles have mass $m_j$ and charge $q_j$.  The $j$-particles may be approximated as all traveling at the species' thermal speed $w_j$.  When a pair of $j$-particles collide, kinetic energy is temporarily converted into electric potential energy.  Assuming (very crudely) that this conversion is complete,
\begin{equation}
2\left(\frac{1}{2} m_j w_j^2\right) = \frac{q_j^2}{x_{\min}},
\end{equation}
where $x_{\min}$ is the particles' distance of closest approach.  Consequently,
\begin{equation}
\sigma \equiv \pi x_{\min}^2 = \frac{\pi q_j^4}{m_j^2 w_j^4}
\end{equation}
is the scattering cross-section for collisions among $j$-particles.

We now consider a volume $V$ containing $N_j$ of the $j$-particles.  The average time $t_j$ that a $j$-particle goes between collisions is roughly equal to the time that it takes to sweep out $1/N_j$ of the total volume.  Taking $\sigma$ to be the particle's effective cross-sectional area,
\begin{equation}
\frac{1}{n_j} = \frac{V}{N_j} = \sigma w_j t_j,
\end{equation}
where $n_j$ is the number density of $j$-particles.  Thus,
\begin{equation}\label{eqn:col:est}
t_j = \frac{1}{n_j w_j \sigma} = \frac{m_j^2 w_j^3}{\pi q_j^4 n_j} = \frac{2^{3/2}m_j^{1/2}\left(k_{\mathrm B} T_j\right)^{3/2}}{\pi q_j^4 n_j}.
\end{equation}

Though Equation~(\ref{eqn:col:est}) was derived from a na\"ive treatment of Coulomb collisions, it can be used to approximate the collisionality of a species such as protons.  For example, at $r = 1\,{\rm au}$ from the Sun, $n_{\mathrm p} \sim 3\,{\rm cm}^{-3}$ and $T_{\mathrm p} \sim 10^5\,{\rm K}$.  These correspond to a proton collisional timescale of $t_{\mathrm p} \sim 10^8\,{\rm s}$, which is substantially longer than the solar wind's typical expansion time to this distance; see Equation~(\ref{expansion_time}).  In contrast, in the middle corona (see Fig.~\ref{fig:radial_profiles}), $n_{\mathrm p}\sim10^8\,{\rm cm}^{-3}$ and $T_{\mathrm p}\sim10^6\,{\rm K}$, which give $t_{\mathrm p}\sim 350\,{\rm s}$.  These estimates, though very rough, reveal that collisional effects have substantially more impact on coronal versus solar-wind plasma.

The stark difference in collisionality between the solar corona and solar wind forms the basis of \emph{exospheric models} of the heliosphere.  Although these models fall beyond the scope of this review, they warrant some mention.  Since the early work on exospheric models by \citet{jockers1968,jockers1970} and \citet{lemaire1971b,lemaire1971a}, they have been shown to account for some features of the interplanetary solar wind.  For example, the preferential heating of minor ions in a coronal exosphere can lead to the preferential acceleration of these ions \citep{pierrard2004}.  \citet{maksimovic2005} offer a more complete overview of exospheric models, and the reviews by \citet{marsch1994a} and \citet{echim2011} provide an even more detailed treatment of the subject.

\subsection{Kinetic theory of collisions}\label{sec:col:theory}

A full treatment of the kinetic theory of collisions in plasmas is beyond the scope of this review.  Instead, this section serves as a brief description of how the collisional term of the Boltzmann equation is used to derive collision rates for particle moments.  More complete presentations of the theory are given by \citet{spitzer1956}, \citet{longmire1963}, \citet{braginskii1965}, \citet{wu1966}, \citet{burgers1969}, \citet[][Chapters~6 and~7]{krall1973}, \citet{schunk1975,schunk1977}, \citet[][Chapter~4]{lifshitz1981},  \citet{klimontovich1997}, and \citet{fitzpatrick2015}.

\subsubsection{The collision term}

Discussions of particle collisions in gases usually begin with the Boltzmann equation (\ref{boltzmann}) since the effects of collisions are neatly grouped into the \emph{collision term} on the right-hand side of the equation:
\begin{equation}\label{eqn:boltz}
\frac{\partial f_j}{\partial t}+\vec v\cdot \frac{\partial f_j}{\partial \vec x}+\vec a\cdot \frac{\partial f_j}{\partial \vec v}=\left(\frac{\delta f_j}{\delta t}\right)_{\mathrm c},
\end{equation}
where the derivative $\left(\delta/\delta t\right)_{\mathrm c}$ is known as the \emph{collision operator}.  The separation of the collision term from the terms on the left-hand side becomes somewhat murky for plasmas.  Coulomb collisions occur through the interaction of the \emph{particle electric fields}, but the plasma's \emph{background electric field} contributes to the acceleration $\vec{a}$.  The particle electric field is the field generated by a single particle, while the background electric field is the collective result of all neighboring charged particles. Ultimately, the distinction between collisions and the effects of the background fields is phenomenological.  Under the \emph{molecular chaos hypothesis} (or \emph{sto\ss{}zahlansatz}), collisions among particles are assumed to be uncorrelated and to occur randomly \citep{maxwell1867}.

\begin{figure}
\begin{center}
\includegraphics{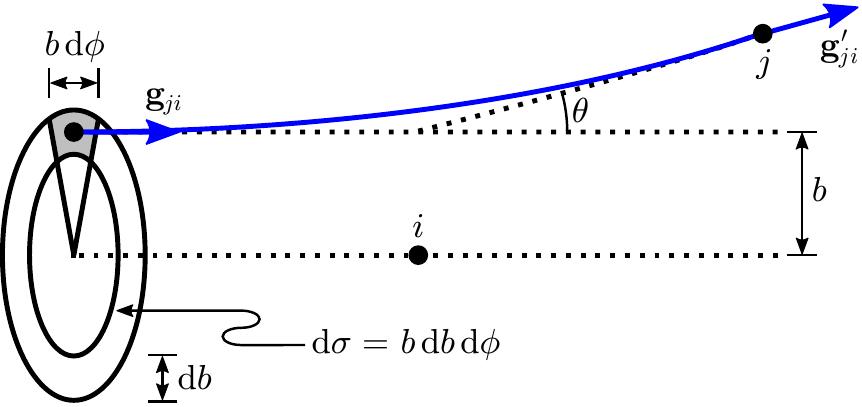}
\caption{\label{fig:scatter} Diagram of a $j$-particle scattering off of an $i$-particle via the electric force in the $i$-particle's reference frame, in which the $j$-particle has an initial velocity $\vec{g}_{ji}$ and a final velocity $\vec{g}'_{ji}$; see Equations~(\ref{eqn:col:relvel:i}) and (\ref{eqn:col:relvel:f}).}
\end{center}
\end{figure}

To derive an expression for the collisional term, we consider the Coulomb scattering of a $j$-particle off of an $i$-particle via the electric force.  We define the particles' initial velocities as $\vec{v}_j$ and $\vec{v}_i$, their final velocities as $\vec{v}^{\prime}_j$ and $\vec{v}^{\prime}_i$, their masses as $m_j$ and $m_i$, and their charges as $q_j$ and $q_i$.  We note that the $j$- and $i$-particles may be of the same species.  The center-of-mass velocity of the two particles is
\begin{equation}
\vec{u}_{ji} \equiv \frac{m_j\vec{v}_j + m_i\vec{v}_i}{m_j + m_i} = \frac{m_j\vec{v}^{\prime}_j + m_i\vec{v}^{\prime}_i}{m_j + m_i} \equiv \vec{u}^{\prime}_{ji},
\end{equation}
which is unchanged by the collision.  
Figure~\ref{fig:scatter} depicts this scattering event in the $i$-particle's frame of reference, in which the $j$-particle has an initial velocity
\begin{equation}\label{eqn:col:relvel:i}
\vec{g}_{ji} \equiv \vec{v}_j - \vec{v}_i 
\end{equation}
and a final velocity
\begin{equation}\label{eqn:col:relvel:f}
\vec{g}^{\prime}_{ji} \equiv \vec{v}^{\prime}_j - \vec{v}^{\prime}_i.
\end{equation}
We denote the impact parameter as $b$  and the scattering angle as $\theta$.  In a Coulomb collision, these two quantities are related by
\begin{equation}\label{eqn:btheta}
\tan\left(\frac{\theta}{2}\right) = \frac{q_jq_i}{m_{ji}g_{ji}^2 b},
\end{equation}
where
\begin{equation}
m_{ji} \equiv \frac{m_jm_i}{m_j+m_i}
\end{equation}
is the reduced mass of the two particles \citep[see, e.g.,][]{thornton2004,fitzpatrick2015}.
We consider an infinitesimal portion of the impact-parameter plane (see Fig.~\ref{fig:scatter}) as
\begin{equation} \label{eqn:dsigma}
{\mathrm d}\sigma = b\,{\mathrm d}b\,{\mathrm d}\phi.
\end{equation}
All $j$-particles that originate from this region are scattered into an infinitesimal solid-angle centered on $\theta$:
\begin{equation} \label{eqn:domega}
{\mathrm d}\Omega = \sin\theta\,{\mathrm d}\theta\,{\mathrm d}\phi.
\end{equation}
To derive the differential cross-section for a Coulomb collision, we assume that the colliding particles only interact electrostatically.  Then, when we combine Equations~(\ref{eqn:dsigma}) and~(\ref{eqn:domega}) with that for the Coulomb force, we arrive at the \emph{Rutherford cross-section} \citep{rutherford1911,geiger1913}:
\begin{equation}
\frac{{\mathrm d}\sigma}{{\mathrm d}\Omega} = \frac{q_j^2 q_i^2}{4 m_{ji}^2 g_{ji}^4\sin^4(\theta/2)}.
\end{equation}

Now, we consider all $i$-particles in the infinitesimal volume of phase space ${\mathrm d}^3\vec v_i$ that is centered on $\vec{v}_i$.  The rate (i.e., the number of particles per unit time) at which $j$-particles, originating from ${\mathrm d}\sigma$, collide with $i$-particles in ${\mathrm d}^3\vec v_i$ is
\begin{equation}
f_i(\vec{v}_i) g_{ji}\,{\mathrm d}\sigma\,{\mathrm d}^3\vec v_i = f_i(\vec{v}_i)g_{ji}\frac{{\mathrm d}\sigma}{{\mathrm d}\Omega}{\mathrm d}\Omega\,{\mathrm d}^3\vec v_i.
\end{equation}
Thus, the rate of decrease in the value of $f_j(\vec{v}_j)$ due to collisions with $i$-particles in all regions of phase space is
\begin{equation}\label{eqn:dfj:dt:cip}
\left(\frac{\delta f_j}{\delta t}\right)_{{\mathrm c},i,-} = - \int {\mathrm d}^3\vec v_i \int {\mathrm d}\Omega\,f_j(\vec{v}_j) f_i(\vec{v}_i) g_{ji} \frac{{\mathrm d}\sigma}{{\mathrm d}\Omega}.
\end{equation}
The above expression is negative because it only accounts for the decrease in $f_j(\vec{v}_j)$ due to $j$-particles of velocity $\vec{v}_j$ being scattered to other velocities by $i$-particles.  The value of $f_j(\vec{v}_j)$ can also increase as collisions scatter $j$-particles of other velocities to $\vec{v}_j$.  Indeed, Coulomb collisions are symmetric: if $j$- and $i$-particles of initial velocities $\vec{v}^{\prime}_j$ and $\vec{v}^{\prime}_i$ collide at an impact parameter $b$, their final velocities will be $\vec{v}_j$ and $\vec{v}_i$.  Thus, the rate of increase in $f_j(\vec{v}_j)$ due to collisions with $i$-particles is
\begin{equation}\label{eqn:dfj:dt:cin}
\left(\frac{\delta f_j}{\delta t}\right)_{{\mathrm c},i,+} = \int {\mathrm d}^3\vec v_i \int {\mathrm d}\Omega\,f_j(\vec{v}^{\prime}_j) f_i(\vec{v}^{\prime}_i) g_{ji} \frac{{\mathrm d}\sigma}{{\mathrm d}\Omega}.
\end{equation}
We note that, in the above equation, $\vec{v}^{\prime}_j$ and $\vec{v}^{\prime}_i$ are functions of $\vec{v}_j$, $\vec{v}_i$, and $\theta$.  The net rate of change in $f_j(\vec{v}_j)$ due to collisions with $i$-particles is
\begin{multline} \label{eqn:dfj:dt:ci}
\left(\frac{\delta f_j}{\delta t}\right)_{{\mathrm c},i}  = \left(\frac{\delta f_j}{\delta t}\right)_{{\mathrm c},i,+} + \left(\frac{\delta f_j}{\delta t}\right)_{{\mathrm c},i,-}\\ 
= \int {\mathrm d}^3\vec v_i \int {\mathrm d}\Omega \left[ f_j(\vec{v}^{\prime}_j) f_i(\vec{v}^{\prime}_i) - f_j(\vec{v}_j) f_i(\vec{v}_i) \right] g_{ji} \frac{{\mathrm d}\sigma}{{\mathrm d}\Omega}.
\end{multline}
Finally, the net rate of change in $f_j(\vec{v}_j)$ due to collisions with all species (i.e., the full collision term) is
\begin{multline} \label{eqn:col:term}
\left(\frac{\delta f_j}{\delta t}\right)_{\mathrm c} = \sum_{i} \left(\frac{\delta f_j}{\delta t}\right)_{{\mathrm c},i} \\
= \sum_{i} \int {\mathrm d}^3\vec v_i \int {\mathrm d}\Omega \left[ f_j(\vec{v}^{\prime}_j) f_i(\vec{v}^{\prime}_i) - f_j(\vec{v}_j) f_i(\vec{v}_i) \right] g_{ji}\frac{{\mathrm d}\sigma}{{\mathrm d}\Omega}.
\end{multline}
This includes Coulomb collisions of $j$-particles with other $j$-particles, so the above sum must include $i = j$.

\subsubsection{The Landau collision integral}

Evaluating Equation~(\ref{eqn:col:term}) is highly non-trivial but it is helped by the fact that the dominant contribution comes from small-angle collisions: those that produce small $\theta$-values.  Before invoking the small-$\theta$ limit, it is convenient to express the particles' initial and final velocities in terms of the center-of-mass velocity $\vec{u}_{ji} = \vec{u}^{\prime}_{ji}$ as
\begin{equation}
\vec{v}_j=\vec{u}_{ji}+\frac{m_{ji}}{m_j} \vec{g}_{ji},
\end{equation}
\begin{equation}
\vec{v}^{\prime}_j=\vec{u}_{ji}+\frac{m_{ji}}{m_j}\vec{g}^{\prime}_{ji},
\end{equation}
\begin{equation}
\vec{v}_i=\vec{u}_{ji}-\frac{m_{ji}}{m_i}\vec{g}_{ji},
\end{equation}
and
\begin{equation}
\vec{v}^{\prime}_i=\vec{u}_{ji}-\frac{m_{ji}}{m_i}\vec{g}^{\prime}_{ji}.
\end{equation}
Thus,
\begin{equation}\label{eqn:col:chng1}
\vec{v}^{\prime}_j = \vec{v}_j + \frac{m_{ji}}{m_j}\Delta\vec{g}_{ji}
\end{equation}
and
\begin{equation} \label{eqn:col:chng2}
\vec{v}^{\prime}_i = \vec{v}_i - \frac{m_{ji}}{m_i}\Delta\vec{g}_{ji},
\end{equation}
where
\begin{equation}
\Delta\vec{g}_{ji} \equiv \vec{g}^{\prime}_{ji} - \vec{g}_{ji}.
\end{equation}
In the small-$\theta$ limit, $\left\lvert\Delta\vec{g}_{ji}\right\rvert$ is also small, so Equations~(\ref{eqn:col:chng1}) and~(\ref{eqn:col:chng2}) can be used as the basis for a Taylor expansion of  $f_j$ and $f_i$ about $\vec{v}=\vec{v}_j$ and $\vec{v}=\vec{v}_i$, respectively.  Retaining terms through the second order gives
\begin{equation}
f_j(\vec{v}^{\prime}_j) \approx f_j(\vec{v}_j) + \frac{m_{ji}}{m_j}\Delta\vec{g}_{ji}\cdot\frac{\partial f_j}{\partial\vec{v}_j}+ \frac{m_{ji}^2}{2 m_j^2} \Delta\vec{g}_{ji}\,\Delta\vec{g}_{ji}:\frac{\partial^2 f_j}{\partial\vec{v}_j\partial\vec{v}_j}
\end{equation}
and
\begin{equation}
f_i(\vec{v}^{\prime}_i) \approx f_i(\vec{v}_i) - \frac{m_{ji}}{m_i} \Delta\vec{g}_{ji}\cdot\frac{\partial f_i}{\partial\vec{v}_i} + \frac{m_{ji}^2}{2 m_i^2}\Delta\vec{g}_{ji}\,\Delta\vec{g}_{ji}:\frac{\partial^2 f_i}{\partial\vec{v}_i\partial\vec{v}_i}.
\end{equation}
These approximations can be substituted into Equation~(\ref{eqn:dfj:dt:ci}), which, after considerable simplification \citep[see, e.g.,][]{hellinger2009,fitzpatrick2015}, yields the \emph{Landau collision integral/operator} \citep{landau1936,landau1937}:
\begin{multline}\label{eqn:col:landau}
\left(\frac{\delta f_j}{\delta t}\right)_{{\mathrm c},i} \approx  \frac{2 \pi q_j^2 q_i^2}{m_j} \ln\Lambda_{ji}\\
\times  \frac{\partial}{\partial \vec{v}_j}\cdot\left[ \int {\mathrm d}^3\vec v_i \frac{\mathsf I_3\,g_{ji}^2-\vec{g}_{ji} \vec{g}_{ji}}{g_{ji}^3}\cdot\left( \frac{f_i(\vec{v}_i)}{m_j}\frac{\partial f_j}{\partial\vec{v}_j} - \frac{f_j(\vec{v}_j)}{m_i}\frac{\partial f_i}{\partial\vec{v}_i} \right) \right],
\end{multline}
where $\ln\Lambda_{ji}$ is the \textit{Coulomb logarithm}, which is the subject of Sect.~\ref{sec:col:theory:log} and is given in Equation~(\ref{eqn:col:log}).

Although Equation~(\ref{eqn:col:landau}) is an improvement over Equation~(\ref{eqn:dfj:dt:ci}), actually calculating the Landau collision integral remains a daunting task even for relatively simple scenarios.  Often, additional approximations are introduced, and numerical methods are employed.  An alternative approach is the \emph{BGK operator}, which explicitly models the departure of a particle species' distribution function from its equilibrium state \citep{bhatnagar1954}.  This method was later generalized for the case of magnetized plasmas \citep[][and references therein]{dougherty1964}.  \citet{pezzi2015} present a numerical comparison of the Landau and Dougherty collision operators.

\subsubsection{The Coulomb logarithm}\label{sec:col:theory:log}

The factor $\ln \Lambda_{ji}$ in Equation~(\ref{eqn:col:landau}) is known as the \textit{Coulomb logarithm}:
\begin{equation}\label{eqn:col:log}
\ln\Lambda_{ji} \equiv \int\limits _{b_{ji,\min}}^{b_{ji,\max}} \frac{{\mathrm d}b}{b} = \ln\left(\frac{b_{ji,\max}}{b_{ji,\min}}\right).
\end{equation}
It arises from the $\Omega$-integral in Equation~(\ref{eqn:dfj:dt:ci}) via the relationship between $b$ and $\theta$ according to Equation~(\ref{eqn:btheta}).  Even though the derivation of Equation~(\ref{eqn:col:landau}) would seemingly imply that all $b$ from $0$ to $\infty$ should be considered, the Coulomb logarithm diverges at both of these limits.  As a result, the integral in Equation~(\ref{eqn:col:log}) has been given the more restrictive limits $b_{ji,\min}$ and $b_{ji,\max}$, which are discussed below. Though there is some degree of arbitrariness in how these limits are defined, Equation~(\ref{eqn:col:log}) is relatively insensitive to their particular values.  In practice, $b_{ji,\min} \ll b_{ji,\max}$, so the logarithm of their ratio only changes appreciably when they are varied by orders of magnitude.

The integral in Equation~(\ref{eqn:col:log}) diverges at small $b$ due to the breakdown of the small-$\theta$ limit used to derive Equation~(\ref{eqn:col:landau}): as the value of $b$  decreases, the value  of $\theta$ increases until it can no longer be considered small.  In reality, collisions with small $b$ have a minimal effect on the distribution function because of their relative rarity.  As a result, collisions with $\theta > \theta_{\max}$ are negligible and may be safely disregarded.  A typical choice is $\theta_{\max} = 90^{\circ}$, which, by Equation~(\ref{eqn:btheta}), corresponds to
\begin{equation}
b_{ji,\min} = \frac{q_jq_i}{m_{ji}\overline{g}_{ji}^2},
\end{equation}
where $\overline{g}_{ji}$ is the average speed of a $j$-particle relative to an $i$-particle.  The quantity $m_{ji}\overline{g}_{ji}^2$ roughly reflects the average kinetic energy of $j$- and $i$-particles in the plasma frame.  As a result,
\begin{equation}
b_{ji,\min} = \frac{q_jq_i}{k_{\mathrm B}T_{ji}},
\end{equation}
where $T_{ji}$ is the average temperature of the $j$- and $i$-particles.

The divergent behavior of Equation~(\ref{eqn:col:log}) at high $b$ stems from a more subtle reason.  The analysis above begins by considering the scattering of a single particle by another.  Effectively, the motion of each particle is modeled as a series of hard scatters, between which the particle's velocity remains constant.  In reality, Coulomb collisions are soft scatters, and each plasma particle is simultaneously colliding with many other particles.  As a result, each particle is partially shielded from the influence of distant particles by the particles closer to it.  An appropriate choice, then, for $b_{ji,\max}$ is the Debye length $\lambda_{\mathrm D}$  \citep{cohen1950,spitzer1956} as defined in Equation~(\ref{intro:debye_total}).  Taking into account all the particle species in the plasma,
\begin{equation}
b_{ji,\max} = b_{\max} \equiv \left( \frac{4\pi}{k_{\mathrm B}} \sum\limits _\ell \frac{q_\ell^2 n_\ell}{T_\ell} \right)^{-1/2},
\end{equation}
where $q_\ell$, $n_\ell$, and $T_\ell$ are the charge, number density, and temperature of each species in the plasma.  As a result of this choice, the value of $b_{ji,\max}$ is the same for all pairs of particle species.

This discussion of $b_{ji,\max}$  raises some concern over the use of binary collisions at all.  In principle, a more accurate approach would be to use an analysis of Markovian processes to derive the collision operator from the Fokker--Planck equation \citep{fokker1914,planck1917}.  Nevertheless, \citet[][Sects. 2-6]{wu1966} notes that both analyses produce the same result, Equation~(\ref{eqn:col:landau}), in the limit of small-angle scattering.

\subsubsection{Rosenbluth potentials}

An alternative expression for the Landau collision integral in Equation (\ref{eqn:col:landau}) can be obtained by using the \emph{Rosenbluth potentials} \citep{rosenbluth1957}, which are defined as
\begin{equation}\label{eqn:rosenbluth1}
G_i(\vec{v}_j) \equiv \int \left|\vec g_{ji} \right|  f_i(\vec{v}_i)\, {\mathrm d}^3 \vec v_i 
\end{equation}
and
\begin{equation}\label{eqn:rosenbluth2}
H_i(\vec{v}_j) \equiv \int \frac{1}{\left|\vec g_{ji}\right|} f_i(\vec{v}_i)\, {\mathrm d}^3\vec v_i.
\end{equation}
Likewise, we define flux densities associated with friction
\begin{equation}
\vec{A}_{ji} \equiv \frac{4 \pi q_j^2 q_i^2}{m_i}\ln\Lambda_{ji}\frac{\partial H_i}{\partial \vec{v}_j} 
\end{equation}
and with diffusion
\begin{equation}
\mathsf{D}_{ji} \equiv \frac{2 \pi q_j^2 q_i^2}{m_j} \ln\Lambda_{ji} \frac{\partial^2 G_i}{\partial \vec{v}_j\partial \vec{v}_j}.
\end{equation}
With these quantities defined, we express the Landau collision operator as the velocity divergence of the sum of these fluxes \citep[see][]{montgomery1964,marsch2006,fitzpatrick2015}, casting it in terms of a Fokker--Planck advection-diffusion equation in velocity space:
\begin{equation}
\left(\frac{\delta f_j}{\delta t}\right)_{{\mathrm c},i} \approx - \frac{1}{m_j}\frac{\partial}{\partial\vec{v}_j}\cdot \left( \vec{A}_{ji} - {\mathsf{D}}_{ji} \cdot \frac{\partial}{\partial\vec{v}_j} \right) f_j.
\end{equation}

\subsubsection{Collisional timescales}

Conceptually, a \emph{collisional timescale} is the time required for collisions to significantly reduce a non-equilibrium feature such as a drift or anisotropy (for examples of non-equilibrium kinetic features in the solar wind, see Sects.~\ref{sec:ion-prop} and \ref{sec:elec-prop}).  Each specific type of non-equilibrium feature has its own expression for its collisional timescale that depends on the conditions in the plasma.  These timescales are derived from moments of the Boltzmann collision term, similar to the procedure described in Sect.~\ref{sec:moments}. This requires that assumptions be made about the particular form of the distribution function of each particle species involved.

As an example, we discuss the \emph{collisional slowing time} for two particle species, $j$ and $i$.\footnote{We note that $j$ and $i$ may refer to two different components of the same particle species (e.g., the proton core and proton beam, or the electron core and the electron halo).}  These species' differential flow is
\begin{equation}
\Delta \vec{U}_{ji} \equiv \vec{U}_j - \vec{U}_i,
\end{equation}
where $\vec{U}_j$ and $\vec{U}_i$ are the bulk velocities of species $j$ and $i$, respectively.  Then, the rate of change in the differential flow due to collisions is
\begin{equation}
\left( \frac{\delta \left(\Delta \vec{U}_{ji}\right)}{\delta t} \right)_{\rm c} = \left( \frac{\delta \vec{U}_j}{\delta t} \right)_{\rm c} - \left( \frac{\delta \vec{U}_i}{\delta t} \right)_{\rm c}.
\end{equation}
We express the bulk velocities $\vec{U}_j$ and $\vec{U}_i$ as moments of $f_j$ and $f_i$, the distribution functions of the $j$- and $i$-particles, according to Equation~(\ref{bulkvel}) and find
\begin{multline}
\left( \frac{\delta \left(\Delta \vec{U}_{ji}\right)}{\delta t} \right)_{\mathrm c} = \left[ \frac{\delta}{\delta t} \left( \frac{1}{n_j} \int {\mathrm d}^3\vec v\,\vec{v} f_j(\vec{v}) \right) \right]_{\mathrm c} - \left[ \frac{\delta}{\delta t} \left( \frac{1}{n_i} \int {\mathrm d}^3\vec v\,\vec{v}\,f_i(\vec{v}) \right) \right]_{\rm c} \\
= \int {\mathrm d}^3\vec v \, \vec{v} \left[ \frac{1}{n_j} \left( \frac{\delta f_j}{\delta t} \right)_{\mathrm c} - \frac{1}{n_i} \left( \frac{\delta f_i}{\delta t} \right)_{\mathrm c} \right].
 \end{multline}
To continue this analysis, we must make a choice for the form of the collision terms and for the distribution functions.  Once these are set, the result, to first order, has the form
\begin{equation}
\left( \frac{\delta \left(\Delta \vec{U}_{ji}\right)}{\delta t} \right)_{\mathrm c} = - \nu_{{\rm s},ji} \, \Delta \vec{U}_{ji},
\end{equation}
where $\nu_{{\rm s},ji}$ is the \emph{collision frequency} for the slowing of $j$ particles by $i$ particles.  The corresponding collisional timescale is defined to be
\begin{equation}
\tau_{{\rm s},ji} \equiv \frac{1}{\nu_{{\rm s},ji}} .
\end{equation}
Collisional timescales are most commonly derived and used for the relaxation of temperature anisotropy $T_{\perp j}/T_{\parallel j}$, unequal temperatures $ T_{j}/T_{i}$, and differential flow $\Delta \vec{U}_{ji}$.

Specific expressions for these collisional timescales have been computed and/or compiled by \citet{spitzer1956}, \citet{schunk1975,schunk1977}, \citet{hernandez1985}, \citet{huba2016}, and \citet{wilson2018}.  Typically, only one type of non-equilibrium feature is considered in each collisional timescale but formul\ae{} derived by \citet{hellinger2009,hellinger2010} consider all three of the features listed above. \citet{hellinger2016} uses observations from the Wind spacecraft to demonstrate that they result in substantially different collision and heating rates.  Likewise, although most derivations assume Maxwellian or bi-Maxwellian distribution functions, \citet{marsch1985} derive timescales for $\kappa$-distributions.

\subsubsection{Coulomb number and collisional age}\label{sec:col:theory:age}

The majority of the heating and acceleration that gives rise to the solar wind's non-equilibrium properties occurs in and around the solar corona.  Beyond that region, the solar wind's bulk velocity $\vec{U}$ remains approximately constant and radial \citep[see, e.g.,][]{hellinger2011,hellinger2013}.  Thus, the time required for a parcel of plasma to travel from the photosphere to a distance $r$ is approximately the \emph{expansion time} according to Equation~(\ref{expansion_time}):
\begin{equation}
\tau = \frac{r}{U_r}.
\end{equation}
The \emph{Coulomb number} of the parcel of plasma is then defined as
\begin{equation}\label{Acoll1}
N_{\mathrm c} \equiv \frac{\tau}{\tau_{\mathrm c}} = \frac{r}{U_r \tau_{\mathrm c}},
\end{equation}
where $\tau_{\rm c}$ is a collisional timescale.  Notwithstanding the caveats noted below, the Coulomb number essentially approximates the number of collisional timescales that elapsed in a parcel of plasma during its journey from the Sun to an observer.  In \emph{collisionally old} ($N_{\mathrm c} \gg 1$) plasma, collisional equilibration has proceeded much farther than in \emph{collisionally young} ($N_{\mathrm c} \ll 1$) plasma.

Although the Coulomb number has seen wide use in the analysis of solar-wind observations (see Sect.~\ref{sec:col:relax}), the concept carries significant limitations.  The above definition for $N_{\mathrm c}$ only allows for a single collision timescale $\tau_{\mathrm c}$.  While the correct formula for $\tau_{\mathrm c}$ can be chosen for the non-equilibrium feature under consideration, accounting for the interactions of multiple departures from equilibrium presents difficulties.  More fundamentally, the expression for $N_{\mathrm c}$ tacitly assumes that $\tau_{\mathrm c}$ remains constant with distance $r$ from the Sun.  In reality, $\tau_{\mathrm c}$ depends on density and temperature, both of which have strong radial trends.

To address some of these issues, various studies \citep{hernandez1987,chhiber2016,kasper2017,kasper2019} employ an integrated Coulomb number of the form
\begin{equation}\label{Acoll2}
A_{\mathrm c}\equiv \int \frac{\mathrm d t}{\tau_{\mathrm c}} = \int \frac{\mathrm d r}{U_r(r) \tau_{\mathrm c}(r)}.
\end{equation}
This formulation directly accounts for the radial dependences of densities, velocities, and temperatures.  These radial trends can either be derived from theoretical expectations (e.g., for quasi-adiabatic expansion) or from empirical observations. Some authors \citep[e.g.,][]{kasper2017} differentiate between the \emph{Coulomb number} $N_{\mathrm c}$ and \emph{collisional age} $A_{\mathrm c}$, with the former defined by Equation~(\ref{Acoll1}) and the latter defined by Equation~(\ref{Acoll2}).\footnote{We adopt the new terminology of \citet{kasper2017}. We note, however, that some earlier publications use the term ``collisional age'' for $N_{\mathrm c}$ \citep{kasper2008,bale2009,maruca2013}.}

\citet{maruca2013} introduce a close alternative to the Coulomb-number analysis, \emph{retrograde collisional analysis}, in which collisional timescales and radial trends are used to ``undo'' the effects of collisions and estimate the state of the solar wind when it was closer to the Sun.

\subsection{Observations of collisional relaxation in the solar wind}\label{sec:col:relax}

This section summarizes observational studies of collisional relaxation's effects on solar-wind plasma as it expands through the heliosphere.

\subsubsection{Ion collisions}

Early observations of solar-wind ions indicate that $\alpha$-particles tend to be significantly faster and hotter than protons (see Sect.~\ref{sec:ion-prop}).  Observations from IMP~6, IMP~7, IMP~8, and OGO~5 \citep{feldman1974,neugebauer1976,neugebauer1979} demonstrate that the values of $\left\lvert \Delta \vec{U}_{\alpha{\mathrm p}} \right\rvert$ and $T_{\alpha}/T_{\mathrm p}$ decrease toward 0 and 1 with increasing $N_{\mathrm c}$.  This negative correlation indicates that $\alpha$-particles are first preferentially accelerated and heated in the corona and then partially equilibrate with protons as the plasma expands through the inner heliosphere.  Later studies using observations from Helios \citep{marsch1982a,marsch1983a,livi1986}, ISEE~3 \citep{klein1985}, Prognoz~7 \citep{yermolaev1989,yermolaev1991,yermolaev1990}, Ulysses \citep{neugebauer1994}, and Wind \citep{kasper2008,kasper2017,maruca2013,hellinger2016} confirm these early results.  Interplanetary coronal mass ejections (ICMEs) are a notable exception to this overall trend in that they exhibit enhancements in $T_{\alpha}/T_{\mathrm p}$, which arise from ongoing heating during expansion \citep{liu2006}.

Measurements of $T_{\perp {\mathrm p}}$ and $T_{\parallel {\mathrm p}}$ from Wind reveal that the average value of the anisotropy ratio $ T_{\perp {\mathrm p}}/T_{\parallel {\mathrm p}}\rightarrow 1$ as the Coulomb number increases \citep{kasper2008,kasper2017}. Further observations \citep{bale2009} show that both Coulomb collisions and kinetic microinstabilities (see Sect.~\ref{sec:inst}) have roles in limiting proton temperature anisotropy.  Numerical models confirm this interplay of collisional and wave--particle effects \citep{tam1999,hellinger2010,matteini2012}.

\begin{figure}
\begin{center}
\includegraphics[height=6.50in]{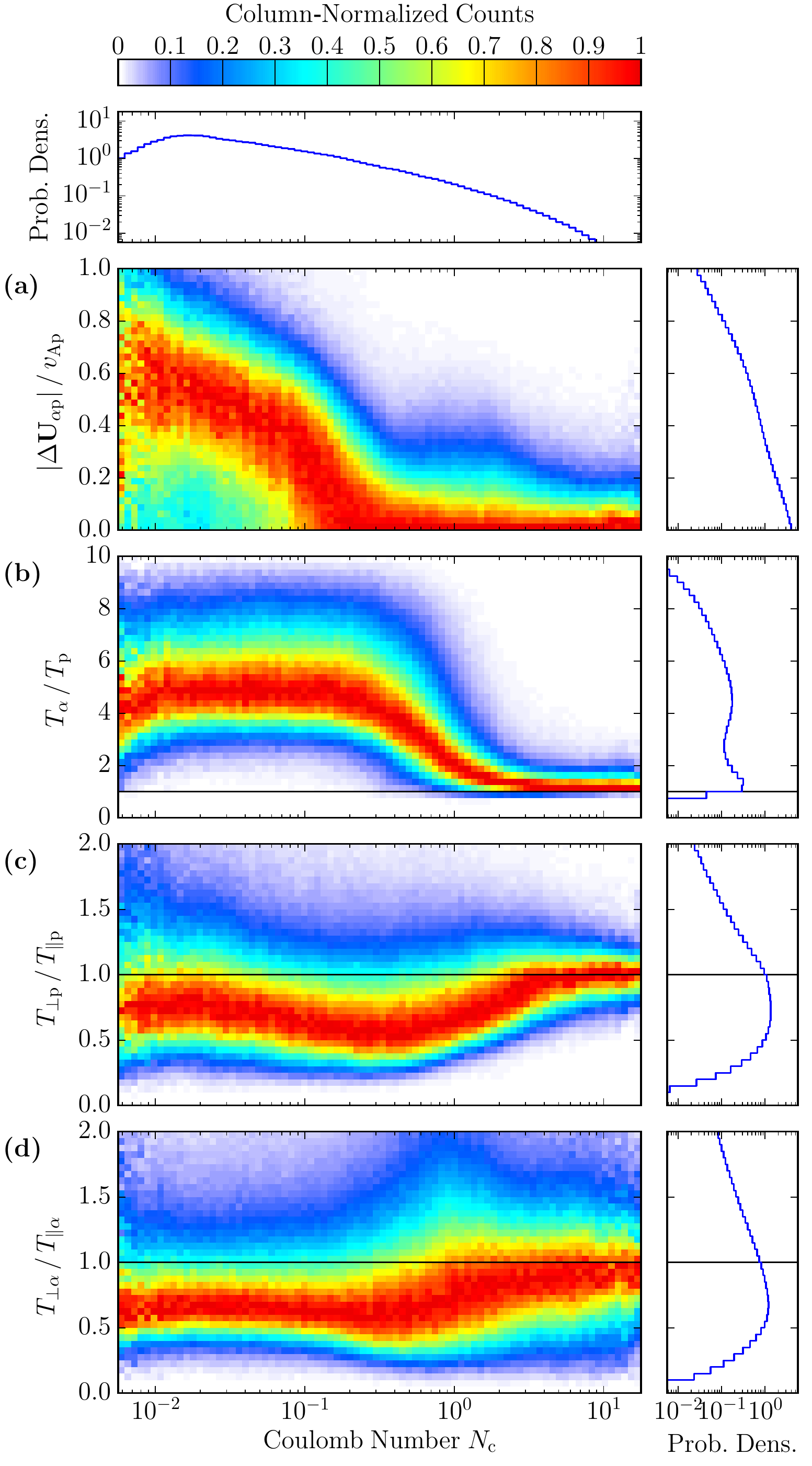}
\caption{\label{fig:col:ion} Trends in four parameters with Coulomb number $N_{\mathrm c}$: (a) $\alpha$--proton differential flow normalized to the proton Alfv\'en speed, (b) $\alpha$-to-proton relative temperature, (c) proton temperature anisotropy, and (d) $\alpha$-particle temperature anisotropy.  The dataset, compiled by \citet{maruca2012,maruca2013}, consists of 2.1-million data from the Wind/SWE Faraday cups.  The color scale is linear, and red indicates the most-likely parameter value for a given $N_{\mathrm c}$-value.  The probability densities of Coulomb number (top) and of each of the four parameters (right) are also shown. After \citet{kasper2008,kasper2017}.}
\end{center}
\end{figure}
Figure~\ref{fig:col:ion} shows trends in four parameters with Coulomb number $N_{\mathrm c}$ in a dataset of 2.1-million data from the Wind/SWE Faraday cups compiled by \citet{maruca2012,maruca2013}.  The values of $N_{\mathrm c}$ are calculated using the expression derived by \citet{maruca2013}, which is based on the proton ``self-collision time'' described by \citet{spitzer1956}.  For each parameter $P$, the $(N_{\mathrm c},P)$-plane is divided into 80 logarithmically spaced $N_{\mathrm c}$-bins and 40 linearly spaced $P$-bins.  Once the data are binned, the grid is column-normalized: the number of counts in each bin is divided by the number of counts in the most-populated bin in its column.  Thus, the color of each bin in Fig.~\ref{fig:col:ion} indicates the relative likelihood of a $P$-value for a given $N_{\mathrm c}$-value.  Each of the four parameters in Fig.~\ref{fig:col:ion} is an indicator of a departure from local thermal equilibrium.  As $N_{\mathrm c}$ increases, the most-likely $P$-value approaches its equilibrium state: $0$ for $\left\lvert\Delta\vec{U}_{\alpha{\mathrm p}}\right\rvert/v_{\mathrm {Ap}}$ and $1$ for $T_{\alpha}/T_{\mathrm p}$, $T_{\perp\mathrm p}/T_{\parallel\mathrm p}$, and $T_{\perp\alpha}/T_{\parallel\alpha}$.  Each parameter reaches equilibrium at a different $N_{\mathrm c}$-value because the formula for $N_{\mathrm c}$ uses the same self-collision time as a generic collisional timescale rather than the specific collisional timescale for each parameter $P$.

Column-normalizing plots (as has been done, e.g., for those in Figs.~\ref{fig:col:ion} and~\ref{fig:sh}) is a powerful and well established technique for exploring collisional effects in solar-wind plasma.  It represents a refinement of the method used in some of the earliest studies of collisional relaxation \citep[e.g.,][]{feldman1974,neugebauer1976}, in which data were divided into logarithmically uniform $N_{\mathrm c}$-intervals, and the average $T_{\alpha}/T_{\mathrm p}$-value was plotted for each interval.  Nevertheless, some caution is warranted in producing and interpreting column-normalized plots in general. First, the procedure of column-normalization modifies the weights of different data points and thus may cause an overemphasis or underemphasis of bins in a statistical data set. Second, the very act of column-normalization imposes causality: the parameter on the vertical axis becomes a function of that on the horizontal axis.  Though this is usually justified in collisionalization studies because of the strong theoretical motivation for such a causal relationship, column-normalization is not appropriate for all correlation studies.  Third, determining which parameters to plot is complicated by the many correlations that exist among particle moments \citep[e.g., the well established temperature--speed relationship for protons;][]{lopez1986}.  Even so, parameters such as $T_{\alpha}/T_{\mathrm p}$ and $\left|\Delta \vec U_{\alpha\mathrm p}\right|$ have been qualitatively \citep{kasper2008} and quantitatively \citep{maruca2013} demonstrated to be more strongly correlated with $N_{\mathrm c}$ than with $n_{\mathrm p}$, $U_{\mathrm p r}$, or $T_{\mathrm p}$ (all three of which $N_{\mathrm c}$ depends on).

Observations also give insight into collisional effects on minor ions.  ISEE~3 and SOHO/CELIAS data show that, while mass-proportional temperatures are most common, the effects of collisional thermalization are apparent at low solar-wind speeds \citep{bochsler1985,hefti1998}.  Interestingly, \citet{vonsteiger1995} and \citet{vonsteiger2006} find no indications of a departure from mass-proportional temperatures at any solar-wind speed.  This may be due to the limited number of data from very slow wind or from the ongoing heating of heavy ions.  Coulomb-number analyses of heavy-ion observations from ACE/SWICS show similar negative trends in the ion-to-proton temperature ratio with Coulomb number \citep{tracy2015,tracy2016}. 


Although most observational studies of ion--ion collisions focus on the effects of collisions on particle moments, some consider how collisions affect the structure of ion distribution functions.  \citet{marsch1983} note that the value of the collision term in Equation~(\ref{eqn:col:term}) varies across phase space and is highest for particles traveling at the bulk speed of the plasma.  This finding is consistent with proton distribution functions observed by Helios, which show Maxwellian cores surrounded by non-Maxwellian tails.  A kinetic model of the collisional effects on proton distribution functions counter-intuitively reveals  that collisional isotropization can actually generate proton beams \citep{livi1987}, which themselves would then be ultimately eroded by collisions.

\subsubsection{Electron collisions}\label{sec:col:elec}

Collisions involving electrons, due to their higher rates \citep[see, e.g.,][]{wilson2018}, are thought to play an even more important role in solar-wind thermodynamics than collisions involving only ions.  As noted in Sect.~\ref{sec:elec-prop}, electron distribution functions in the solar wind typically exhibit a three-component structure consisting of a core, halo, and strahl.  Many theories \citep[e.g.,][]{scudder1979a,scudder1979b,liesvendsen1997,liesvendsen2000} for the origin of these electron populations rely on the transition from highly collisional plasma in the lower corona to weakly collisional plasma in the upper corona.

Beyond the corona, numerous studies find that Coulomb collisions among electrons continue to affect them in the interplanetary solar wind.  An analysis of Mariner~10 data \citep{ogilvie1978} reveals that collisions have the greatest influence on the electron core while the electron halo remains weakly collisional. Electron distribution functions observed by Helios show that Coulomb collisions have a significant impact on the phase-space location of the core--halo boundary  \citep{pilipp1987,pilipp1987a,pilipp1987b}.  Kinetic simulations suggest that the interplay of collisions and expansion in the solar wind can give rise to the electron core, halo, and beam \citep{landi2010,landi2012}. Moreover, a kinetic model for the radial evolution of the strahl developed by \citet{horaites2018} indicates that Coulomb collisions provide a significant source of pitch-angle scattering for this population.

Solar-wind electrons typically exhibit less temperature anisotropy than ions \citep[][Fig.~1]{chen2016a}, which is at least partially ascribed to the higher rate of electron versus ion collisions.  Analytical models that account for electron expansion and collisions in the interplanetary solar wind agree well with ISEE~3 and Ulysses observations of electron temperature anisotropy \citep{phillips1989a,phillips1990,phillips1993}.  A study of Wind observations by \citet{salem2003} finds that electron temperature anisotropy is strongly correlated with Coulomb number, with collisionally old electrons being most likely to exhibit isotropy.  As is the case for protons, data from Helios, Cluster, and Ulysses show that both Coulomb collisions and kinetic microinstabilities play significant roles in isotropizing solar-wind electrons \citep{stverak2008,stverak2015}.

\begin{figure}
  \includegraphics[width=\textwidth]{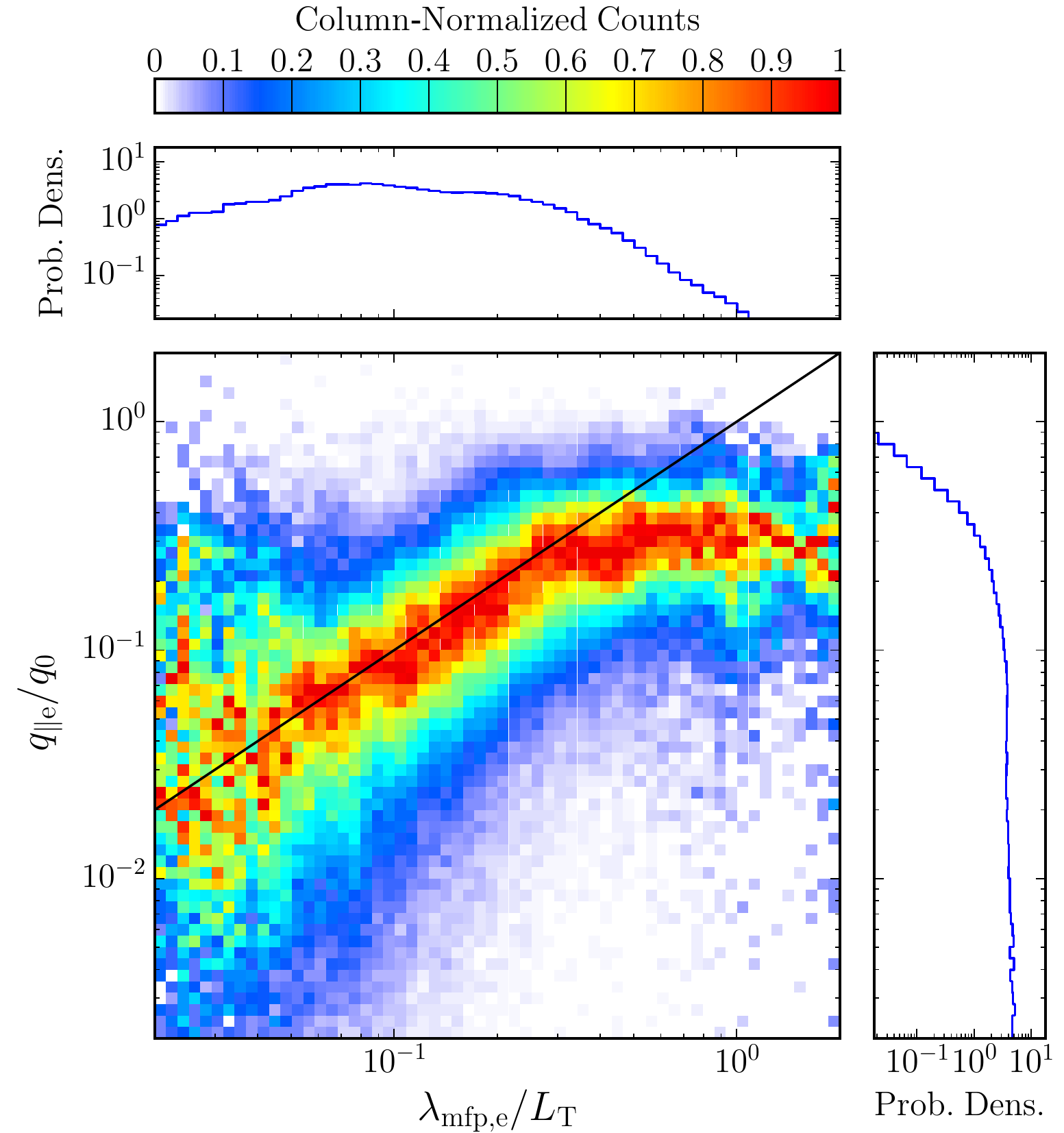}
\caption{Column-normalized distribution of Wind/3DP electron data as a function of the parallel heat flux $q_{\parallel\mathrm e}$ and the electron mean free path $\lambda_{\mathrm{mfp,e}}$.  The Spitzer-H\"arm prediction in this normalization is given by $q_{\parallel\mathrm e}/q_0=1.07\lambda_{\mathrm{mfp,e}}/L_{\mathrm T}$ and is shown  as a black line. We use $\alpha=2/7$. The probability densities for $\lambda_{\mathrm{mfp,e}}/L_{\mathrm T}$ (top) and $q_{\parallel\mathrm e}/q_0$ (right) are also shown. After \citet{salem2003} and \citet{bale2013} and using data provided by C.~Salem. }
\label{fig:sh}       
\end{figure}

Collisions also significantly affect electron heat flux.  According to \emph{Spitzer-H\"arm theory} \citep{spitzer1953}, the electron heat flux is proportional to the timescale of electron--electron collisions.  Statistical analyses of Wind electron measurements show that this relationship holds true but only in highly collisional plasma  \citep{salem2003,bale2013}.  Figure~\ref{fig:sh} shows the distribution of Wind/3DP electron data in the plane of the normalized parallel heat flux versus the normalized electron mean free path in the solar wind. We normalize $q_{\parallel\mathrm e}$ to the free-streaming saturation heat flux $q_0\equiv 3n_{\mathrm e}k_{\mathrm B}T_{\mathrm e}w_{\mathrm e}/2$ and $\lambda_{\mathrm{mfp,e}}$ to the temperature gradient $L_{\mathrm T}\equiv r/\alpha$, where $r$ is the heliocentric distance of the measurement and $\alpha$ describes the observed temperature profile through $T_{\mathrm e}\propto r^{-\alpha}$. The dimensionless quantity $\lambda_{\mathrm{mfp,e}}/L_{\mathrm T}$ is called the \emph{Knudsen number}. The black line shows the Spitzer-H\"arm prediction. The heat flux follows this prediction at large collisionality but deviates in the collisionless limit.
Spitzer-H\"arm theory is found to overestimate electron heat flux in moderately and weakly collisional plasma, which is consistent with results from the kinetic simulations of \citet{landi2012} and \citet{landi2014}.

Occasionally, a parcel of solar-wind plasma is found to have an especially low or high rate of Coulomb collisions, which offers insight into the most extreme effects of collisions on electrons.  In a study of several periods of very-low-density solar wind, each period exhibits an unusually narrow electron strahl \citep{ogilvie2000}. This likely results from the combination of a low collision rate and the conservation of the first adiabatic invariant, given in Equation~(\ref{cgl1}), to first order as suggested by \citet{fairfield1985}. Conversely, data from ISEE~1 and ISEE~3 exhibit several \emph{heat-flux dropouts} \citep{fitzenreiter1992}: periods of very low electron heat flux. The weak electron halos observed during these dropouts likely result, at least in part, from enhanced electron collisionality.  Likewise, \citet{larson2000} and \citet{farrugia2002}, using the Wind and ACE spacecraft, identify weak halos in particularly dense and cold magnetic clouds and find them to be consistent with collisional effects.

\section{Plasma waves}\label{sec:waves}

\emph{Plasma waves} are important processes for the transport and dissipation of energy in a plasma. They can accelerate plasma flows and heat plasma by damping.  Section~\ref{sec:waves:lin} introduces basic concepts to describe plasma waves. Section~\ref{sec:damping} describes damping and dissipation mechanisms, and Sect.~\ref{sect:waves} then presents types of plasma waves that are relevant to the multi-scale evolution of the solar wind. For more details on the broad topic of plasma waves, we refer to the excellent textbooks by \citet{stix1992} and \citet{swanson2003}. 

\subsection{Plasma waves as self-consistent electromagnetic and particle fluctuations}\label{sec:waves:lin}

Waves are periodic or quasi-periodic spatio-temporal fluctuations which arise through the action of a \emph{restoring force}.  The self-consistent electromagnetic interactions in a plasma provide additional restoring forces that do not occur in a neutral gas. Therefore, a plasma can exhibit many more types of wave modes than a neutral gas. In this section, we introduce the linear theory of plasma waves. For further details on linear theory, we refer the reader to the general review on solar-wind plasma waves by \citet{ofman2010} and the textbooks by \citet{stix1992}, \citet{brambilla1998}, and \citet{swanson2003}. 

Linear wave theory considers a wave to be a fluctuating perturbation on an equilibrium state.
We assume that any physical quantity $A$ of the system can be written as 
\begin{equation}
A(\vec x,t)=A_0+\delta A(\vec x,t),
\end{equation}
where $A_0$ is the constant background equilibrium, and $\delta A$ is the fluctuating perturbation of $A$.  Moreover, we assume that the fluctuating quantities in a wave behave like 
\begin{equation}\label{dAfour}
\delta A(\vec x,t)= \mathrm{Re}\left[ A(\vec k,\omega)\exp\left(i\vec k\cdot \vec x-i\omega t\right)\right],
\end{equation}
where $ A(\vec k,\omega)$ is the complex Fourier amplitude of $A$, the wavevector $\vec k$ is real, and the frequency $\omega$ is complex. We define the \emph{real frequency} as
\begin{equation} 
\omega_{\mathrm r}\equiv \mathrm{Re}\,\omega
\end{equation}
and the \emph{growth or damping rate} as 
\begin{equation}\label{damprate}
\gamma \equiv \mathrm{Im}\,\omega.
\end{equation}
The \emph{linear dispersion relation} is a mathematical expression based on a self-consistent set of linearized equations for the plasma particles and the electromagnetic fields. It connects the wavevector $\vec k$ with the frequency $\omega$ in such a way that its solutions represent self-consistent waves in the plasma. If multiple solutions exist for a given $\vec k$, then each corresponds to a distinct \emph{mode}. According to Equations~(\ref{dAfour}) and (\ref{damprate}), the amplitude of the fluctuations decreases exponentially with time if $\gamma< 0$. As a solution to the linear dispersion relation, we describe such a wave as being \emph{linearly damped} (see Sect.~\ref{sec:ql}). Likewise, if $\gamma>0$, the wave amplitude increases exponentially with time and the wave is \emph{linearly unstable} (see Sect.~\ref{sec:inst}). 

Neglecting any background electric field $\vec E_0$, we rewrite the electric and magnetic fields according to Equation~(\ref{dAfour}) as
\begin{equation}\label{deltaE}
\vec E(\vec x,t)=\delta \vec E(\vec x,t)=\mathrm{Re}\left[\vec E(\vec k,\omega)\exp\left(i\vec k\cdot \vec x-i\omega t\right)\right]
\end{equation}
and 
\begin{equation}\label{deltaB}
\vec B(\vec x,t)=\vec B_0+\delta\vec B(\vec x,t)=\vec B_0+\mathrm{Re}\left[\vec B(\vec k,\omega)\exp\left(i\vec k\cdot \vec x-i\omega t\right)\right],
\end{equation}
using the complex Fourier amplitudes $\vec E(\vec k,\omega)$ and $\vec B(\vec k,\omega)$. In the following, we write the Fourier amplitudes without their arguments $(\vec k,\omega)$ and assume that $|\delta \vec B|\ll |\vec B_0|$. Substituting Equations~(\ref{deltaE}) and (\ref{deltaB}) into Maxwell's equations~ (\ref{Maxw1full}) through (\ref{Maxw5full}), we find in Fourier space
\begin{equation}\label{Maxw1}
\vec k\cdot \vec E=-4\pi i \rho_{\mathrm c},
\end{equation}
\begin{equation}
\vec k\cdot \vec B=0,
\end{equation}
\begin{equation}\label{Maxw3}
\vec k\times \vec E-\frac{\omega}{c}\vec B=0,
\end{equation}
and
\begin{equation}\label{Maxw4}
\vec k\times \vec B+\frac{\omega}{c}\vec E=-\frac{4\pi i}{c}\vec j,
\end{equation}
where 
\begin{equation}\label{chdens1}
\rho_{\mathrm c}=\sum \limits_j \rho_{\mathrm cj}=\sum \limits_j q_jn_j
\end{equation}
is the charge density and
\begin{equation}\label{jdens1}
\vec j=\sum\limits_j \vec j_j 
\end{equation}
is the current density. In Equations~(\ref{chdens1}) and (\ref{jdens1}), the sums are carried over all particle species $j$ in the plasma.
The left-hand sides of Equations~(\ref{Maxw1}) through (\ref{Maxw4}) represent the  interactions between the electric and magnetic fields, while the right-hand sides represent the self-consistent effects of the particles on the fields. 

We define the \emph{plasma susceptibility tensor} $\vec{\chi}_j$ of species $j$ through
\begin{equation}
\vec{\chi}_j\cdot \vec E\equiv \frac{4\pi i}{\omega}\vec j_j
\end{equation}
and the \emph{dielectric tensor} $\vec{\epsilon}$ as
\begin{equation}
\vec{\epsilon}\equiv \vec 1 +\sum\limits_j \vec{\chi}_j.
\end{equation}
The dielectric tensor is additive in the contributions from each plasma species $j$ and reflects the interaction between fields and particles. With these definitions, we find
\begin{equation}
\vec{\epsilon}\cdot \vec E=\vec E+\frac{4\pi i}{\omega}\vec j
\end{equation}
and, by using Equation~(\ref{Maxw4}),
\begin{equation}\label{Maxw4b}
\vec k\times \vec B+\frac{\omega}{c}\vec{\epsilon}\cdot \vec E=0.
\end{equation}
Combining Equation~(\ref{Maxw3}) with Equation~(\ref{Maxw4b}) leads to the \emph{wave equation}:
\begin{equation}
\vec n \times \left(\vec n \times \vec E\right)+\vec{\epsilon}\cdot \vec E =\vec{\mathcal D}\cdot \vec E=0,
\end{equation}
where $\vec n\equiv \vec kc/\omega$ is the \emph{refractive index} and
\begin{equation}
\vec{\mathcal D}\equiv \begin{pmatrix} \epsilon_{xx}-n_z^2 & \epsilon_{xy} & \epsilon_{xz}+n_xn_z \\  \epsilon_{yx} & \epsilon_{yy}-n_x^2-n_z^2 & \epsilon_{yz} \\  \epsilon_{zx}+n_zn_x & \epsilon_{zy} & \epsilon_{zz}-n_x^2 \end{pmatrix}
\end{equation}
is the \emph{dispersion tensor}. The \emph{phase velocity} of a solution is given by $\omega\vec k/k^2$. Non-trivial solutions to the wave equation fulfill 
\begin{equation}\label{dispersion_relation}
\mathrm{det}\,\left[\vec{\mathcal D}(\vec k,\omega)\right]=0,
\end{equation}
which is the mathematical dispersion relation. The identification of plasma waves then involves the calculation of a proper dielectric tensor for the plasma conditions at hand as well as the derivation of the roots of Equation~(\ref{dispersion_relation}). 

If the calculation of $\vec {\epsilon}$ is based on the linearized Vlasov equation \citep{gary1993}, Equation~(\ref{dispersion_relation}) leads to the full \emph{hot-plasma dispersion relation}, which is a standard-tool in the calculation of plasma waves \citep{roennmark1982,klein2015,verscharen2018b,verscharen2018}. In this model, Equation~(\ref{vlasov}) is linearized for each plasma species $j$ to first order in $\delta f_j$, under the assumption that  $f_j=f_{0j}+\delta f_j$, as
\begin{equation}
\frac{\partial \delta f_j}{\partial t}+ \vec v\cdot \frac{\partial \delta f_j}{\partial \vec x}+\Omega_j \left(\vec v\times \hat{\vec b} \right) \cdot \frac{\partial \delta f_{j}}{\partial \vec v}=-\frac{q_j}{m_j}\left(\delta \vec E+\frac{1}{c}\vec v\times \delta \vec B\right)\cdot \frac{\partial f_{0j}}{\partial \vec v},
\end{equation}
where the left-hand side describes the change of $\delta f_j$ along the zeroth-order particle trajectory, $\Omega_j$ is calculated based on the background magnetic-field magnitude $B_0$, and $\hat{\vec b}\equiv \vec B_0/B_0$. The resulting solutions for $\delta f_j$ from integration along the particle trajectories then define $\rho_c$ and $\vec j$  according to Equations~(\ref{chargedens}) and (\ref{currentdens}).
We refer to the textbooks by \citet{melrose1991}, \citet{stix1992}, and \citet{gary1993} for more details on the calculation of $\vec{\epsilon}$.  

In our discussion of wave modes in Sect.~\ref{sect:waves}, we present analytical results for wave dispersion and polarization relations based on different models and in different limits, which we identify whenever necessary. Fluid models and kinetic models often lead to different predictions in the dispersion relation and polarization properties of linear waves \citep[see, e.g.,][]{verscharen2017,wu2019}. These differences result from differences in the models' underlying assumptions (e.g., the closure of the hierarchy of moment equations; see Sect.~\ref{sec:moments}). Furthermore, analytical calculations of the dispersion relation often rely on mathematical approximations in certain limits (e.g., taking $m_{\mathrm e}\rightarrow 0$ or  $T_j\rightarrow 0$). Before we discuss the wave modes further, we describe damping and dissipation mechanisms in the following section.

\subsection{Damping and dissipation mechanisms}\label{sec:damping}

The damping and dissipation of plasma waves are important for the global behavior of the plasma because these processes transfer energy between the electromagnetic fields and the particles and are also candidates for the dissipation of turbulent plasma fluctuations in the solar wind (see Sect.~\ref{sec:turbulence}).

For our discussion, we distinguish between \emph{damping} as a reduction in the amplitude of field fluctuations (i.e., $\gamma<0$) and \emph{dissipation} as an irreversible increase in entropy of a plasma species (i.e., $\mathrm dS_j>0$, where $S_j$ is the entropy of species $j$). Lastly, we define \emph{heating} as an increase of the plasma's thermal energy. 
In this section, we address three important damping and dissipation mechanisms for plasma waves: (1) \emph{quasilinear diffusion} from Landau-resonant or cyclotron-resonant wave--particle interactions, (2) \emph{nonlinear phase mixing}, and (3) \emph{stochastic heating}. 
So long as the Boltzmann equation~(\ref{boltzmann}) is valid, dissipation in the sense of entropy generation can only occur through particle--particle collisions. 
Even if collisions are not frequent enough to bring the plasma distribution function into local thermodynamic equilibrium, phase-space structures in the velocity distribution function can become small enough that collisions lead to dissipation (cf Sect.~\ref{sec:col:theory}). When we study the dissipation of ``collisionless'' plasma waves, we, therefore, assume that collisions only affect small-scale structures in the distribution function and investigate the processes that create these small-scale structures, which in turn generate entropy through collisions. 
We note that deviations of velocity distributions from local thermodynamic equilibrium (see Sects.~\ref{sec:ion-prop} and~\ref{sec:elec-prop}) can affect the polarizations,  transport ratios, and damping rates of the plasma normal modes, as well as the heating mechanisms \citep{chandran2013,kasper2013,klein2015,tong2015,kunz2018}.

\subsubsection{Quasilinear diffusion}\label{sec:ql}

Quasilinear diffusion describes the evolution of the distribution function as velocity-space diffusion that arises from the resonant interaction between waves and particles \citep{marsch2006}. Quasilinear theory assumes the presence of a superposition of non-interacting and randomly phased waves that are solutions to linear plasma-wave theory as described in Sect.~\ref{sec:waves:lin}. The force term in the Vlasov equation is then averaged over the gyro-phases of the unperturbed particle orbits so that a diffusion term for the background distribution $f_{0j}$ in $v_{\perp}$ and $v_{\parallel}$ results, independent of the gyro-phase of the particles. This process is quasilinear in the sense that the fluctuations are solutions to the linear dispersion relation (Sect.~\ref{sec:waves:lin}), which closes the system of equations, but the field amplitudes enter the equations quadratically. In quasilinear theory, the background distribution $f_{0j}$ evolves slowly compared to the timescale of the fluctuations $1/\omega_{\mathrm r}$. Under the assumption of small wave amplitudes and $|\gamma/\omega_{\mathrm r}|\ll1$, quasilinear diffusion follows the equation \citep{shapiro1962,kennel1966,rowlands1966,stix1992}
\begin{equation}\label{qldiff}
\frac{\partial f_{0j}}{\partial t}=\frac{q_j^2}{8\pi^2m_j^2}\lim\limits_{V\rightarrow \infty} \frac{1}{V} \sum \limits_{n=-\infty}^{+\infty}\int \mathrm d^3k\frac{1}{v_{\perp}}\hat Gv_{\perp}\delta\left(\omega_{\mathrm r}-k_{\parallel}v_{\parallel}-n\Omega_j\right)\left|\psi_{n}\right|^2\hat G f_{0j},
\end{equation}
where the pitch-angle operator is defined as
\begin{equation}\label{Goperator}
\hat G\equiv \left(1-\frac{k_{\parallel}v_{\parallel}}{\omega_{\mathrm r}}\right)\frac{\partial }{\partial v_{\perp}}+\frac{k_{\parallel}v_{\perp}}{\omega_{\mathrm r}}\frac{\partial }{\partial v_{\parallel}},
\end{equation}
and
\begin{equation}
\psi_n\equiv \frac{1}{\sqrt{2}}\left[E_{\mathrm r}e^{i\phi}J_{n+1}(\sigma_j)+E_{\mathrm l}e^{-i\phi}J_{n-1}(\sigma_j)\right]+\frac{v_{\parallel}}{v_{\perp}}E_zJ_{n}(\sigma_j).
\end{equation}
We define the wavevector components perpendicular and parallel to the background magnetic field as $k_{\perp}$ and $k_{\parallel}$, respectively.
The right-handed and left-handed components of the Fourier-transformed electric field are $E_{\mathrm r}\equiv\left(E_x-iE_y\right)/\sqrt{2}$ and $E_{\mathrm l}\equiv\left(E_x+iE_y\right)/\sqrt{2}$, respectively, $J_n$ is the $n$\textsuperscript{th} order Bessel function of the first kind, $\sigma_j\equiv k_{\perp}v_{\perp}/\Omega_j$, $\phi$ is the azimuthal angle of $\vec k$, and $V$ is the spatial volume under consideration. Since Equation~(\ref{qldiff}) is a second-order differential equation in $v_{\perp}$ and $v_{\parallel}$, it indeed corresponds to a diffusion in velocity space. The $\delta$-function in Equation~(\ref{qldiff}) guarantees that the only particles that participate in the resonant interactions are those for which $v_{\parallel}$ is equal to the \emph{resonance speed}:
\begin{equation}\label{rescond}
v_{\mathrm{res}}\equiv \frac{\omega_{\mathrm r}-n\Omega_j}{k_{\parallel}}.
\end{equation}
Due to the form of $\hat G$, the diffusive flux of particles is tangent to semicircles in the $v_{\parallel}$-$v_{\perp}$ plane defined by 
\begin{equation}\label{qldiffcirc}
\left(v_{\parallel}-\frac{\omega_{\mathrm r}}{k_{\parallel}}\right)^2+v_{\perp}^2=\mathrm{constant}
\end{equation}
and directed from larger to smaller values of $f_{0j}$ \citep{verscharen2013}. During the diffusion, the particles gain kinetic energy if ($v_{\perp}^2+v _{\parallel}^2$) increases and lose it if this quantity decreases. The energy gained or lost by the particles is taken from or given to the wave at the resonant $k_{\parallel}$ and $\omega_{\mathrm r}$ so that this wave's amplitude changes. The $n=0$ term in the sum in Equation~(\ref{qldiff}) corresponds to \emph{ Landau damping} (\citeyear{landau1946}) and \emph{transit-time damping}, and the $n\neq 0$ terms correspond to \emph{cyclotron damping}.

\begin{figure}
  \includegraphics[width=\textwidth]{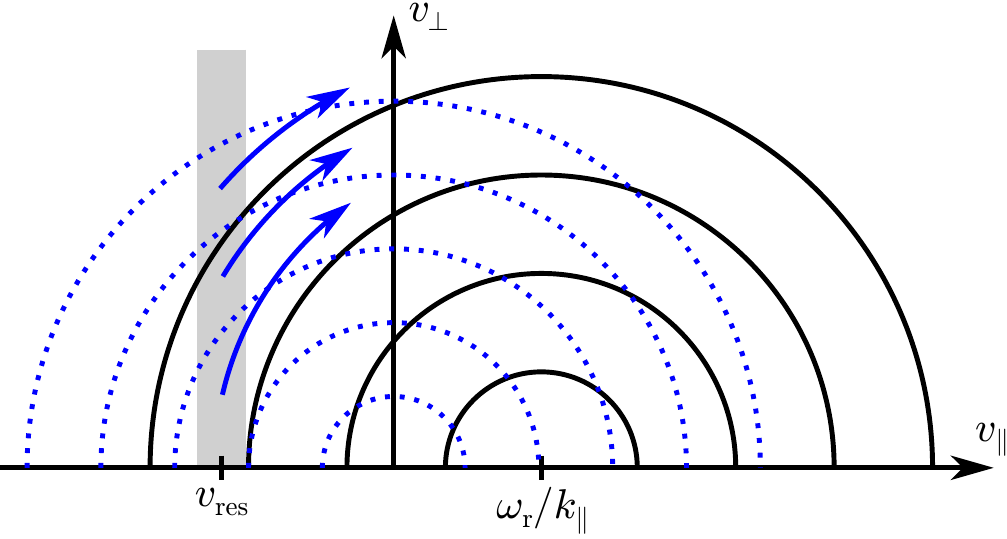}
\caption{\emph{Quasilinear diffusion} in the cyclotron-resonant damping of particles with $v_{\parallel}=v_{\mathrm{res}}<0$ (gray shaded area) with waves of parallel phase speed $\omega_{\mathrm r}/k_{\parallel}$. The blue dotted circles represent isocontours of the background distribution function $f_{0j}$. The diffusion paths (blue arrows) are locally tangential to circles around the point $(v_{\perp},v_{\parallel})=(0,\omega_{\mathrm r}/k_{\parallel})$ (black circles). In this example, the resonant particles gain kinetic energy, which corresponds to an increase in $(v_{\perp}^2+v_{\parallel}^2)$. This energy is removed from the waves at $\omega_{\mathrm r}$ and $k_{\parallel}$, which are thus damped. }
\label{fig:ql_diffusion}       
\end{figure}

We illustrate the quasilinear diffusion process for a cyclotron-damped wave in Fig.~\ref{fig:ql_diffusion}. In this example, cyclotron-resonant particles with $v_{\parallel}=v_{\mathrm{res}}<0$ interact with waves with $\omega_{\mathrm r}$ and $k_{\parallel}$ and diffuse in velocity space. The cyclotron-resonant damping of left-handed waves propagating parallel to $\vec B_0$ exhibits these characteristics. We illustrate the case of quasilinear diffusion for a cyclotron-resonant instability in Fig.~\ref{fig:instability} in Sect.~\ref{sec:inst}.

\subsubsection{Entropy cascade and nonlinear phase mixing}\label{sec:entropy}

Since dissipation, by definition, is irreversible, all dissipation processes cause entropy to increase. In a plasma with low collisionality, wave turbulence (see Sect.~\ref{sec:waveturb}) is associated with fluctuations in entropy\footnote{These largely reversible fluctuations in entropy do not violate the second law of thermodynamics which only applies to the total entropy of a closed system.} that cascade to small scales, where collisions have greater effects and ultimately dissipate these fluctuations. Applying Boltzmann's $H$-theorem to Equation~(\ref{boltzmann}), we obtain the entropy relation
\begin{equation}\label{entropy}
\frac{\mathrm d S_j}{\mathrm dt}=\frac{\mathrm d}{\mathrm dt}\left(-\int\frac{\mathrm d^3\vec r}{V}\int \mathrm d^3\vec v \,f_j\ln f_j\right)=-\int\frac{\mathrm d^3\vec r}{V}\int \mathrm d^3\vec v \,\left(\frac{\delta f_j}{\delta t}\right)_{\mathrm c}\,\ln f_j,
\end{equation}
where $S_j$ is the entropy of species $j$,  and $V$ is the spatial volume under consideration. Equation~(\ref{entropy}) shows that entropy only increases in the presence of particle--particle collisions. We now separate $f_j$ into its equilibrium part $f_{0j}$ and its fluctuating part $\delta f_j$ as 
\begin{equation}
f_j(\vec x, \vec v,t)=f_{0j}(\vec v)+\delta f_j(\vec x,\vec v,t).
\end{equation}
We assume that the collision frequency is of order $\omega_{\mathrm r}$,\footnote{In gyrokinetic theory, the collision frequency and $\omega_{\mathrm r}$ are both ordered to the intermediate timescale. This ordering does not prevent us from considering the collisionless and collisional limits and justifies the assumption of a Maxwellian $f_{0j}=f_{\mathrm M}$ \citep{howes2006,schekochihin2008}.} and $f_{0j}$ is a Maxwellian as in Equation~(\ref{maxwellian}) with temperature $T_{0j}$. After averaging over the timescales greater than the typical fluctuation time $\sim 1/\omega_{\mathrm r}$ and summing over all species, we describe the evolution of the generalized energy through the energy equation with the help of the expression for the entropy from Equation (\ref{entropy}) as \citep{schekochihin2008}
\begin{multline}\label{scheken}
\frac{\mathrm dW}{\mathrm dt}=\frac{\mathrm d}{\mathrm dt}\int \frac{\mathrm d^3\vec r}{V}\left(\frac{E^2+B^2}{8\pi}+\sum\limits_j \int \mathrm d^3\vec v \frac{k_{\mathrm B}T_{0j}\delta f_j^2}{2f_{0j}}\right)\\
=\epsilon+\int \frac{\mathrm d^3\vec r}{V}\sum \limits_j \int \mathrm d^3\vec v\frac{k_{\mathrm B}T_{0j}\delta f_j}{f_{0j}}\left(\frac{\delta f_j}{\delta t}\right)_{\mathrm c},
\end{multline}
where $W$ is the generalized energy and $\epsilon$ is the externally supplied power (e.g., through large-scale driving by shears or compressions).\footnote{Although Equation~(\ref{scheken}) was derived under the assumption of a Maxwellian background distribution, \citet{kunz2018} derive an expression for $\mathrm dW/\mathrm dt$ assuming a drifting bi-Maxwellian $f_{0j}=f_{\mathrm{bM}}$.}  

The  \emph{entropy cascade} constitutes the redistribution of generalized energy from electromagnetic fluctuations ($E^2+B^2$) to entropy fluctuations ($\delta f_j^2/f_{0j}$) according to Equation~(\ref{scheken}). These fluctuations in entropy then cascade to smaller scales in velocity space through a combination of \emph{linear} and \emph{nonlinear phase mixing}. Linear phase mixing corresponds to Landau damping, which we describe in Sect.~\ref{sec:ql}. The spread in parallel velocity of the particle distribution leads to a dependency of the Landau-resonant interactions between particles and the electric field on the particles' parallel velocity.
 
Nonlinear phase mixing often serves as a faster mechanism of entropy cascade.
A particle with a greater $v_{\perp}$ has a greater $\rho_j$ and thus experiences a slower $\vec E\times \vec B$ drift than a particle with smaller $v_{\perp}$ \citep{dorland1993}. 
Two particles of the same species $j$ but distinct perpendicular velocities $v_{\perp}$ and $v_{\perp}^{\prime}$  experience  spatially decorrelated fluctuations in the electric and magnetic fields if the difference between the particles' gyro-radii $v_{\perp}/|\Omega_j|$ and $v_{\perp}^{\prime}/|\Omega_j|$ is greater than the perpendicular correlation length $1/k_{\perp}$ of the field fluctuations \citep{schekochihin2008}. In kinetic theory, this process leads to spatial perpendicular mixing of ion distributions with different gyro-centers and hence to the creation of small-scale structure in the gyro-center distribution.  Small-scale structure in the fields in physical space thus leads to small-scale structure in the distribution function in velocity space perpendicular to $v_{\perp}$ as the result of this nonlinear phase mixing  \citep{tatsuno2009,banon2011,kawamori2013,navarro2016,cerri2018}.  Once these velocity-space structures are small enough, collisions can efficiently smooth them -- see Equation~(\ref{eqn:col:term}) and the associated discussion -- and thereby increase entropy and the perpendicular temperature of the ions.

\subsubsection{Stochastic heating}\label{sec:stoc}

Stochastic heating is a non-resonant energy-diffusion process. It arises from field fluctuations with spatial variations on the gyro-radius scale of the diffusing particles ($k_{\perp}\rho_j\sim 1$) and frequencies that are small compared to the gyro-frequency ($\omega_{\mathrm r}\ll |\Omega_j|$) in a constant background magnetic field $\vec B_0$ \citep{mcchesney1987,chen2001,johnson2001,chaston2004,fiksel2009}. 

If these fluctuations are low in amplitude, they induce only small perturbations in the particles' otherwise circular orbits. With increasing amplitude, however, the fluctuations increasingly distort the gyro-orbits. If the amplitude of the gyro-scale fluctuations is so large that the orbits become stochastic in the plane perpendicular to $\vec B_0$, particles experience stochastic increases and decreases in their kinetic energy due to the fluctuations' electric fields. Consequently, the particles diffuse in $v_{\perp}^2$, which corresponds to perpendicular heating \citep{chandran2010b,klein2016}. This process is consistent with observations of solar-wind protons \citep{bourouaine2013a,martinovic2019} and minor-ion temperatures and drifts \citep{chandran2010,wang2011,chandran2013}.

\begin{figure}
  \includegraphics[width=\textwidth]{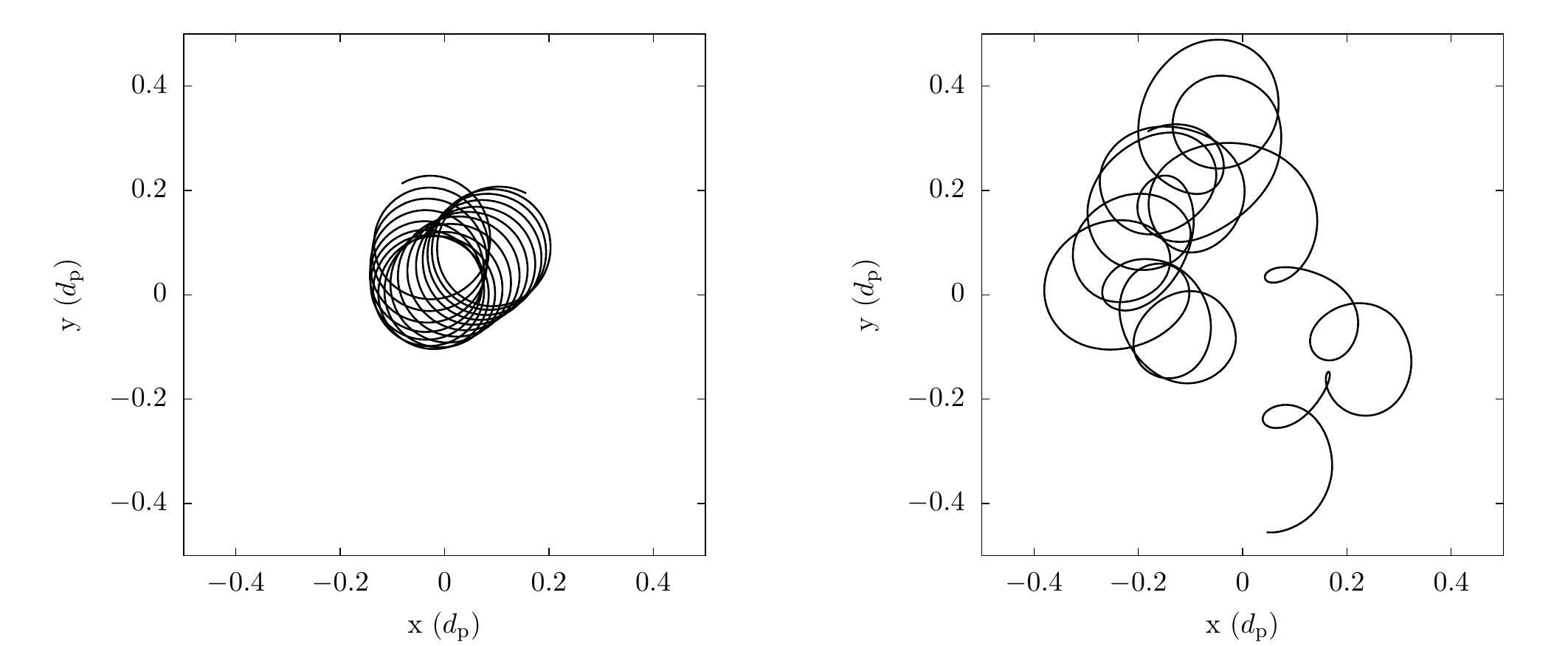}
\caption{Trajectories of test particles in the plane perpendicular to $\vec B_0$. We use a setup similar to the kinetic-Alfv\'en-wave (KAW) simulations of stochastic heating described by \citet{chandran2010b}. In the left panel, we show solutions for a thermal-proton trajectory when the amplitude of the Alfv\'enic fluctuations at $k_{\perp}\rho_{\mathrm p}\approx 1$ is small. The proton drifts due to the large-scale Alfv\'enic fluctuations, but its gyro-motion is still circular to first order. In the right panel, we show the same solutions but with an amplitude of the gyro-scale KAW fluctuations that is by a factor of five greater than in the left panel. The gyro-motion is strongly perturbed and becomes stochastic, creating the conditions for \emph{stochastic heating}. }
\label{fig:stoc_heat}       
\end{figure}

Figure~\ref{fig:stoc_heat} shows the orbits of two thermal protons in test-particle simulations of stochastic heating based on a superposition of randomly-phased kinetic Alfv\'en waves (KAWs; see Sect.~\ref{sec:KAWs}). If the amplitude of the gyro-scale fluctuations is small (left panel), the magnetic moment is conserved and the particle trajectory corresponds to a drifting quasi-circular motion. If the amplitude of the gyro-scale fluctuations is large (right panel), the magnetic moment is no longer conserved. As a result, the particle's trajectory becomes stochastic, which corresponds to stochastic heating through the waves' electric fields.

The mechanisms of stochastic proton heating are different in the low-$\beta_{\mathrm p}$ regime and in the high-$\beta_{\mathrm p}$ regime.
In plasmas with low $\beta_{\mathrm p}$, the proton orbits become stochastic mainly due to spatial variations in the electrostatic potential, and the protons gain primarily energy from the slow temporal variations in the electrostatic potential associated with the fluctuations \citep{chandran2010b}.
In plasmas with high $\beta_{\mathrm p}$, the proton orbits become stochastic mainly due to spatial variations in the magnetic field, and the protons primarily gain energy from the solenoidal component of the electric field \citep{hoppock2018}. 
Despite these differences, stochastic heating remains a universal candidate process to explain ion heating in the direction perpendicular to $\vec B_0$ in weakly collisional plasmas.

\subsection{Wave types in the solar wind}\label{sect:waves}

In this section, we discuss large-scale \emph{Alfv\'en waves}, \emph{kinetic Alfv\'en waves}, \emph{Alfv\'en/ion-cyclotron waves}, \emph{slow modes}, and \emph{fast modes}, which are the most important wave types for the multi-scale dynamics of the solar wind. We note that the nomenclature of wave types is not universal and that different names are commonly used for waves of the same type depending on their location in wavevector space \citep[e.g.,][Fig.~1]{tenbarge2012}.

\subsubsection{Large-scale Alfv\'en waves}\label{sec:alfven}

Alfv\'en waves are electromagnetic plasma waves for which magnetic tension serves as the restoring force \citep{alfven1942,alfven1943}. To first order, these waves are non-compressive.  At large scales (i.e., $kd_{\mathrm p}\ll1$ and $k\rho_{\mathrm p}\ll 1$), Alfv\'en waves obey the linear dispersion relation 
\begin{equation}\label{Alfdisp}
\omega =  \pm |k_{\parallel}|v_{\mathrm A}^{\ast},
\end{equation}
where the upper (lower) sign corresponds to propagation parallel (anti-parallel) to $\vec B_0$, and $v_{\mathrm A}^{\ast}\equiv B_0/\sqrt{4\pi \rho}$ is the MHD Alfv\'en speed.
The group-velocity vector is parallel or anti-parallel to $\vec B_0$, and large-scale Alfv\'en waves are only weakly damped in a plasma with Maxwellian distribution functions. The fluctuating magnetic-field vector $\delta \vec B$ is perpendicular to $\vec k$ and $\vec B_0$.
Alfv\'en waves are characterized by negligible fluctuations in $n_j$ (i.e., they are non-compressive) and $B\equiv |\vec B|$, but an (anti-)correlation between velocity fluctuations $\delta \vec U_j$ and magnetic-field fluctuations $\delta \vec B$. In the MHD approximation, this polarization property is given by
\begin{equation}\label{MHDAlf}
\frac{\delta \vec U}{v_{\mathrm A}^{\ast}}=\mp \frac{\delta \vec B}{B_0}.
\end{equation}

\begin{figure}
  \includegraphics[width=\textwidth]{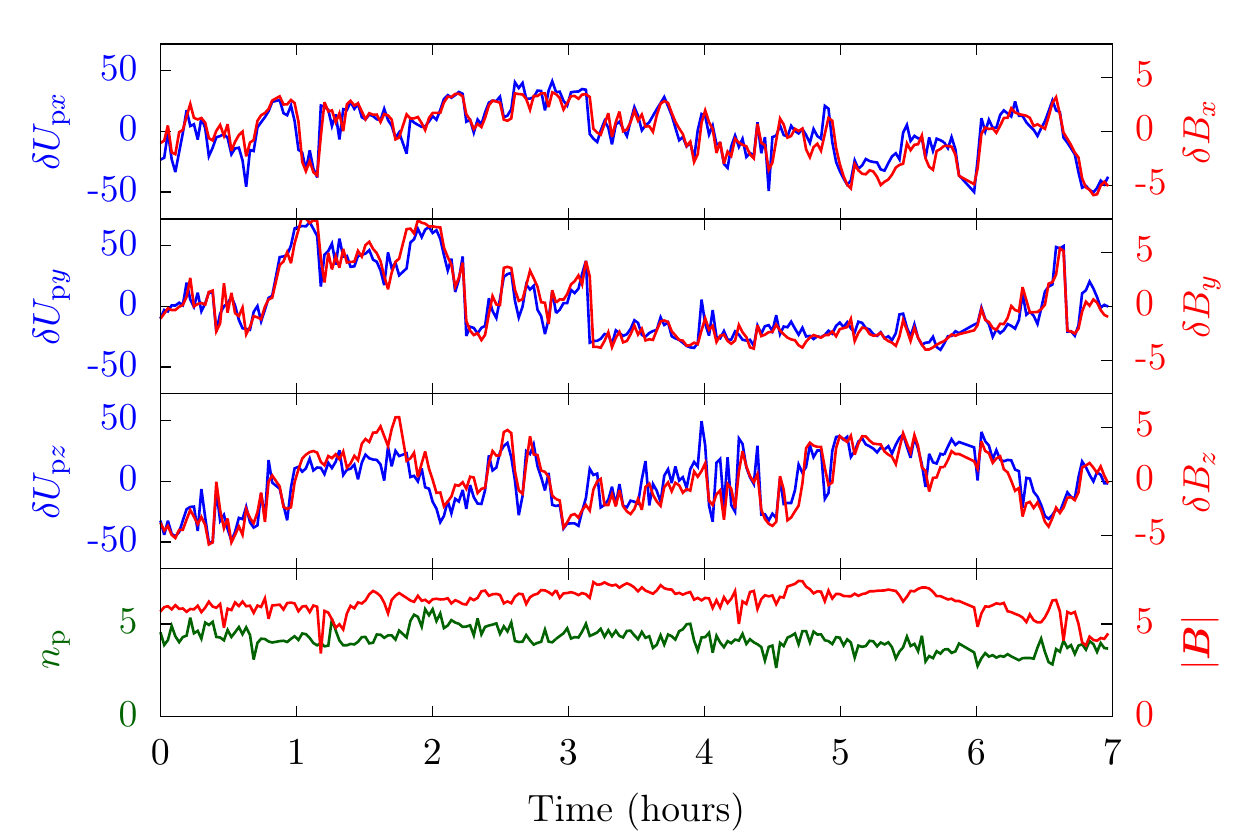}
\caption{Alfv\'enic correlations between $\delta \vec U_{\mathrm p}$ and $\delta \vec B$. We show data from the Wind spacecraft's SWE and MFI instruments starting at 18:01:59 on 2018-05-06 for a total duration of seven hours.  The top three panels show the three components of the vector velocity (km/s; blue) and magnetic-field (nT; red) fluctuations. The vector components are positively correlated in this example. The bottom panel shows that the density (cm$^{-3}$; green) and the absolute value of the magnetic field (nT; red) stay approximately constant.  }
\label{fig:alfvenic_correlation}       
\end{figure}

In the solar wind, the center-of-mass frame, in which we define $\omega$ and $\vec k$, is dominated by the proton flow so that  $\vec U\approx \vec U_{\mathrm p}$ and $\rho\approx n_{\mathrm p}m_{\mathrm p}$. Therefore, Equation~(\ref{MHDAlf}) is approximately $\delta \vec U_{\mathrm p}/v_{\mathrm {Ap}}\approx \mp \delta \vec B/B_0$. Observations of the vector components of the plasma velocity and the magnetic field in the solar wind often exhibit this polarization \citep{unti1968,belcher1969,belcher1971,bruno1985,velli1997,chandran2009,boldyrev2012,he2012,he2012a,podesta2012a}, and we illustrate one such example in Fig.~\ref{fig:alfvenic_correlation}. 
In fact, since this polarization characterizes the majority of the solar wind's large-scale fluctuations, its large-scale turbulence is believed to be Alfv\'en-wave-like turbulence (see Sect.~\ref{sec:waveturb}). At large scales, the amplitudes of the Alfv\'enic fluctuations in the solar wind are often so large that their behavior becomes nonlinear. Their polarization fulfills $B=\mathrm{constant}$, while the magnetic-field and velocity vectors often show a spherical or arc-like polarization \citep{tsurutani1994,riley1996,vasquez1996}. Although Alfv\'en waves predominantly occur in the fast solar wind, \citet{damicis2015} identify a type of slow wind that also carries large-amplitude Alfv\'en waves and shows many other characteristics usually associated with fast wind \citep{damicis2019}.

We note that left-circularly polarized and parallel-propagating Alfv\'en waves are a solution of the full nonlinear MHD and multi-fluid equations \citep{marsch2011a}.
At large scales, these waves follow a polarization relation that follows directly from the multi-fluid equations:
\begin{equation}\label{Alfvpol}
\frac{\delta \vec U_{j}}{v_{\mathrm A}^{\ast}}=\mp \left(1 \mp \frac{U_{\parallel j}}{v_{\mathrm A}^{\ast}}\right)\frac{\delta \vec B}{B_0},
\end{equation}
where the upper and lower signs describe the propagation direction as in Equation~(\ref{Alfdisp}). 
Equation~(\ref{Alfvpol}) shows that a particle species with $U_{\parallel j}\approx v_{\mathrm A}^{\ast}$ does not participate in the bulk-velocity polarization motion associated with parallel-propagating large-scale Alfv\'en waves: in the reference frame of these particles, the wave has no electric field. Observations confirm that $\alpha$-particles (see Sect.~\ref{sec:ion-prop}) with $U_{\parallel \alpha}\approx v_{\mathrm A}^{\ast}$ exhibit $\delta \vec U_{\alpha}\approx 0$, which is an effect known as \emph{surfing $\alpha$-particles}  \citep{marsch1982a,goldstein1995,matteini2015}.

 There are two extensions of the Alfv\'en wave to smaller scales: the kinetic Alfv\'en wave (KAW) at $k_{\perp}\rho_{\mathrm p} \gtrsim 1$ and $k_{\perp}\gg k_{\parallel}$, and the Alfv\'en/ion-cyclotron (A/IC) wave at $k_{\parallel}d_{\mathrm p}\gtrsim 1$ and $k_{\perp }\ll k_{\parallel}$. Although KAWs and A/IC waves belong to the Alfv\'en-wave family \citep{andre1985,yoon2008,klein2015}, we discuss them separately in the following two sections due to their great importance for the physics of the solar wind.

\subsubsection{Kinetic Alfv\'en waves}\label{sec:KAWs}

Kinetic Alfv\'en waves (KAWs) are the short-wavelength extension of the Alfv\'en-wave branch for $k_{\perp}\gg k_{\parallel}$. This type of wave has received much attention since large-scale turbulence in the solar wind is Alfv\'en-wave-like and supports a cascade with increasing anisotropy toward $k_{\perp}\gg k_{\parallel}$ (see Sect.~\ref{sec:waveturb}). Thus, KAWs are the prime candidate for extending the Alfv\'enic cascade to small scales. 

When $k_{\perp}\rho_{\mathrm p}\gtrsim 1$, finite-Larmor-radius effects modify the properties of the Alfv\'en wave. The linear KAW dispersion relation in the gyrokinetic limit with isotropic temperatures is given by \citep{howes2006}
\begin{equation}
\omega=\pm  \frac{|k_{\parallel}|v_{\mathrm {Ap}}k_{\perp}\rho_{\mathrm p}}{\sqrt{\beta_{\mathrm p}+{\displaystyle\frac{2}{ 1+T_{\mathrm e}/T_{\mathrm p}}}}}.
\end{equation}
KAWs are electromagnetic, are elliptically right-hand polarized, and have a frequency $\ll\Omega_{\mathrm p}$ in this limit. While large-scale Alfv\'en waves are non-compressive, KAWs exhibit fluctuations in the particle density $n_j$ and the magnetic-field strength $B$. Observations of polarization properties of proton-scale and sub-proton-scale fluctuations in the solar wind and other space plasmas often find an agreement with  the predicted KAW polarization  \citep{bale2005,salem2012,chen2013,podesta2013,roberts2013,klein2014,safrankova2019,zhu2019}. 

The compressive behavior of KAWs introduces fluctuations in the parallel electric field, allowing KAWs to experience Landau damping (see Sect.~\ref{sec:ql}). Hybrid fluid-gyrokinetic simulations suggest that KAW turbulence leads to preferential electron heating at low $\beta_{\mathrm p}$ and to preferential ion heating at high $\beta_{\mathrm p}$ \citep{kawazura2019}. At low $\beta_{\mathrm p}$, thermal protons do not satisfy the Landau-resonance condition according to Equation~(\ref{rescond}) with $n=0$. In this case, the KAW turbulence cascades to even smaller scales, ultimately leading to preferential electron heating through electron Landau damping and subsequent collisions. At the same time, nonlinear phase mixing of the ions (see Sect.~\ref{sec:entropy}) creates smaller structures in the ions' $v_{\perp}$ distribution, which eventually dissipate via collisions and perpendicularly heat the ions. At high $\beta_{\mathrm p}$, KAWs efficiently dissipate through proton Landau damping and subsequent collisions, which result in preferential parallel proton heating \citep{quataert1998,leamon1999,howes2010,plunk2013,tenbarge2013,he2015,told2015,hughes2017,howes2018}. 
Under certain conditions, KAW turbulence approaches the local ion-cyclotron frequency in the plasma frame, at which point perpendicular ion heating through cyclotron-resonant processes (see Sect.~\ref{sec:ql}) occurs \citep{arzamasskiy2019}.

In their stochastic-heating model (see Sect.~\ref{sec:stoc}), \citet{chandran2010b} determine the proton heating rate for stochastic heating by KAWs in low-$\beta_{\mathrm p}$ plasma to be
\begin{equation}\label{stoc1}
Q_{\perp}=c_1\frac{\left(\delta v_{\rho}\right)^3}{\rho_{\mathrm p}}\exp\left(-\frac{c_2}{\bar{\epsilon}}\right),
\end{equation}
where the empirical factors $c_1$ and $c_2$ are constants, $\delta v_{\rho}$ is the amplitude of the gyro-scale fluctuations in the $\vec E\times \vec B$ velocity, and $\bar{\epsilon}\equiv\delta v_{\rho}/w_{\perp \mathrm p}$. Test-particle simulations using plasma parameters consistent with low-$\beta_{\mathrm p}$ solar-wind streams suggest that $c_1\approx 0.75$ and $c_2\approx 0.34$ \citep{chandran2010b}, while reduced MHD simulations suggest larger values for $c_1$ and smaller values for $c_2$ \citep{xia2013}. 

In intermediate- to high-$\beta_{\mathrm p}$ plasma ($1\lesssim \beta_{\mathrm p}\lesssim 30$), the stochastic KAW proton heating rate is given by \citep{hoppock2018}
\begin{equation}\label{stoc2}
Q_{\perp}=\sigma_1\frac{\left(\delta v_{\rho}\right)^3}{\rho_{\mathrm p}}\sqrt{\beta_{\mathrm p}}\exp\left(-\frac{\sigma_2}{\bar{\delta}}\right),
\end{equation}
where $\sigma_1$ and $\sigma_2$ are constants,  $\bar{\delta}\equiv \delta B_{\rho}/B_0$, and $\delta B_{\rho}$ is the amplitude of gyro-scale fluctuations in the magnetic field. Test-particle simulations suggest that $\sigma_1=5$ and $\sigma_2=0.21$.\footnote{The use of $\bar{\epsilon}$ in Equation~(\ref{stoc1}) and $\bar{\delta}$ in Equation~(\ref{stoc2}) reflects the importance of the two different stochastization mechanisms discussed in Sect.~\ref{sec:stoc}: the electrostatic potential in low-$\beta_{\mathrm p}$ plasmas and the magnetic field in high-$\beta_{\mathrm p}$ plasmas.}

\subsubsection{Alfv\'en/ion-cyclotron waves}

Alfv\'en/ion-cyclotron (A/IC) waves are the short-wavelength extension of the Alfv\'en-wave branch for $k_{\parallel}\gg k_{\perp}$. The anisotropic Alfv\'enic turbulent cascade on its own cannot generate A/IC waves. However, A/IC waves have received considerable attention due to their ability to heat ions preferentially in the direction perpendicular to $\vec B_0$ through cyclotron resonance  \citep[see Sect.~\ref{sec:ql};][]{dusenbery1981,isenberg1983,gomberoff1991,hollweg1999,araneda2009,rudakov2012}.

The linear dispersion relation for quasi-parallel A/IC waves in the cold-plasma limit (i.e., $\beta_j\rightarrow 0$) is given by \citep{verscharen:phd}
\begin{equation}\label{AICdisp}
\frac{\omega_{\mathrm r}}{\Omega_{\mathrm p}}=\pm \frac{k^2d_{\mathrm p}^2}{2}\left(\sqrt{1+\frac{4}{k^2d_{\mathrm p}^2}}-1\right).
\end{equation}
In this regime, the A/IC wave is also known as the \emph{L-mode}.
The frequency is always less than $\Omega_{\mathrm p}$, and the quasi-parallel A/IC wave is almost fully left-circularly polarized -- the same sense of rotation as the cyclotron motion of positively charged particles. This polarization accounts for the frequency cutoff at the proton cyclotron frequency, above which plasmas are opaque to A/IC waves. For finite-temperature plasmas, $\omega_{\mathrm r}$ asymptotes to an even smaller value than $\Omega_{\mathrm p}$ since, with increasing temperature, an increasing number of particles resonate with the Doppler-shifted wave frequency in their reference frame.

The amplitudes of the perpendicular components of the fluctuating proton and electron bulk velocities are equal in the limit of $k\rightarrow 0$. The amplitude of the perpendicular proton bulk velocity then increases as $\omega_{\mathrm r}\rightarrow \Omega_{\mathrm p}$, while the amplitude of the perpendicular electron bulk velocity remains approximately constant. Therefore, the proton contribution to the polarization current increases with $\omega_{\mathrm r}$, until the protons carry most of the current.

The inherent ambiguities of single-spacecraft measurements (see Sect.~\ref{sec:meas:multi}) complicate the identification of A/IC waves within background solar-wind turbulence. However, \emph{A/IC-storms} have been observed as enhancements in the magnetic-field power spectrum at $\omega_{\mathrm r}\lesssim \Omega_{\mathrm p}$ with predominantly left-handed polarization \citep{jian2009,jian2010,he2011,jian2014,boardsen2015,wicks2016}.

A/IC waves damp on particles that fulfill the cyclotron-resonance condition according to Equation~(\ref{rescond}) in Sect.~\ref{sec:ql} with $n=+1$,
\begin{equation}
\omega_{\mathrm r}=k_{\parallel} v_{\parallel } + \Omega_{\mathrm p}.
\end{equation}
This effect heats ions very efficiently in the perpendicular direction. More specifically, the quasilinear pitch-angle diffusion through the $n=+1$ resonance creates a characteristic \emph{plateau} along pitch-angle gradients, which has often been observed in the fast solar wind \citep{cranmer2001,isenberg2001,marsch2001,tu2001,hollweg2002,gary2005,kasper2013,cranmer2014,woodham2018}.
These observations strongly support the A/IC-heating scenario, but difficulties remain in explaining the origin of these waves in the solar wind. Microinstabilities may play an important role in the generation of A/IC waves as we discuss in Sect.~\ref{sec:inst}.

\subsubsection{Slow modes}\label{slow_waves}

Although most solar-wind fluctuations are non-compressive, about 2\% of the fluctuating power is in compressive modes in the inertial range \citep{chen2016,safrankova2019}. Due to its polarization properties, the slow mode is a major candidate to explain these compressive fluctuations.

The linear dispersion relation of slow modes in the MHD limit is given by
\begin{equation}
\omega_{\mathrm r}=\pm kC_{-},
\end{equation}
where 
\begin{equation}\label{magsonspeed}
C_{\pm}\equiv v_{\mathrm A}^{\ast} \left[\frac{1}{2}\left(1+\frac{\kappa}{2}\beta_{\mathrm p}\right)\pm\frac{1}{2}\sqrt{\left(1+\frac{\kappa}{2}\beta_{\mathrm p}\right)^2-2\kappa\beta_{\mathrm p}\cos^2\theta}\right]^{1/2}
\end{equation}
is the fast (upper sign; see Sect.~\ref{sec:fast_modes}) and slow (lower sign) magnetosonic speed, $\kappa$ is the polytropic index, and $\theta$ is the angle between $\vec k$ and $\vec B_0$.  Oblique MHD slow modes at $\beta_{\mathrm p}<2/\kappa$ are characterized by an anti-correlation between fluctuations in density $\delta n_{j}$ and magnetic-field strength $\delta |\vec B|$. In this limit, the mode is largely acoustic in nature, and the mode's velocity perturbation is closely aligned with $\vec B_0$. In the high-$\beta_{\mathrm p}$ limit, the MHD slow mode is largely tensional in nature, and the mode's velocity perturbation $\delta \vec U$ is then predominantly (anti-)parallel to $\vec B_0$. In both of these limits of the MHD slow wave, the vector $\delta \vec B$ lies in the $\vec k$-$\vec B_0$ plane. In the limit of $\theta=0^{\circ}$, the MHD slow wave is either a pure acoustic wave with  $\delta \vec B=0$ when $\beta_{\mathrm p}<2\kappa$ or degenerate with the Alfv\'en wave when $\beta_{\mathrm p}>2\kappa$. In the limit of $\theta=90^{\circ}$, the slow mode does not propagate. 

Polarization properties are often more useful than phase speeds in defining the type of plasma wave. Therefore, we more generally define slow modes as the solutions to the dispersion relation that exhibit the anti-correlation between $\delta n_{j}$ and $\delta |\vec B|$ that characterizes the MHD slow mode's low-$\beta_{\mathrm p}$ limit. 
In kinetic theory, two solutions exhibit this anti-correlation.\footnote{In fact, kinetic linear theory has an infinite number of solutions with this anti-correlation. However, almost all of them are so heavily damped with $|\gamma|\gg |\omega_{\mathrm r}|$ that they are irrelevant for all practical purposes to the solar wind.} We consequently identify both of them with the \emph{kinetic slow mode}  \citep{verscharen2017}. 

The first solution is the \emph{ion-acoustic wave} \citep{narita2015}, which obeys the linear dispersion relation
\begin{equation}
\omega_{\mathrm r}=\pm |k_{\parallel}|\sqrt{\frac{3k_{\mathrm B}T_{\parallel \mathrm p}+k_{\mathrm B}T_{\parallel \mathrm e}}{m_{\mathrm p}}}
\end{equation}
which can be obtained in the gyrokinetic limit \citep{verscharen2017}.
The phase speed of this wave is the ion-acoustic speed, which indicates that the parallel pressures of protons and electrons provide this mode's restoring force, while the proton mass provides its inertial force. The protons behave like a one-dimensional adiabatic fluid since $\kappa_{\mathrm p}=3$, while the electrons behave like an isothermal fluid since $\kappa_{\mathrm e}=1$, where $\kappa_j$ is the polytropic index of species $j$.

The second type of kinetic slow mode is the \emph{non-propagating mode},\footnote{The non-propagating kinetic slow mode is sometimes called the kinetic \emph{entropy mode} in reference to the non-propagating MHD entropy mode. Although both modes share this non-propagating behavior, the MHD entropy mode is different from the kinetic slow mode in the sense that it does not exhibit variations in $\delta |\vec B|$.} which obeys the linear dispersion relation
\begin{equation}
\omega_{\mathrm r}=0.
\end{equation}
If any plasma species has a sufficiently strong temperature anisotropy with $T_{\perp j}>T_{\parallel j}$, the non-propagating mode can become unstable and then gives rise to the mirror-mode instability (see Sect.~\ref{sec:inst:tani}).

\begin{figure}
  \includegraphics[width=\textwidth]{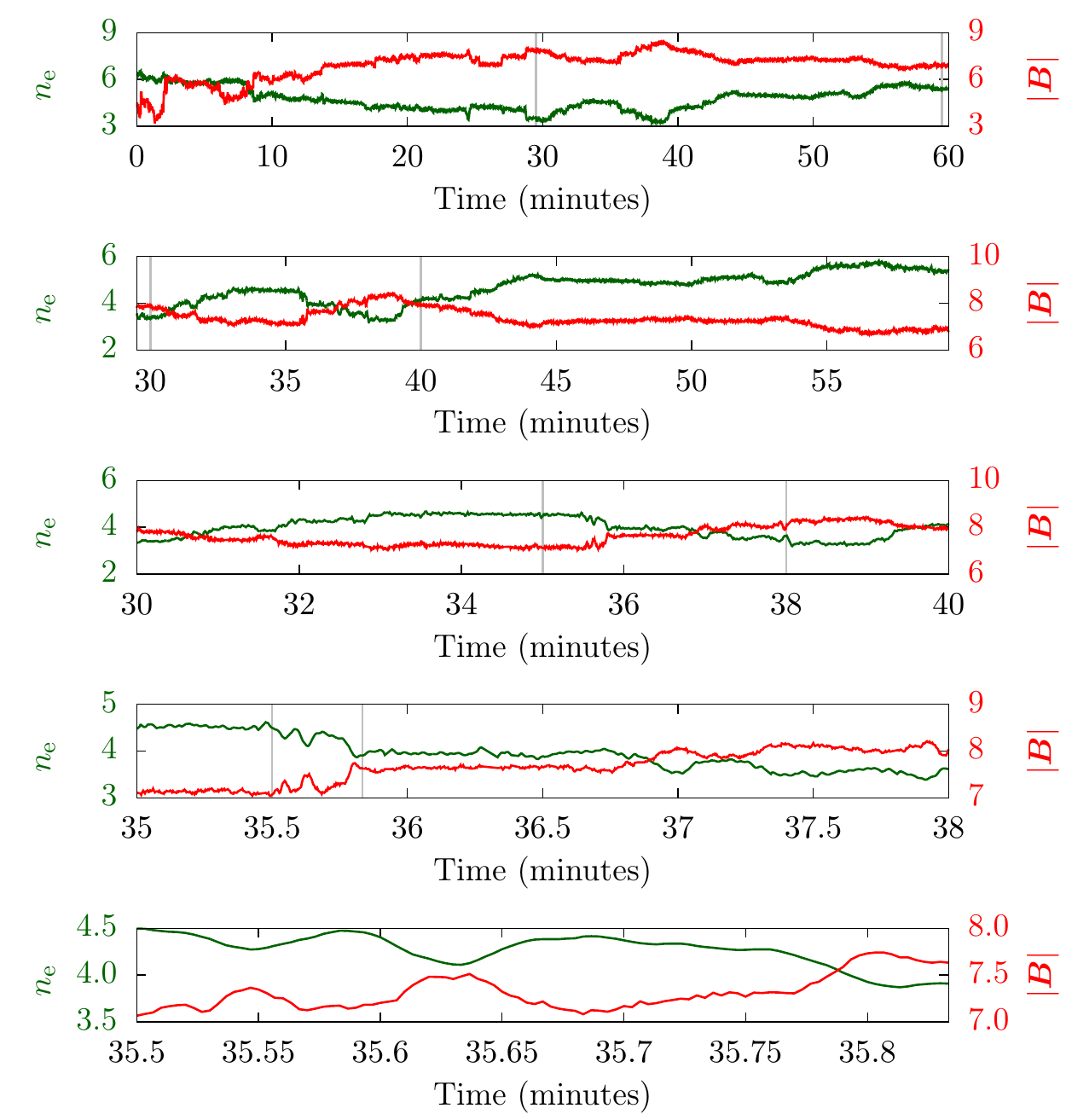}
\caption{Time series of $n_{\mathrm e}$ (cm$^{-3}$; green) and $|\vec B|$ (nT; red) in the solar wind on multiple scales, each of which has fluctuations that clearly exhibit the anti-correlation between $\delta n_{\mathrm e}$ and $\delta |\vec B|$ that characterizes slow waves. These panels show data from the Cluster EFW and FGM instruments measured for one hour starting at 22:30:00 on 2001-04-05. Following the technique by \citet{yao2011}, we show from top to bottom decreasing interval lengths. The gray lines in each plot indicate the start and end points of the interval shown in the plot immediately below it. We use a running average to filter the spacecraft spin tones from the data. }
\label{fig:compressions}       
\end{figure}

The anti-correlation of $\delta n_j$ and $\delta |\vec B|$, which defines slow modes, is frequently observed in the solar wind \citep{yao2011,kellogg2005,chen2012a,howes2012,klein2012,roberts2017,yang2017,roberts2018}.   Figure~\ref{fig:compressions} shows a period of solar-wind measurements that exemplify this anti-correlation over a wide range of scales.

Ion-acoustic waves mainly damp through Landau damping \citep{barnes1966}. Since the mode's phase speed is of order the proton thermal speed  (unless $T_{\parallel\mathrm e}\gg T_{\parallel \mathrm p}$), the ion-acoustic mode predominantly heats ions in the field-parallel direction. We note that the damping rate of slow modes is significant even at scales $\gg d_{\mathrm p}$. On this basis, slow modes have at times been rejected as candidates for the compressive fluctuations in the solar wind. Nevertheless, at very large angles between $\vec k$ and $\vec B_0$, the damping rate decreases significantly, and the ion-acoustic wave and the MHD slow wave no longer propagate. Instead, they become non-propagating structures that exhibit pressure balance,
\begin{equation}
P_{\mathrm{tot}}\equiv P+\frac{B^2}{8\pi}=\mathrm{constant}.
\end{equation}
These \emph{pressure-balanced structures} have been observed often and across many scales both in the solar wind and in plasma simulations  \citep{burlaga1970,marsch1990,marsch1993,tu1994,bavassano2004,verscharen2012,yao2013,yao2013a}. 
A recent study suggests that slow modes also play an important role in how low-frequency, low-$\beta_j$ plasma turbulence partitions heating between ions and electrons \citep{schekochihin2019}.

\subsubsection{Fast modes}\label{sec:fast_modes}

Fast modes are another type of compressive fluctuation, although they are non-compressive in parallel propagation. Their linear dispersion relation in the MHD approximation is given by
\begin{equation}
\omega_{\mathrm r}=\pm kC_+,
\end{equation}
where $C_+$ is the fast magnetosonic speed according to Equation~(\ref{magsonspeed}).
Oblique MHD fast modes at $\beta_{\mathrm p}<2/\kappa$ are characterized by a positive correlation between fluctuations in density $\delta n_{j}$ and magnetic-field strength $\delta |\vec B|$. In this limit, the mode's restoring force is a combination of the total-pressure-gradient force and the magnetic-tension force, and its velocity perturbation $\delta \vec U$ lies in the $\vec k$-$\vec B_0$ plane. In the high-$\beta_{\mathrm p}$ limit, the MHD fast mode is largely acoustic in nature, and the mode's velocity perturbation $\delta \vec U$ is mainly parallel to $\vec k$. In the limit of $\theta=0^{\circ}$, the MHD fast wave is either degenerate with the Alfv\'en wave when $\beta_{\mathrm p}<2\kappa$ or a purely acoustic wave with its velocity perturbation $\delta \vec U$ parallel to $\vec k$ when $\beta_{\mathrm p}>2\kappa$. In the limit of $\theta=90^{\circ}$, the MHD fast mode is a magnetoacoustic pressure wave. In the MHD fast wave, the vector $\delta \vec B$ lies in the $\vec k$-$\vec B_0$ plane.
Analogous to the case of generalized slow modes, we define fast modes as the solutions to the linear dispersion relation that exhibit a characteristic positive correlation between $\delta n_j$ and $\delta |\vec B|$ known from the low-$\beta_{\mathrm p}$ limit of the MHD fast mode. 

On smaller scales, the fast-mode family includes the \emph{whistler mode}, the \emph{lower-hybrid mode}, and the \emph{kinetic magnetosonic mode}. We refer to all modes of this family as \emph{fast-magnetosonic/whistler (FM/W) waves}. In the limit $k d_{\mathrm e}\ll 1$ in a cold plasma with quasi-parallel direction of propagation, the linear FM/W-wave  dispersion relation is approximately given by
\begin{equation}
\frac{\omega_{\mathrm r}}{\Omega_{\mathrm p}}=\pm \frac{k^2d_{\mathrm p}^2}{2}\left(\sqrt{1+\frac{4}{k^2d_{\mathrm p}^2}}+1\right),
\end{equation}
which connects to the Alfv\'en-wave branch at small $k$ as in Equation~(\ref{AICdisp}). The quasi-parallel FM/W wave is also known as the \emph{R-mode}. In the limit $kd_{\mathrm p}\gg 1$ and allowing for oblique propagation with $\cos^2 \theta \gtrsim m_{\mathrm e}/m_{\mathrm p}$, the cold-plasma FM/W-wave dispersion relation can be approximated by
\begin{equation}
\frac{\omega_{\mathrm r}}{|\Omega_{\mathrm e}|}\approx \pm \frac{k|k_{\parallel}|d_{\mathrm e}^2}{1+k^2d_{\mathrm e}^2}.
\end{equation}
In the limit $k\rightarrow \infty$, this dispersion relation  asymptotes toward $\sim |\Omega_{\mathrm e}|\cos\theta$. In this regime, the FM/W wave is known as the whistler wave.  The amplitudes of the perpendicular components of the fluctuating proton and electron bulk velocities are equal in the limit of $k\rightarrow 0$.  The amplitude of the fluctuations in the perpendicular electron bulk velocity then increases as $\omega_{\mathrm r}\rightarrow |\Omega_{\mathrm e}|$ while the amplitude of the fluctuations in the perpendicular proton bulk velocity decreases until the proton bulk velocity is almost zero. Therefore, the electron contribution to the polarization current increases with $\omega_{\mathrm r}$ until the electrons carry most of the current.
The electrons remain magnetized at these frequencies, while the protons are unmagnetized. The phase speed of whistler waves is proportional to $k$, so waves with a higher frequency travel faster than waves with a lower frequency. This strongly dispersive behavior of whistler waves is responsible for their name since they were first discovered as whistling sounds with decreasing pitch in radio measurements of ionospheric disturbances caused by lightning \citep{barkhausen1919,storey1953}.

In the highly-oblique limit ($\cos^2 \theta \lesssim m_{\mathrm e}/m_{\mathrm p}$), the FM/W wave corresponds to the lower-hybrid wave. A useful approximation for its linear dispersion relation in the cold-plasma limit is \citep{verdon2009}
\begin{equation}
\frac{\omega_{\mathrm r}^2}{\omega_{\mathrm{LH}}^2}\approx \frac{1}{1+\omega_{\mathrm e}^2/k^2c^2}\left(1+\frac{m_{\mathrm p}}{m_{\mathrm e}}\frac{\cos^2\theta}{1+\omega_{\mathrm {pe}}^2/k^2c^2}\right),
\end{equation}
where
\begin{equation}
\omega_{\mathrm{LH}}\equiv \frac{\omega_{\mathrm {pp}}}{\sqrt{1+\displaystyle\frac{\omega_{\mathrm {pe}}^2}{\Omega_{\mathrm e}^2}}}
\end{equation}
is the \emph{lower-hybrid frequency}. Under typical solar-wind conditions,  $\beta_{\mathrm p}\gtrsim 10^{-3}$, and the lower-hybrid wave is very strongly Landau-damped. However, this mode may be driven unstable by certain electron configurations and thus account for some of the electrostatic noise observed in the solar wind \citep{marsch1982b,lakhina1985,migliuolo1985,mcmillan2006}.

Quasi-parallel FM/W waves are right-hand polarized -- the same sense of rotation as the cyclotron motion of electrons. This polarization results in a frequency cutoff at the electron gyro-frequency.
FM/W waves are almost undamped at ion scales ($kd_{\mathrm e}\ll 1$). When they reach the electron scales, they cyclotron-resonate with thermal electrons very efficiently through the $n=-1$ resonance (see Sect.~\ref{sec:ql}). This leads to efficient perpendicular electron heating. Oblique FM/W modes can resonate with ions through other resonances, including the Landau resonance with $n=0$.

Quasi-perpendicular FM/W waves have been an alternative candidate to KAWs for explaining the observed solar-wind fluctuations at $k_{\perp}\rho_{\mathrm p}\gtrsim 1$  \citep{coroniti1982,he2012a,sahraoui2012,narita2016}. However, their existence is unlikely to result from the large-scale Alfv\'enic cascade since this scenario would necessitate a transition from Alfv\'enic modes to fast modes at some point in the cascade. 
The solar wind only rarely exhibits pronounced time intervals with a positive correlation between $\delta n_j$ and $\delta |\vec B|$ at large scales \citep{klein2012}. However, a number of observations of polarization properties of fluctuations reveal occasional consistency with the predictions for FM/W waves \citep{beinroth1981,marsch2011,chang2014,gary2016,narita2016}. FM/W modes may be the result of a class of microinstabilities (see Sects.~\ref{sec:inst:tani} and~\ref{sec:inst:beam}) and thus may be important for the thermodynamics of the solar wind beyond the turbulent cascade.


\section{Plasma turbulence}\label{sec:turbulence}

After a brief introduction to the phenomenology of plasma turbulence in Sect.~\ref{sec:waves:phen}, we discuss the important concepts of wave turbulence in Sect.~\ref{sec:waveturb} and critical balance in Sect.~\ref{sec:critical}. Section~\ref{sec:advanced} closes our description of turbulence with a brief discussion of more advanced topics. There are many excellent textbooks and review articles on plasma turbulence \citep[e.g.,][]{tu1995,bavassano1996,petrosyan2010,bruno2013}. We refer the reader to this literature for a deeper discussion of the topic.

\subsection{Phenomenology of plasma turbulence in the solar wind}\label{sec:waves:phen}

Turbulence is a state of fluids in which their characteristic quantities such as their velocity or density fluctuate in an effectively unpredictable way.\footnote{We use the term ``unpredictable'' here to refer to the statistic nature of turbulence and the notion of randomness \citep{leslie1973}. The fluctuations in these quantities are still bound within certain limits and exhibit  correlations.} 
Fluids with low viscosity transition easily into a turbulent flow pattern. Turbulence is inherently a multi-scale phenomenon. Energy enters the system at large scales. Nonlinear interactions between fluctuations on comparable scales then transfer the energy to fluctuations on different scales with a net transfer of energy to smaller and smaller scales. This cascade of energy occurs through the interaction of neighboring eddies in the fluid that break up into smaller eddies. At the smallest scales, the fluctuations eventually dissipate into heat through collisions and raise the medium's entropy. In a neutral fluid, the injection at large scales may represent a slow (compared to the characteristic time associated with the turbulent cascade) stirring mechanism.  The dissipation is a consequence of the viscous interaction, which strengthens with decreasing scale. Turbulence in a plasma, however, is different from turbulence in a neutral fluid due to the additional, electromagnetic interactions and  the presence of additional, non-viscous dissipation channels at the characteristic plasma scales ($\rho_j$, $d_j$, $\lambda_j$, etc.). The solar wind, due to its low collisionality, exemplifies such a turbulent plasma.

\begin{figure}
  \includegraphics[width=\textwidth]{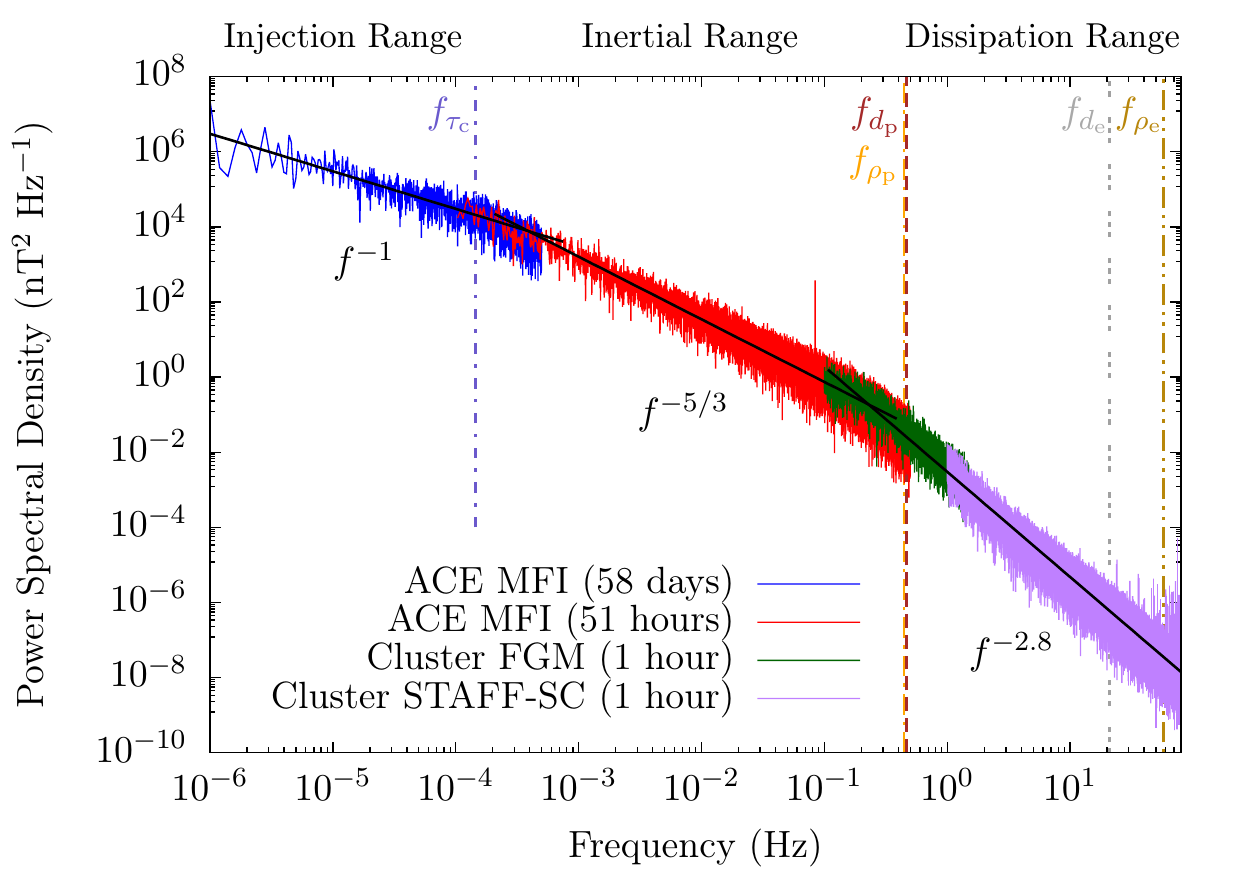}
\caption{Power spectral density of magnetic-field fluctuations in the solar wind during a time interval with $\beta_{\mathrm p}\sim 1$. The black lines show power laws with the power indices $-1$, $-5/3$, and $-2.8$, which are characteristic of the injection, inertial, and dissipation ranges, respectively. The frequency is measured in the spacecraft reference frame. The average plasma parameters are $B=4.528\,\mathrm{nT}$, $n_{\mathrm p}=1.02\,\mathrm{cm}^{-3}$, $n_{\mathrm e}=1.12\,\mathrm{cm}^{-3}$, $T_{\mathrm p}=1.26\,\mathrm{MK}$, $T_{\mathrm e}=0.138\,\mathrm{MK}$, and $U_{\mathrm p}=658\,\mathrm{km/s}$. After \citet{kiyani2015}. }
\label{fig:spectrum}       
\end{figure}

The multi-scale nature of turbulence leads to a broad power-law in the power spectral density of the fluctuating quantities. For fluid turbulence, a dimensional scale analysis shows that the power spectral density in the \emph{inertial range}, which is the range of scales between the large \emph{injection scales} and the small \emph{dissipation scales}, follows a power law in wavenumber $k$ (see also Fig.~\ref{fig:spectrum}). \citet{kolmogorov1941,kolmogorov1941a} estimates the power index of the power spectral density of the fluid velocity fluctuations by employing the following dimensional analysis. He identifies the dissipation rate with the constant rate  of energy transfer $\epsilon$ in the inertial range under steady-state conditions. For an eddy of size $\ell$ and velocity difference $\delta U_{\ell}$ across its extent, the characteristic time to turn over is approximately $\tau_{\mathrm{nl}}\sim \ell/\delta U_{\ell}$. The transfer rate of energy density for this eddy, on the other hand, is related to the energy density $\mathcal E$ through $\epsilon\sim \mathcal E/\tau_{\mathrm{nl}}=\mathrm{constant}$, where $\mathcal E\sim \left(\delta U_{\ell}\right)^2$. Combining these relations, we find $\mathcal E\sim \left(\epsilon \ell \right)^{2/3}$. Relating scale and wavenumber through $\ell\sim 1/k$ and defining the power spectral density as $E(k)\sim \mathcal E/k$ then leads to
\begin{equation}\label{kolm}
E(k)\sim \epsilon^{2/3}k^{-5/3}.
\end{equation}
Such a power law in $k$ is characteristic of turbulent fluids. Indeed, spectra of the solar wind's magnetic field, which have been measured in progressively greater detail for decades, often exhibit this power law  \citep{coleman1968,kiyani2015}. We show an examplar power spectrum of solar-wind magnetic fluctuations in frequency in Fig.~\ref{fig:spectrum}, which spans almost eight orders of magnitude in frequency \citep[for other examples, see][]{leamon1998,alexandrova2009,sahraoui2010,bruno2017}. We use the same instruments and data intervals in January and February of 2007 as \citet{kiyani2015} and compose a spectrum based on a direct fast Fourier analysis of a 58-day interval from ACE MFI, a 51-hour interval from ACE MFI, a 1-hour interval from Cluster 4 FGM, and the same 1-hour interval from Cluster 4 STAFF-SC. These time intervals are nested: each interval lies within the next longer time interval. 

When a single spacecraft measures a time series of a fluctuating quantity, it cannot distinguish between local temporal variations and variations due to the convection of spatial structures over the spacecraft with the solar-wind speed. Even purely spatial variations appear as temporal variations, so  a power spectrum in frequency reflects the combined effects of temporal and spatial variations \citep{taylor1938}.  More precisely, the Doppler shift connects the observed frequency $f_{\mathrm{sc}}$ of fluctuations in the spacecraft frame to the  wavevector $\vec k$ and the frequency $f_{0}$ of the fluctuations in the plasma frame through
\begin{equation}\label{doppler}
f_{\mathrm{sc}}=f_0+\frac{1}{2\pi}\vec k\cdot \Delta \vec U,
\end{equation}
where $\Delta \vec U$ is the velocity difference between the spacecraft frame and the plasma frame. For low-frequency fluctuations (i.e., $f_0\ll \vec k\cdot \Delta \vec U$), \emph{Taylor's hypothesis} simplifies the Doppler-shift relationship in Equation~(\ref{doppler}) to 
\begin{equation}
f_{\mathrm{sc}}\approx \frac{1}{2\pi} \vec k\cdot \Delta \vec U,
\end{equation} which is often used in the analysis of solar-wind fluctuations \citep[for a more detailed discussion of its applicability, see][]{howes2014a,klein2014a,klein2015a,bourouaine2018}.
 In Fig.~\ref{fig:spectrum}, we use Taylor's hypothesis to convert the convected frequencies associated with the scales $d_{j}$ and $\rho_j$ as $f_{d_j}\equiv U_{\mathrm p}/2 \pi d_j$ and $f_{\rho_j}\equiv U_{\mathrm p}/2 \pi \rho_j$, respectively, based on the average Cluster 4 FGM, CIS, and PEACE measurements during the 1-hour time interval used in this analysis. 
Figure~\ref{fig:spectrum} shows all three of the typical ranges observed in the solar wind. At the lowest frequencies ($f_{\mathrm{sc}}\lesssim 10^{-4}\,\mathrm{Hz}$), is the injection range, which follows a power law with $f_{\mathrm{sc}}^{-1}$. For comparison, we note that the expansion time of $\tau=2.4\,\mathrm d$ corresponds to a frequency of about $5\times 10^{-6}\,\mathrm{Hz}$, while the solar rotation period $\tau_{\mathrm{rot}}=25\,\mathrm d$ corresponds to a frequency of about $5\times 10^{-7}\,\mathrm{Hz}$ (see Sect.~\ref{sec:scales}). The nature and origin of fluctuations in the injection range are not well understood \citep{matthaeus1986,verdini2012,consolini2015}. The fluctuations exhibit Alfv\'enic polarization properties (see Sect.~\ref{sec:alfven}) and $B\approx \mathrm{constant}$ \citep{matteini2018,bruno2019}. 

At intermediate frequencies ($10^{-4}\,\mathrm{Hz}\lesssim f _{\mathrm{sc}}\lesssim 1\,\mathrm{Hz}$),  the inertial range of magnetic fluctuations approximately follows a power law with $f_{\mathrm{sc}}^{-5/3}$, which roughly agrees with Kolmogorov's theory according to Equation~(\ref{kolm}). 
Fluctuations in other quantities, such as bulk velocity \citep{boldyrev2011} and density \citep{kellogg2005}, have similar but not identical spectral indices compared to the magnetic fluctuations. The differences between the magnetic-field and velocity spectra are interpreted as  resulting from significant residual energy being generated at large scales. 
At high frequencies ($f_{\mathrm{sc}}\sim 1\,\mathrm{Hz}$), the magnetic-field spectrum steepens again toward a power law approximately following  $f_{\mathrm{sc}}^{-2.8}$, which may indicate the beginning of the \emph{dissipation range}. The power index at small scales varies, however, and the origin of this break is still unclear. Recent work suggests that there is a further transition at the electron scales toward an even steeper slope of the spectrum \citep{alexandrova2009,sahraoui2009}. The e-folding de-correlation time of the 51-hours time interval is $\tau_{\mathrm c}=18.3\,\mathrm {min}$, and we define $f_{\tau_{\mathrm c}}\equiv 1/ 2\pi \tau_{\mathrm c}$ as the spacecraft frequency associated with the e-folding de-correlation length. Like most properties of the solar wind, the fluctuations change with distance from the Sun. For instance, solar-wind expansion causes the overall level of fluctuation amplitudes to decrease with distance \citep{bavassano1982,burlaga1984}. The power of the large-scale magnetic-field fluctuations beyond a few tens of $R_{\odot}$ decreases approximately $\propto r^{-3}$ as predicted by WKB theory \citep{belcher1974,hollweg1974}. Moreover, the positions of the spectral breakpoints vary with distance \citep{matthaeus1982,bavassano1986,roberts1987}. The spacecraft-frame frequency $f_{\mathrm b1}$ of the breakpoint between the injection range and the inertial range decreases with distance $r$ from the Sun as $f_{\mathrm b1}\propto r^{-1.5}$ \citep{bruno2009}, while the frequency $f_{\mathrm b2}$ of the breakpoint between the inertial range and the dissipation range decreases as $f_{\mathrm b2}\propto r^{-1.09}$ \citep{bruno2014}.

The importance of damping and dissipation of plasma turbulence in the solar wind is underlined by the finding that the energy cascade rate through the inertial range  in solar-wind turbulence \citep[e.g.,][]{macbride2008} is typically sufficient to explain the observed heating of the solar wind (see Sect.~\ref{open_questions}). These studies are based on the relationship found by \citet{politano1998}, which estimates the energy transfer rate assuming isotropy, incompressibility, homogeneity, and equipartition between magnetic and kinetic energies. However, it is as yet unclear what underlying physics mechanisms heat the plasma through the damping and dissipation of the turbulent fluctuations.

\subsection{Wave turbulence and its composition} \label{sec:waveturb}

In order to understand the effects of solar-wind turbulence on the multi-scale evolution of the plasma, we must determine the nature of the fluctuations. \citet{iroshnikov1963} and \citet{kraichnan1965} suggest that \emph{MHD turbulence} in a strongly magnetized medium is a manifestation of nonlinear  collisions between counter-propagating Alfv\'en-wave packets. According to their statistically isotropic theory, the Alfv\'en-wave-collision mechanism leads to a power law of the magnetic-field spectrum with 
\begin{equation}\label{iro}
E(k)\sim k^{-3/2}
\end{equation}
 in the inertial range.  This work introduced the framework of \emph{wave turbulence}  \citep[see also][]{howes2014} into plasma-turbulence research. Wave turbulence accounts for the fact that a plasma, unlike a neutral fluid, carries plasma waves as linear normal modes for the system (see Sect.~\ref{sec:waves:lin}). The linear response of the system still plays a role in the dynamics of the turbulence, even though the evolution of the turbulence is nonlinear. Therefore, fluctuations in wave turbulence retain certain characteristics of the plasma's linear normal modes such as propagation and polarization properties. In the wave-turbulence framework, the identification of the nature of plasma turbulence is thus informed by the identification of the dominant wave modes of the turbulence.  As a caveat to this picture, we note that nonlinear interactions may generate fluctuations that are not (linear) normal modes of the system as those described in Sect.~\ref{sect:waves}. These driven modes may behave unexpectedly, and linear theory does not predict their properties.

There are two important timescales associated with fluctuations in wave turbulence: the \emph{linear time} $\tau_{\mathrm{lin}}$ and the \emph{nonlinear time} $\tau_{\mathrm{nl}}$. The linear time is associated with the evolution of the plasma's dominant wave modes due to propagation along $\vec B_0$. It is related to the wave frequency through 
\begin{equation}\label{tlin}
\tau_{\mathrm{lin}}\sim \frac{1}{\omega_{\mathrm r}}.
\end{equation}
The nonlinear time is associated with the nonlinear interaction between the modes perpendicular to the field direction, which leads to the nonlinear cascade process. It is related to the perpendicular wavenumber $k_{\perp}$ and the perpendicular fluctuations in velocity $\delta U_{\perp}$ through 
\begin{equation}\label{tnl}
\tau_{\mathrm{nl}}\sim \frac{1}{k_{\perp}\,\delta U_{\perp}}.
\end{equation}
Turbulence is called \emph{strong} when $\tau_{\mathrm{lin}} \gtrsim \tau_{\mathrm{nl}}$ and \emph{weak} when $\tau_{\mathrm{lin}}\ll \tau_{\mathrm{nl}}$. Wave turbulence can exist in the strong and in the weak regime, and we emphasize that the terms \emph{wave turbulence} and \emph{weak turbulence} are not interchangeable.

In the weak-turbulence paradigm,  the collision of two waves with frequencies $\omega_1$ and  $\omega_2$  and with wavevectors $\vec k_1$ and $\vec k_2$ most efficiently leads to a resultant wave with frequency \citep{montgomery1981,shebalin1983,montgomery1995}
\begin{equation}\label{wkt1}
\omega_3=\omega_1+\omega_2
\end{equation}
and wavevector 
\begin{equation}\label{wkt2}
\vec k_3=\vec k_1+\vec k_2.
\end{equation}
 Assuming Alfv\'en waves with $\omega=\pm k_{\parallel}v_{\mathrm A}^{\ast}$ (see Sect.~\ref{sec:alfven}), where $k_{\parallel}\equiv \vec k \cdot \vec B_0/B_0$, these wave--wave resonances cannot feed an MHD Alfv\'en-wave triad with $\omega_3\neq 0$. Although $k_{\perp}$ can increase, these triads lead to a situation with  $k_{\parallel}\rightarrow 0$, where $k_{\perp}\equiv |\vec k-k_{\parallel}\vec B_0/B_0|$. This weak-turbulence process plays an important role in the onset of plasma turbulence because it creates increasingly perpendicular wavevectors. Indeed, spacecraft observations show a strong wavevector anisotropy with $k_{\perp}\gg k_{\parallel}$ in the solar wind for the majority of turbulent fluctuations \citep{dasso2005,hamilton2008,tessein2009,macbride2010,wicks2010,chen2011,ruiz2011,chen2012,horbury2012,oughton2015,lacombe2017}. 

Indirect measurements of the \emph{two-point correlation function} 
\begin{equation}
R(\vec r)\equiv \left\langle \vec B(\vec x)\cdot \vec B(\vec x+\vec r) \right\rangle
\end{equation}
and the \emph{magnetic helicity}
\begin{equation}
H\equiv \int \vec A\cdot \vec B\,\mathrm d^3\vec x,
\end{equation}
where $\langle \cdots \rangle$ indicates the average over many positions $\vec x$, and $\vec A$ is the magnetic vector potential, independently reveal the existence of two highly-anisotropic components of turbulence \citep{matthaeus1990,tu1993,bieber1996,podesta2011,he2012}. The first component consists of highly-oblique fluctuations with $k_{\perp}\gg k_{\parallel}$. The second component consists of fluctuations that are more field-aligned ($k_{\perp}\ll k_{\parallel}$) and have lower amplitudes. This discovery led to the notion of the simultaneous existence of two-dimensional ($k_{\parallel}\simeq 0$) turbulent fluctuations and slab ($k_{\perp}\simeq 0$) wave-like fluctuations. Although this \emph{slab+2D model} successfully reproduces the bimodal nature of the fluctuations in the solar wind, it does not account for a broader distribution of power in three-dimensional wavevector space.

Since waves and turbulence are interlinked through the concept of wave turbulence,  a good understanding of the linear properties of plasma waves (Sect.~\ref{sect:waves}) is important to understand the nature of the fluctuations and their dissipation mechanisms. By combining these concepts, we achieve a deeper insight into the dissipation mechanisms of turbulence.  Working in the framework of wave turbulence, however, we emphasize again that we refer to waves as both the classical linear wave modes and the carriers of the turbulent fluctuations in wave turbulence.

\subsection{The concept of critical balance}\label{sec:critical}

\emph{Critical balance} describes the state of strong wave turbulence in which the linear and the nonlinear timescales from Equations~(\ref{tlin}) and (\ref{tnl}) are of the same order \citep{sridhar1994,goldreich1995,lithwick2007} :
\begin{equation}\label{cbcond}
\omega_{\mathrm r}(k_{\parallel},k_{\perp})\sim k_{\perp}\,\delta U_{\perp}.
\end{equation}
The physics justification for critical balance is based on a causality argument \citep{howes2015}. Initially, a weak-turbulence interaction of two counter-propagating plasma waves as quantified in Equations~(\ref{wkt1}) and (\ref{wkt2}) generates a pseudo-wave packet with $k_{\parallel}\simeq 0$ and with $k_{\perp}$ greater than that of either of the first two waves. However, causality forbids the final state of the turbulence from being completely two-dimensional. If it were, two planes at different locations along the background magnetic field would have to be identical if truly $k_{\parallel}=0$, which precludes any structure along $\vec B_0$ \citep{montgomery1982}. These two arbitrary planes, though, can only be identical if they are able to causally communicate with each other, which occurs via the exchange of Alfv\'en waves between them. This interplay between the generation of smaller $k_{\parallel}$ through weak-turbulence interactions and the requirement of causal connection along $\vec B_0$ creates a situation in which the timescale of the nonlinear interactions in one plane (i.e., $\tau_{\mathrm{nl}}$) is of order the timescale of the communication between the two planes (i.e., $\tau_{\mathrm{lin}}$). This describes the critical-balance condition in Equation~(\ref{cbcond}). 
In this model, the wave collision creates a pseudo-wave packet with $k_{\parallel}\simeq 0$, which then interacts with another propagating wave from the pool of fluctuations. This results in a new propagating wave with an even higher $k_{\perp}$. This multi-wave process, mediated by pseudo-wave packets and propagating wave packets, generates anisotropy while still satisfying causality through the field-parallel propagating waves. This process fills the \emph{critical-balance cone}, which is the wavevector space satisfying Equation~(\ref{cbcond}), as it distributes power in three-dimensional wavevector space at increasing wavenumbers. Turbulence in the critical-balance state is still strong turbulence (rather than weak), notwithstanding that it retains properties of the associated plasma normal modes according to the wave-turbulence paradigm. 

Although the justification of critical balance is still under debate \citep{matthaeus2014,zank2017}, there is a growing body of evidence from spacecraft measurements for the existence of conditions consistent with critical balance and wave turbulence in the solar wind \citep[for a summary, see][]{chen2016}. We note, however, that the fluctuations in the solar wind do not consist of only one prescribed type of fluctuations (quasi-parallel waves, non-propagating structures and vortices, critically balanced wave turbulence, etc.) but rather a combination of these.

The concept of critical balance can be further illustrated in the MHD approximation (see Sect.~\ref{sec:MHD}), which has a long and successful history in plasma-turbulence research.
For \emph{incompressible MHD turbulence} ($\nabla \cdot \vec U=0$) consisting of transverse ($\delta \vec B\perp \vec B_0$ and $\delta \vec U\perp \vec B_0$) fluctuations, the \citet{elsasser1950} formulation of the MHD equations is a useful parameterization, {which has been applied successfully to solar-wind measurements \citep{grappin1990,marsch1990a}}. We define the \emph{Elsasser variables} 
\begin{equation}
\vec z^{\pm}\equiv \delta \vec U\mp\frac{\delta \vec B}{\sqrt{4\pi \rho}}
\end{equation}
for forward (upper sign) and backward (lower sign) propagating Alfv\'en waves with respect to the background field $\vec B_0$. Using these variables, we rewrite the MHD momentum equation (\ref{MHDmoment}) and Faraday's law (\ref{MHDfaraday}) as
\begin{equation}\label{elsasser}
\frac{\partial \vec z^{\pm}}{\partial t}\pm \left( \vec v_{\mathrm A}^{\ast}\cdot \nabla\right)\vec z^{\pm}=-\left(\vec z^{\mp}\cdot\nabla\right)\vec z^{\pm}-\frac{1}{\rho}\nabla P_{\mathrm{tot}},
\end{equation}
where $\vec v_{\mathrm A}^{\ast}\equiv \vec B_0/\sqrt{4\pi\rho}$ is the MHD Alfv\'en speed and $P_{\mathrm{tot}}\equiv P+B^2/8\pi$. The terms on the left-hand side of Equation~(\ref{elsasser}) represent the linear behavior of $\vec z^{\pm}$, while the terms on the right-hand side represent their nonlinear behavior. The linear terms are responsible for propagation effects, while the nonlinear terms are responsible for the cross-scale interactions, which are the building blocks of Alfv\'en-wave turbulence. Using Equations~(\ref{tlin}) and (\ref{tnl}), we estimate the frequencies associated with the linear timescale $\tau_{\mathrm{lin}}$ and the nonlinear timescale $\tau_{\mathrm{nl}}$ from the spatial operators on $\vec z^{\pm}$ in Equation~(\ref{elsasser}) as
\begin{equation}\label{elspar}
\frac{1}{\tau_{\mathrm{lin}}} \sim  \left(\vec v_{\mathrm A}^{\ast}\cdot \nabla\right)\sim \frac{v_{\mathrm A}^{\ast}}{\ell_{\parallel}}
\end{equation}
and 
\begin{equation}\label{elsperp}
\frac{1}{\tau_{\mathrm{nl}}}\sim \left(\vec z^{\mp}\cdot \nabla\right)\sim \frac{\delta U}{\ell_{\perp}},
\end{equation}
where we define the characteristic scales $\ell_{\parallel}$ and $\ell_{\perp}$ parallel and perpendicular with respect to $\vec B_0$. In critical balance, $\tau_{\mathrm{lin}}\sim \tau_{\mathrm{nl}}$ so that
\begin{equation}
\frac{\delta U}{\ell_{\perp}}\sim \frac{v_{\mathrm A}^{\ast}}{\ell_{\parallel}},
\end{equation}
which corresponds to $k_{\perp }\delta U\sim k_{\parallel}v_{\mathrm A}^{\ast}$ as in Equation~(\ref{cbcond}). Critical balance predicts that the inertial-range power spectrum of magnetic-field fluctuations in the direction perpendicular to $\vec B_0$  follows the Kolmogorov slope given by Equation~(\ref{kolm}), where $k$ is replaced by $k_{\perp}$. The inertial-range power spectrum of magnetic fluctuations in the direction parallel to $\vec B_0$ then follows $E(k_{\parallel})\sim k_{\parallel}^{-2}$. 

The phenomenological model of \emph{dynamic alignment} describes an extension of critical balance \citep{boldyrev2005,boldyrev2006,mallet2015}. In this model, the turbulent velocity fluctuations $\delta \vec U$ increasingly  align their directions with the directions of the mangetic-field fluctuations $\delta \vec B$ as the energy cascades toward smaller scales. This framework predicts two limits depending on the strength of the background magnetic field. If the background field is strong, the turbulent spectrum follows the Iroshnikov--Kraichnan  slope given by Equation~(\ref{iro}), where $k$ is replaced by $k_{\perp}$, in the perpendicular direction. Conversely, if the background field is weak, the perpendicular spectrum follows the Kolmogorov slope given by Equation~(\ref{kolm}), where $k$ is replaced by $k_{\perp}$. This prediction is consistent with MHD simulations of driven turbulence \citep{mueller2003}. In the fully aligned state, either $\vec z^+$ or $\vec z^-$ is exactly zero, so nonlinear interactions cease.

%
%

\subsection{Advanced topics}\label{sec:advanced}

We briefly address three topics of great importance for solar-wind turbulence research that go beyond the direct focus of our review on the multi-scale nature of the solar wind: intermittency, reconnection, and anti-phase-mixing.

\subsubsection{Intermittency}\label{sec:intermittent}

The two-point speed increment is defined as $\delta u(r)\equiv \langle U(x+r)-U(x)\rangle$, where $x$ is the distance along a straight path through a volume of plasma and $\langle\dots\rangle$ is the average over many $x$. Though the probability distribution of $\delta u(r)$ in the solar wind has a Gaussian distribution at larger scales $r$, it exhibits non-Gaussian features at smaller $r$ \citep{marsch1994,sorriso1999,sorriso2001,osman2014a}. Specifically, the distribution develops enhanced tails, which indicate that sharp changes in velocity occur more frequently than predicted by Gaussian statistics. The increments in the magnetic field also exhibit this statistical property. These findings suggest that the solar-wind turbulence is \emph{intermittent} (i.e., exhibiting bursty patches of increased turbulence) and forms localized regions of enhanced fluctuations. 

The diagnostic called \emph{Partial Variance of Increments (PVI)} is defined as \citep{greco2008}
\begin{equation}
\mathrm{PVI}\equiv \frac{\left|\delta \vec B(t,\tau)\right|}{\sqrt{\left \langle \left |\delta \vec B(t,\tau )\right|^2\right \rangle}},
\end{equation}
where $\delta \vec B(t,\tau)\equiv \vec B(t+\tau)-\vec B(t)$ is the magnetic-field increment in a time-series measurement of $\vec B(t)$ \citep{greco2018}. PVI enables the identification of intermittency and allows for the statistical comparison of intermittency in plasma simulations and solar-wind observations \citep{wang2013,greco2016}. Large PVI values indicate \emph{coherent structures}, which are organized and persistent turbulent flow patterns and are believed to be the building blocks of intermittency. Because non-linearities are locally quenched inside these coherent structures, they survive longer than the surrounding turbulence.
The slow solar wind exhibits greater enhancements in PVI values than the fast solar wind \citep{servidio2011,greco2012}, which demonstrates that the slow solar wind contains a greater density of coherent structures than the fast solar wind \citep[see also][]{bruno2003}.  Regions of increased plasma heating and non-Maxwellian features in the particle distribution functions tend to occur in and around coherent structures \citep{osman2011,wan2012,karimabadi2013,wu2013,wan2015,parashar2016,yang2017a}.

Intermittency is a general feature known from fluid turbulence \citep{mccomb1990}. However, it remains unclear how intermittency and wave turbulence interact in the solar wind and what role intermittency plays in the dissipation of turbulence \citep{wang2014,wan2015,wan2016,zhdankin2016,perrone2017,howes2018,mallet2019}. 

\subsubsection{Magnetic reconnection}

\emph{Magnetic reconnection} refers to the rearrangement of the magnetic field in a highly-conducting fluid through resistive diffusion, which leads to a conversion of magnetic-field energy into particle energy.
In regard to plasma turbulence, magnetic reconnection is a process that is closely related to intermittency. Intermittency is associated with localized large gradients in the magnetic field, which, according to Amp\`ere's law in Equation~(\ref{Maxw4full}), corresponds to \emph{current sheets}: localized regions of enhanced current $\vec j$, which are a type of coherent structure as introduced in Sect.~\ref{sec:intermittent} \citep{karimabadi2013,tenbarge2013a,howes2016}. Current sheets are candidate regions for magnetic reconnection, which demonstrates the direct link between turbulence and reconnection \citep{matthaeus1984,servidio2009,servidio2010,osman2014}, and reconnection acts as a dissipation channel for the turbulent fluctuations \citep{retino2007,sundkvist2007,cerri2017,shay2018}. On the other hand, reconnection sites are inherently unstable to the \emph{tearing instability}, which progressively fragments them into smaller and smaller current sheets \citep{loureiro2007,lapenta2008,loureiro2016,tenerani2016}. In this way, reconnection sites generate a cascade to smaller scales by themselves and thus drive turbulence. In these progressively fragmented current sheets, the reconnection time gradually becomes faster than any other timescale, including the nonlinear time \citep{pucci2014}. When this condition is established, reconnection is able to interrupt the cascade of Alfv\'en-wave turbulence \citep{boldyrev2017,loureiro2017,mallet2017}. Therefore, reconnection must be considered when studying turbulence dynamics at small scales.

For further information on the connection between turbulence, coherent structures, and reconnection, we recommend the review article by \citet{matthaeus2011} and the  comprehensive textbook by \citet{frisch1995}.

\subsubsection{Anti-phase-mixing}

In Sects.~\ref{sec:ql} and \ref{sec:entropy}, we discuss the formation of smaller velocity-space structure in the particle distribution function through linear and nonlinear phase mixing. \emph{Anti-phase-mixing}, which is a stochastic variant of the \emph{plasma echo} effect \citep{gould1967}, is a process by which small-scale structure is removed from the distribution function in a turbulent plasma. For electrostatic turbulence, \citet{parker2016} and \citet{schekochihin2016} describe phase mixing and anti-phase-mixing in terms of the flux of energy in Hermite space of the particle distribution function. Phase mixing creates a transfer of energy from small to large Hermite moments. In a turbulent plasma with a low collision rate, a stochastic plasma echo creates a transfer of energy from large to small Hermite moments: effectively from small-scale structure to large-scale structure in velocity space. It therefore suppresses small-scale structure in the distribution function and thus non-Maxwellian features that may have otherwise led to collisional damping after ongoing phase mixing as described in Sect.~\ref{sec:entropy}. Anti-phase-mixing not only counteracts collisionless damping mechanisms but also leads to a fluid-like behavior of fluctuations even at low collisionality because higher-order-moment closures become unnecessary \citep{meyrand2019}. This process is potentially responsible for the observed fluid-like behavior of compressive and KAW-like fluctuations in space plasmas \citep{verscharen2017,wu2019}.

\section{Kinetic microinstabilities}\label{sec:inst}

\emph{Instabilities} are mechanisms that transfer energy from free-energy
sources, such as the non-equilibrium particle distributions described in Sects.~\ref{sec:ion-prop} and \ref{sec:elec-prop} or
large-amplitude waves, to plasma normal modes
that initially have amplitudes at the thermal-noise level  \citep{rosenbluth1965}. The
amplitude of these normal modes then grows exponentially with time as shown in Equation~(\ref{deltaE}),
\begin{equation}\label{Egrowth}
A(\vec x,t)\propto e^{\gamma t},
\end{equation}
where $\gamma>0$ is the growth rate of the instability, out of the
thermal noise during the linear phase of the instability, while it
extracts energy from its free-energy source. After the linear phase,
the normal-mode amplitude reaches some saturation level, at which
point nonlinear behavior  occurs that limits the
exponential growth of the instability. 

In this section, we focus on
small-scale instabilities that have characteristic wavelengths of order the
particle kinetic scales $d_j$ and $\rho_j$ and that affect the large-scale dynamic
evolution of the solar wind. We divide these instabilities into two categories. First, we discuss those associated with non-thermal structure in the particle
velocity distributions, including
temperature anisotropies and beams. These instabilities lead to
\emph{wave--particle interactions} that drive unstable growth. Second, we discuss those instabilities caused by large-amplitude
fluctuations, producing \emph{wave--wave interactions} that drive
unstable growth. This taxonomy provides the organizational
structure for this section.

Generically, both types of instabilities generate small-scale
fluctuations in the electric and/or magnetic field. While the turbulent cascade is dominated by interactions that are local in wavevector space (see Sect.~\ref{sec:waves:phen}),  instabilities  directly inject energy into the fluctuation spectrum at small scales. The scattering of particles on these small-scale field structures acts as an effective viscosity for the large-scale plasma behavior and thereby influences 
the thermodynamic evolution of the solar wind \citep{kunz2011,kunz2014,rincon2015,riquelme2015,riquelme2016,riquelme2017,riquelme2018}. As we focus on the effects of small-scale structure on larger-scale
behavior, we point the interested reader to the complementary review by
\citet{matteini2012} on the complementary effects of large-scale solar-wind
behavior on kinetic-scale phenomena. In particular, the discussion of the effects of background inhomogeneities at larger scales
are left for later editions of this review.

\subsection{Wave--particle instabilities}\label{wpinst}

Wave--particle instabilities are driven by departures of velocity distribution functions from the Maxwellian
equilibrium  given in Equation~(\ref{maxwellian}). Such departures are frequently observed in
the solar wind (see Sect.~\ref{sec:ion-prop} and~\ref{sec:elec-prop}), but not all of the associated energy is available to drive the system unstable. For instance,
unequal temperatures between different plasma species are not known by
themselves to drive wave--particle instabilities, which has major implications for accretion-disk dynamics in astrophysics \citep{begelman1988,narayan2008,sironi2015}. A non-Maxwellian velocity-space
structure must conform to specific conditions in order to drive an
instability: i.e., to transfer energy from the particles to
the electric and magnetic fields. This process simultaneously leads to an exponentially growing mode and drives the system closer to \emph{local thermodynamic equilibrium}.
Once the system no longer meets the conditions for instability, the march toward equilibrium halts, and the system lingers in a state of \emph{marginal stability}; i.e., the conditions for which $\gamma=0$. This
effect has been identified in numerical simulations
\citep{matteini2006,hellinger2008}, but recent work
suggests that dynamic interactions between the ions and electrons may
modify the stability threshold conditions \citep{yoon2017c}.
\citet{gary1993} and \citet{yoon2017a} offer more details into the
theory of unstable wave--particle interactions in the solar wind.

A variety of different schemes are used to classify wave--particle
instabilities 
\citep{krall1973,treumann1997,schekochihin2010,klein2015}.  Most
focus on the spatial scales at which unstable modes are driven:
\emph{macroinstabilities} and
\emph{microinstabilities} respectively drive unstable modes with
wavelengths much greater than and comparable to kinetic scales.  Other classifications focus on the mechanisms that drive the unstable modes:  \textit{configuration-space instabilities} are driven
by the departure of macroscopic quantities from thermodynamic
equilibrium and thus can be modeled by fluid equations, and \textit{kinetic} or \textit{velocity-space
  instabilities} are driven by resonant interactions with structures in the particle velocity distributions.
  
A prototypical macroscopic configuration-space instability is the
Chew-Goldberger-Low (CGL) firehose instability \citep{chew1956}, in
which the pressure $p_{\perp}$ perpendicular to the magnetic field becomes insufficient to counteract the centrifugal force experienced by the particles along a bend in the magnetic field.  Without a sufficiently
robust restoring force, initial magnetic perturbations are not damped
but in fact amplified, leading to the growth of a large-scale unstable
Alfv\'en mode.\footnote{The CGL marginal stability threshold arises at larger pressure anisotropies than those derived  from kinetic theory \citep{klein2015,hunana2017}, which, combined with the limited relevance of a fluid theory to a weakly collisionless system, limits this instability's relevance to the solar wind.}

\begin{figure}
  \centerline{\includegraphics[width = \textwidth]
{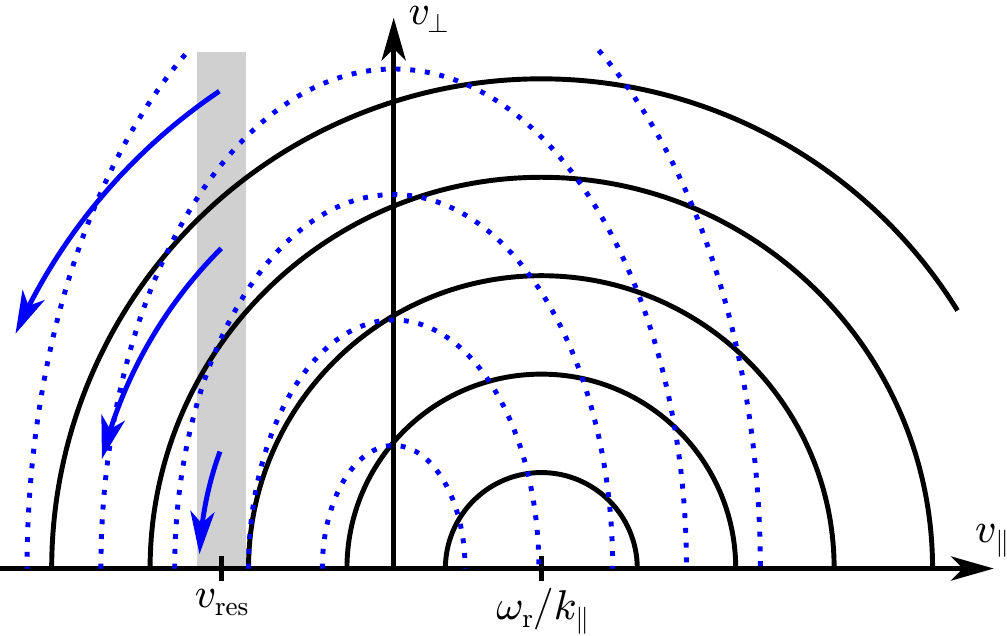}}%
  \caption{Quasilinear diffusion for an anisotropic particle
    distribution $f_{0j}$ (isocontours shown as blue dashed lines) unstable to left-hand circularly polarized ion-cyclotron waves with frequency $\omega_{\mathrm r}$ and parallel wavenumber $k_\parallel$. Unlike the cyclotron-resonant damping case (Fig.~\ref{fig:ql_diffusion}), the velocity-space diffusion along the pitch-angle gradients of $f_{0j}$ (black semi-circles) at $v_{\parallel}=v_{\textrm{res}}$ (gray shaded area) causes
    resonant particles to lose kinetic energy (i.e., to decrease in $v_{\perp}^2+v_{\parallel}^2$), which is transferred to the growing electromagnetic wave. This mechanism drives the kinetic ion-cyclotron instability.}
\label{fig:instability}
\end{figure}

A typical microscopic kinetic instability is the \emph{ion-cyclotron
instability}, which is physically very similar to the cyclotron-resonant damping of A/IC waves discussed in Sect.~\ref{sec:ql} but with $\gamma>0$. A left-hand circularly polarized wave with finite $k_{\parallel}$ may resonantly interact with particles from a
narrow range of parallel velocities $\approx v_{\mathrm{res}}$ that satisfy the resonance condition in Equation~(\ref{rescond}) for $n=+1$. These resonant particles diffuse according to the quasilinear diffusion relation in Equation~(\ref{qldiff}) along trajectories tangent to semi-circles defined by Equation~(\ref{qldiffcirc}) around the point $(v_{\perp},v_{\parallel})=(0,\omega_{\mathrm r}/k_{\parallel})$ in velocity space.  At the same time, quasilinear diffusion demands that the particles diffuse from higher $f_{0j}$ toward lower $f_{0j}$. We discuss the differences between the damped and the unstable cases with the help of Fig.~\ref{fig:instability}, which shows the same situation as Fig.~\ref{fig:ql_diffusion} but a different shape of $f_{0j}$ (blue dashed lines). This new shape of $f_{0j}$ now exhibits a temperature anisotropy with $T_{\perp\mathrm p}>T_{\parallel\mathrm p}$, which causes particles to diffuse toward smaller $v_{\perp}$ in Fig.~\ref{fig:instability} rather than toward larger $v_{\perp}$ as in Fig.~\ref{fig:ql_diffusion}. This change in behavior is a direct consequence of the altered alignment between the diffusion paths (black semi-circles) and the contours of $f_{0j}$ (blue dashed lines). The diffusive particle motion now results in a loss of kinetic energy (i.e., a decrease in $v_{\perp}^2+v_{\parallel}^2$) by the resonant particles, which is transferred to  growing field fluctuations.
Importantly, the direction of the energy flow between the fields and
the particle distribution depends on the local sign of the \emph{pitch-angle
gradient} of $f_{0j}$ at the resonance speed according to Equation~(\ref{Goperator}).  In addition to temperature anisotropies,
drifting populations and other non-Maxwellian features can
lead to pitch-angle gradients that drive resonant instabilities.

Despite their apparent similarity, the macro/micro and configuration/kinetic schemes are not
synonymous. Some instabilities occur at large spatial
scales but are driven by velocity-space effects. For example, the
mirror-mode instability \citep{southwood1993} is driven by the
interaction between the slow-mode-like anti-phase response of bulk
thermal and magnetic fluctuations, $\delta p$ and $\delta |\vec B|$,
and the in-phase response felt by particles with $v_\parallel \sim
0$. This latter population is approximately stationary along the
background magnetic field and gains or loses energy with changes in
the magnetic-field strength. On the other hand, the bulk population,
which does move parallel to the magnetic field in a slow-mode-like
polarized wave (see Sect.~\ref{slow_waves}), is able to effectively conserve energy via
transfer between parallel and perpendicular degrees of freedom.

The numerical evaluation of linear instabilities in kinetic theory follows the same procedure as the numerical evaluation of wave dispersion relations described in Sect.~\ref{sec:waves:lin}: the linearized Vlasov equation is used to calculate the dielectric tensor $\vec{\epsilon}$. Solutions to the dispersion relation in Equation~(\ref{dispersion_relation}) with $\gamma>0$ for a particular wavevector $\vec k$ represent linear kinetic instabilities, which grow with time according to Equation~(\ref{Egrowth}). Following from the linear set of Vlasov--Maxwell equations, these solutions are independent of the fluctuation amplitude. In contrast, the wave--wave instabilities discussed in Sect.~\ref{wwinst} depend on fluctuation amplitude. 

The behavior of instabilities in the inhomogeneous and turbulent solar wind as well as the nonlinear evolution of plasma instabilities are important matters of ongoing research. Most numerical evaluations of linear instabilities assume homogeneous plasma conditions, which are not fulfilled in the solar wind in general. For instance, the expansion of the plasma, the interaction of different plasma streams, and the ubiquitous turbulence create inhomogeneities and temporal variability that call into question the assumption of homogeneity. Nevertheless, the solar wind's parameter space is often observed to be restricted by the linear-instability thresholds, which suggests that linear theory bears some applicability to the solar wind. 

We define the \emph{marginal stability threshold} as a contour of constant maximum growth rate $\gamma_{\mathrm m}$ at any $\vec k$ through parameter space for a given instability. The choice of the relevant $\gamma_{\mathrm m}$ is somewhat arbitrary. Assuming that only a couple of parameters (e.g., $\beta_{\parallel j}$ and $T_{\perp j}/T_{\parallel j}$) have a significant impact on the growth rate of a specific instability, it is possible to construct a parametric model for the instability threshold. The inverse relation between a species'
temperature anisotropy and $\beta_{\parallel j}$ serves as the
prototypical example of such a threshold model, given for instance by
\citet{gary1994b,gary1994a}, \citet{gary1994c}, and \citet{hellinger2006}:
\begin{equation}
  \frac{T_{\perp j}}{T_{\parallel j}} = 1 + \frac{a}{\left(\beta_{\parallel j}-c\right)^b},
  \label{eqn:inverse}
  \end{equation}
where $a$, $b$, and $c$ are constant parameters calculated from fits
to solutions of the hot-plasma dispersion relation.  This form for the
inverse relation is introduced by \citet{hellinger2006} for a bi-Maxwellian proton
background distribution function according to Equation~(\ref{biMax}) and an isotropic Maxwellian electron distribution. The values of
$a$, $b$, and $c$ are different for the four unstable modes that can
be driven by proton temperature anisotropies (i.e., the ion-cyclotron,
parallel firehose, mirror-mode, or oblique firehose instability), as well as the desired maximum growth rates.
\citet{verscharen2016} compare the parameters $a$, $b$, and $c$ for
thresholds depending on maximum growth rates. Table~\ref{table_fit} lists best-fit values for these parameters for three different $\gamma_{\mathrm m}/\Omega_{\mathrm p}$-values for each of the four instabilities driven by proton temperature anisotropy. The growth rates have been calculated for a quasi-neutral plasma consisting of bi-Maxwellian protons and Maxwellian electrons with $T_{\mathrm e}=T_{\parallel\mathrm p}$ and $v_{\mathrm {Ap}}/c=10^{-4}$.
\begin{table}
\caption{Fit parameters for isocontours of constant maximum growth rate $\gamma_{\mathrm m}=10^{-2}\Omega_{\mathrm p}$, $\gamma_{\mathrm m}=10^{-3}\Omega_{\mathrm p}$, and $\gamma_{\mathrm m}=10^{-4}\Omega_{\mathrm p}$ in the $\beta_{\parallel\mathrm p}$-$T_{\perp\mathrm p}/T_{\parallel\mathrm p}$ plane for use in Equation~(\ref{eqn:inverse}). Calculated with the NHDS code \citep{verscharen2018b} and adapted from \citet{verscharen2016}.  \label{table_fit}}
\begin{tabular}{ld{2.3}d{2.3}d{2.3}}
    \noalign{\smallskip}\hline\noalign{\smallskip}    
Instability & $a$ & $b$ & $c$ \\
\noalign{\smallskip}\hline\noalign{\smallskip}
$\gamma_{\mathrm m}=10^{-2}\Omega_{\mathrm p}$ & & & \\
ion-cyclotron  & 0.649 & 0.400 & 0.000 \\
mirror-mode  & 1.040 & 0.633 & -0.012 \\ 
parallel firehose  & -0.647 & 0.583 & 0.713 \\
oblique firehose  & -1.447 & 1.000 & -0.148 \\
\noalign{\smallskip}\hline\noalign{\smallskip}
$\gamma_{\mathrm m}=10^{-3}\Omega_{\mathrm p}$ & & & \\
ion-cyclotron  & 0.437 & 0.428 & -0.003 \\
mirror-mode  & 0.801 & 0.763 & -0.063 \\ 
parallel firehose  & -0.497 & 0.566 & 0.543 \\
oblique firehose  & -1.390 & 1.005 & -0.111 \\
\noalign{\smallskip}\hline\noalign{\smallskip}
$\gamma_{\mathrm m}=10^{-4}\Omega_{\mathrm p}$ & & & \\
ion-cyclotron  & 0.367 & 0.364 & 0.011 \\
mirror-mode  & 0.702 & 0.674 & -0.009 \\ 
parallel firehose  & -0.408 & 0.529 & 0.410 \\
oblique firehose  & -1.454 & 1.023 & -0.178 \\
\noalign{\smallskip}\hline\noalign{\smallskip}
\end{tabular}
\end{table}
The values of $a$, $b$, and $c$ change in the presence of other plasma components, including
beams and minor ion components, which may act as additional sources of free
energy or may stabilize unstable growth
\citep{price1986,podesta2011b,maruca2012,matteini2015a}.  If the
underlying distribution has a shape other than bi-Maxwellian -- e.g., if
the particles have a $\kappa$-distribution according to Equation~(\ref{kappadist}) or a bi-$\kappa$-distribution according to Equation~(\ref{bikappadist}) -- these
threshold curves can be significantly different \citep{summers1991,xue1993,summers1994,xue1996,astfalk2015,astfalk2016}.
The exploration of more general phase-space densities requires direct
numerical integration of the dispersion relation
\citep{dum1980,matsuda1992,astfalk2017,horaites2018a,verscharen2018}.
Such general distributions produce instabilities that are either enhanced or suppressed relative to those associated with bi-Maxwellian particle distributions.

\begin{sidewaystable}
\vspace{0.58\textwidth}
  \caption{\label{tab:instabilities} Wave--particle instabilities
    relevant to the solar wind organized by free-energy source. For each instability, we list its name, classification, name of the unstable normal mode, and further references.}
  \begin{tabular}{llll}
\hline\noalign{\smallskip}
Instability & Classification & Unstable Normal Mode & References \\
    \noalign{\smallskip}\hline\noalign{\smallskip}    
    $T_{\perp i}/T_{\parallel i}>1$\fn{a} & & \\
    ion-cyclotron & micro/resonant & parallel A/IC &
    \citet{kennel1966a,davidson1975}\\
    mirror-mode & macro/resonant & non-propagating oblique kinetic slow mode&
    \citet{tajiri1967,southwood1993} \\
     & & & \citep{kivelson1996}\\
    \noalign{\smallskip}\hline\noalign{\smallskip}    
    $T_{\perp i}/T_{\parallel i}<1$ & & \\ 
    parallel firehose & micro/resonant & parallel FM/W & \citet{quest1996,gary1998}\\
    oblique firehose & micro/resonant & non-propagating oblique Alfv\'en & \citet{hellinger2000}\\
    \noalign{\smallskip}\hline\noalign{\smallskip}    
    $T_{\perp \mathrm e}/T_{\parallel \mathrm e}<1$ & & \\
    parallel electron firehose & micro/resonant & parallel FM/W & \citet{hollweg1970b,gary1985} \\
    oblique electron firehose & micro/configuration & oblique non-propagating Alfv\'en & \citet{li2000,kunz2018}\\
    \noalign{\smallskip}\hline\noalign{\smallskip}    
      $T_{\perp \mathrm e}/T_{\parallel \mathrm e}>1$ & & \\
    whistler anisotropy & micro/resonant & parallel FM/W& \citet{kennel1966a,scharer1967}\\
    \noalign{\smallskip}\hline\noalign{\smallskip}    
    $P_{\perp}/P_{\parallel}<1$\fn{b} & & \\
        CGL firehose & macro/configuration & non-propagating oblique Alfv\'en & \citet{chew1956}\\
    \noalign{\smallskip}\hline\noalign{\smallskip}    
    electromagnetic beam & & \\
    ion/ion RH resonant & micro/resonant & parallel FM/W & \citet{barnes1970}\\
    ion/ion nonresonant & macro/configuration & backward propagating firehose-like & \citet{sentman1981,winske1986} \\
    ion/ion LH resonant  & micro/resonant & parallel A/IC & \citet{sentman1981} \\
    electron/ion & micro/resonant & FM/W and A/IC modes & \citet{akimoto1987} \\
    electron heat flux & micro/resonant  & parallel FM/W &
    \citet{gary1975,gary1994,gary1999} \\
     & & & \citep{gary2000,horaites2018a,tong2018}\\
    ion drift & micro/resonant & parallel \& oblique FM/W & \citet{verscharen2013} \\
    ion drift & micro/resonant & parallel \& oblique A/IC & \citet{verscharen2013} \\
    \noalign{\smallskip}\hline\noalign{\smallskip}    
    ion drift \& anisotropy & micro/resonant & parallel FM/W and A/IC & \citet{verscharen2013b,bourouaine2013}\\
    \noalign{\smallskip}\hline\noalign{\smallskip}    
  \end{tabular}
  \fn{a}{\footnotesize Resonant instabilities due to temperature anisotropies
    can arise for each ion species \citep[index $i$; see][]{maruca2012}.} \\
 \fn{b}{\footnotesize Configuration-space instabilities are triggered by contributions to the total excess pressure from each plasma species \citep{kunz2015,chen2016a}.}
\end{sidewaystable}

Table~\ref{tab:instabilities} lists the wave--particle instabilities that are most important in regulating the large-scale dynamics of the solar wind. Many foundational publications
\citep[e.g.,][]{hollweg1975,schwartz1980,gary1993} provide more complete catalogues.

  Two of the most common free-energy sources are distinct temperatures or pressures perpendicular and parallel to the background magnetic
  field and the presence of faster populations that form a shoulder on or
  a beam distinct from the core population (Fig.~\ref{fig:distr_schematics}).  These two specific cases are
  considered in Sects.~\ref{sec:inst:tani} and \ref{sec:inst:beam}, with particular emphasis on their impact on the macroscale behavior of the solar
  wind. Significant work has been done on the effects of instabilities
  in other space environments such as the magnetosphere and magnetosheath \citep[][and
    references therein]{maruca2018}, but these results lie beyond the
  scope of this work.

\subsubsection{Temperature anisotropy}\label{sec:inst:tani}

\begin{figure}
  \centerline{\includegraphics[width = \textwidth]{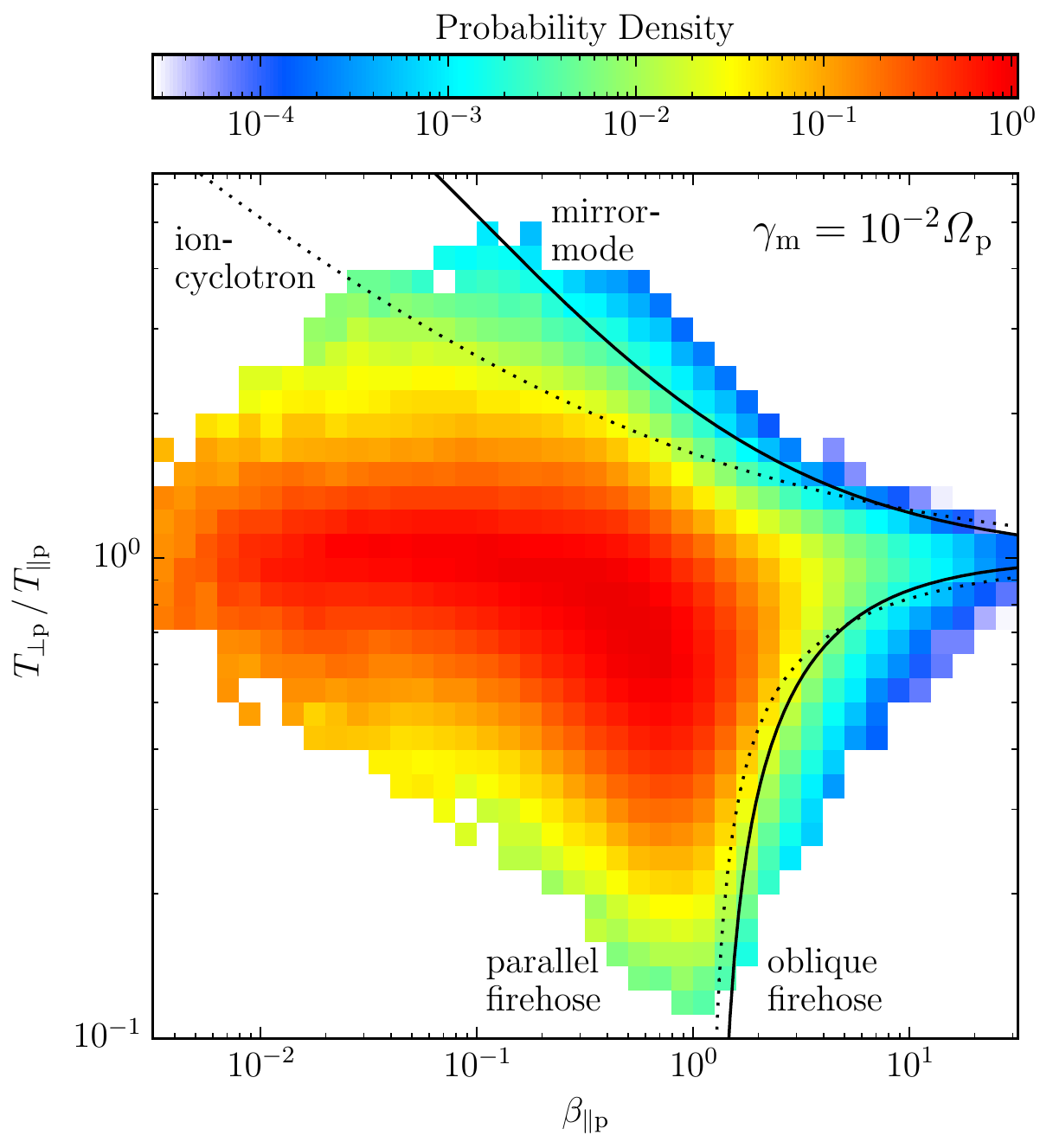}}%
  \caption{Probability distribution of the pristine solar wind in the $\beta_{\parallel \mathrm p}$-$T_{\perp \mathrm p}/T_{\parallel \mathrm p}$ plane. The instability thresholds for the four instabilities associated with proton temperature anisotropy according to Equation~(\ref{eqn:inverse}) and Table~\ref{table_fit} with $\gamma_{\mathrm m}=10^{-2}\Omega_{\mathrm p}$ are plotted for comparison. We only plot bins containing at least 25 counts. A significant fraction of the distribution exceeds the two resonant thresholds (ion-cyclotron and parallel firehose), while the non-resonant mirror-mode and oblique-firehose thresholds set more precise boundaries to the data distribution.}
\label{fig:brazil}
\end{figure}

Wave--particle instabilities  associated with \emph{temperature
  anisotropies} serve as a canonical example for the effects of
wave--particle instabilities on the solar wind's large-scale evolution.
Initial investigations of instability limits on solar-wind proton temperature anisotropy address either the $T_{\perp \mathrm
  p}>T_{\parallel\mathrm p}$ limit or the $T_{\perp \mathrm
  p}<T_{\parallel\mathrm p}$ limit separately.  For the former,
\citet{gary2001} find that the ion-cyclotron stability threshold
limits the maximum anisotropy of observations from the ACE spacecraft. 
For the latter limit, \citet{kasper2002} find that the Wind spacecraft's temperature-anisotropy values are mostly bounded by the parallel firehose instability threshold. Subsequent work \citep{hellinger2006} shows that, for the slow solar wind, the distribution of temperature anisotropies is
well constrained for $T_{\perp\mathrm p}/T_{\parallel \mathrm p}>1$
and $T_{\perp\mathrm p}/T_{\parallel \mathrm p}<1$ by the threshold of each of the configuration-space instabilities: i.e., the mirror-mode and oblique firehose instabilities. The probability distribution of data in the 
$\beta_{\parallel \mathrm p}$-$T_{\perp \mathrm p}/T_{\parallel \mathrm p}$ plane using measurements
from the Wind spacecraft is illustrated in
Fig.~\ref{fig:brazil}.\footnote{Plots of the data distribution in the $\beta_{\parallel
  \mathrm   p}$-$T_{\perp \mathrm p}/T_{\parallel \mathrm p}$ plane have become colloquially known as
  ``Brazil plots'' due to the characteristic shape of the data
  distribution for near-Earth solar wind.} We use the same dataset as described by \citet{maruca2013a}.
Interestingly, as seen in Fig.~\ref{fig:brazil}, the solar wind is not
constrained by all possible temperature-anisotropy thresholds: a significant portion of the $\beta_{\parallel \mathrm  p}$-$T_{\perp \mathrm p}/T_{\parallel \mathrm p}$ distribution extends beyond the ion-cyclotron threshold, which, for $\beta_{\parallel
\mathrm  p}\lesssim 1$, sets a stricter limit on the departure from isotropy than the mirror-mode instability threshold,  as is pointed out by \citet{hellinger2006}. Several
justifications for this apparent inactivity of the ion-cyclotron instability have been proposed: 
low efficiency energy extraction \citep{shoji2009}, 
stabilizing effects of minor ions and/or drifts \citep{maruca:phd,maruca2012}, or quasilinear
flattening of the resonant region \citep{isenberg2013}.

A na\"ive model for the expanding solar wind would have $T_{\perp j}$
and $T_{\parallel j}$ follow the double-adiabatic prediction (see Equations~(\ref{cgl1}) and (\ref{cgl2}) in
Sect.~\ref{sec:moments}). Using data from Helios and Ulysses at different
heliocentric distances, \citet{matteini2007} show that the
distribution in $\beta_{\parallel\mathrm p}$-$T_{\perp\mathrm p}/T_{\parallel \mathrm p}$ space follows
a radial trend, albeit one with a smaller radial gradient than that predicted by double-adiabatic expansion, until the system
encounters the instability thresholds. Then, the
distribution's anisotropy is constrained by the parametric thresholds
to the stable parameter space.

Identifying polarization and other linear quantities
associated with the predicted instabilities allows us to infer the
presence of modes driven by temperature-anisotropy instabilities. For
instance, the signal of strongly peaked magnetic helicity near
parallel ion-kinetic scales \citep{he2011,podesta2011,klein2014}
indicates the presence of parallel-propagating
FM/W or A/IC waves associated
with proton temperature-an\-iso\-tro\-py instabilities. Wind observations provide evidence for enhanced magnetic fluctuations near threshold boundaries \citep{bale2009}, suggesting that instabilities are active near these thresholds in generating unstable
modes which are associated with such fluctuations.  Other quantities
are found to be enhanced in marginally unstable parameter regions:
ion temperature \citep{maruca2011,bourouaine2013} and intermittency
\citep{osman2012a,servidio2014}.   Calculating polarization as a
function of $T_{\perp\mathrm p}/T_{\parallel\mathrm p}$ and $\beta_{\parallel \mathrm p}$ reveals the
presence of a population of A/IC waves in the region
in which they are expected to become unstable \citep{telloni2016}.
The identification of parallel-propagating A/IC
waves \citep[e.g.,][]{jian2009,jian2010,jian2014,gary2016a} that do
not naturally arise from critically balanced turbulence (see
Sect.~\ref{sec:critical}) serves as further, indirect evidence for the action of these instabilities.

We emphasize that caution must be exercised in the analysis of  $\beta_{\parallel
  j}$-$T_{\perp j}/T_{\parallel j}$ plots. \citet{hellinger2014} raise concerns about the effects of projecting the distribution of quantities onto any
reduced parameter space. By partitioning the data into different temperature quartiles and studying the temperature-anisotropy distribution of each, they find that enhanced quantities near the instability thresholds may primarily result from underlying correlations between solar-wind temperatures
and speeds.  Moreover, it is important to carefully account for the
blurring of temperature-anisotropy observations due to the finite time
required to construct a velocity distribution measurement
\citep{verscharen2011,maruca2013a}.

In addition to instabilities triggered by the temperature anisotropy
of the core proton velocity distribution, anisotropic distributions of
the other plasma components, including the electrons
\citep{hollweg1970b,gary1985,li2000,kunz2018} and heavy ions
\citep{ofman2001,maruca2012,bourouaine2013} can lead to resonant
instabilities. We discuss the combined effect of these sources of free energy in Sect.~\ref{sec:multi-sources}.

\subsubsection{Beams and heat flux}\label{sec:inst:beam}

The \emph{relative drift} between plasma components is another common
source of free energy that can drive wave--particle
instabilities. The velocity difference between the two components (of the same or different species) can
contribute to excess parallel pressure or induce non-zero currents, and
the drifting distributions themselves may resonate with unstable
waves \citep[e.g., the parallel propagating beam instability described
  by][]{verscharen2013a}. As with temperature anisotropies, some
thresholds associated with drifts and beams constrain the observed
data distributions in parameter space.

Beam and heat-flux instabilities regulate non-thermal features in the electron distribution function. For instance, \citet{tong2018} find compelling evidence that the heat-flux-driven Alfv\'en-wave instability limits the electron core drift with respect to the halo and the protons. To some degree, this result contradicts the earlier work of \citet{bale2013}, who find that the collisional transport rather than a heat-flux instability is more active in limiting the electron-core drift (see also Sect.~\ref{sec:col:elec}). However, collisions and kinetic instabilities can co-exist in the solar wind and simultaneously regulate the heat flux.  The electron-strahl heat flux can drive oblique instabilities of the lower-hybrid and the oblique FM/W wave \citep{omelchenko1994,shevchenko2010,vasko2019,verscharen2019a}.

\begin{figure}
  \centerline{\includegraphics[width = \textwidth]
{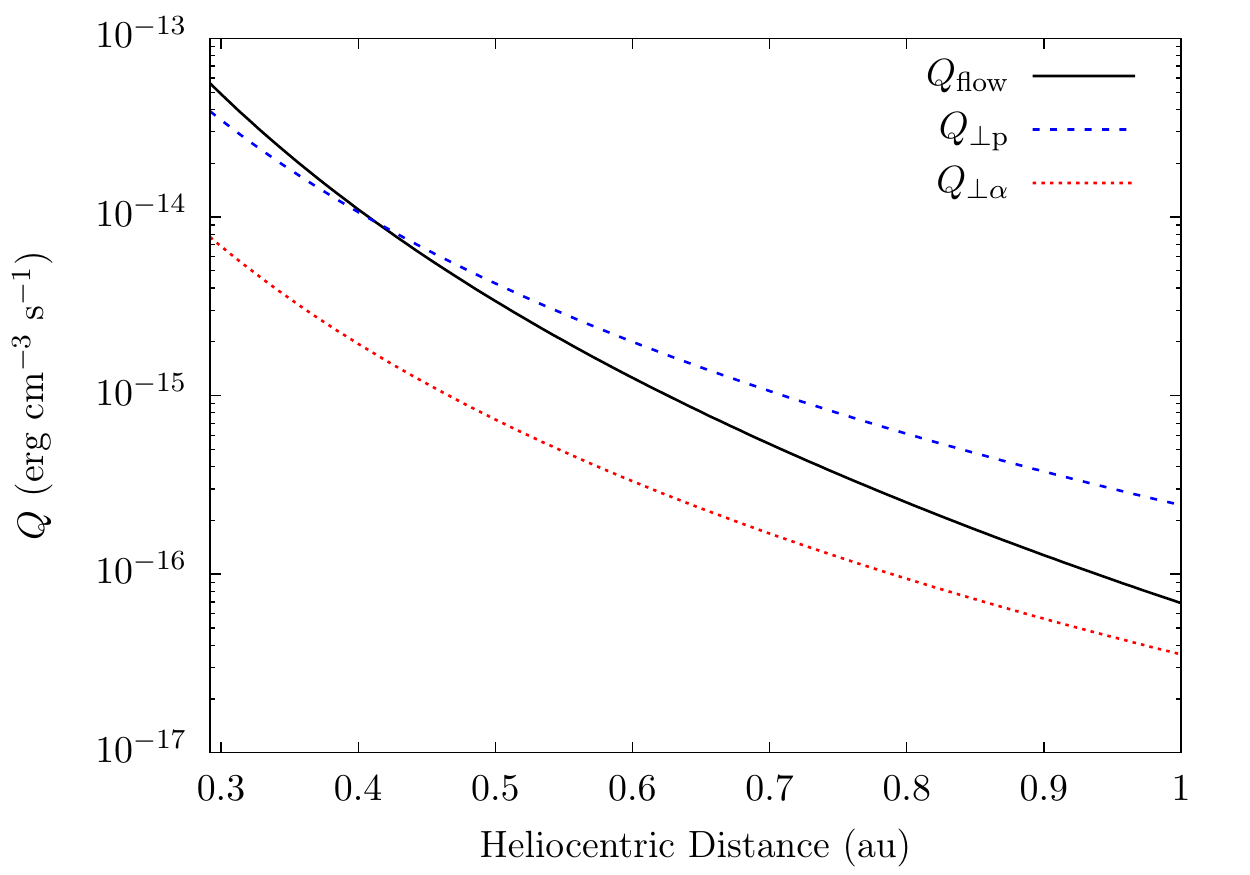}}%
  \caption{Rate of energy release $Q_{\mathrm{flow}}$ from the deceleration of $\alpha$-particles through kinetic microinstabilities as a function of distance in the inner heliosphere. We assume that the $\alpha$-particle drift speed is always fixed to the  local threshold for the FM/W instability based on average fast-solar-wind measurements from Helios. $Q_{\mathrm{flow}}$ then follows from energy conservation. $Q_{\perp\mathrm p}$ and $Q_{\perp\alpha}$ are calculated based on Equation~(\ref{cgl1}), setting $\vec q_{\perp j}=0$ and the right-hand side to $Q_{\perp j}$. Using empirical profiles for $B$, $p_{\perp j}$, $n_j$, and $\vec U_j$ for $j=\mathrm p$ and $j=\alpha$ then gives the empirical heating rates $Q_{\perp\mathrm p}$ and $Q_{\perp\alpha}$. Adapted from \citet{verscharen2015}.}
\label{fig:qflow}
\end{figure}

Likewise, ion beams can drive plasma instabilities. \citet{bourouaine2013} report constraints on the drift of $\alpha$-particles relative to protons through parallel-propagating A/IC and FM/W instabilities. 
These ion-beam instabilities result in a quasi-continuous deceleration of the $\alpha$-particles, which leads to a quasi-continuous release of energy from the $\alpha$-particle kinetic energy into field fluctuations \citep{verscharen2015}. Figure~\ref{fig:qflow} shows, as functions of distance from the Sun, the rate of energy-density release $Q_{\mathrm{flow}}$ derived from energy conservation as well as the empirical perpendicular heating rates $Q_{\perp\mathrm p}$ for protons and $Q_{\perp\alpha}$ for $\alpha$-particles. $Q_{\mathrm{flow}}>Q_{\perp \alpha}$ at distances between 0.3 and 1~au, and $Q_{\mathrm{flow}}>Q_{\perp \mathrm p}$ at distances between 0.3 and 0.4~au. This finding suggests that the energy release through $\alpha$-particle instabilities comprises a significant fraction of the solar wind's energy, and that large-scale solar-wind models must account for $\alpha$-particle thermodynamics.  Due to the lack of in-situ measurements at smaller heliocentric distances, we are unable to compare $Q_{\mathrm{flow}}$ with $Q_{\perp\mathrm p}$ or $Q_{\perp \alpha}$ closer to the Sun yet; however, we expect this trend to continue toward the acceleration region of the solar wind.

\subsubsection{Multiple sources of free energy}\label{sec:multi-sources}

Under typical solar-wind conditions, \emph{multiple sources of free energy}
are simultaneously available to drive distinct unstable modes.  For example, beams,
temperature anisotropies, and anisothermal temperatures between 
species are all frequently and simultaneously present in solar-wind plasma \citep{kasper2008,kasper2017}. The introduction of an additional source of free energy can act either to enhance an instability's growth rate or act to
stabilize the system.  

The thresholds of configuration-space instabilities (i.e., the mirror-mode and the oblique firehose instabilities) depend on the total free energy in the system \citep{chen2016a}. 
The threshold of the oblique firehose instability limits the observed plasma to the stable parameter space, when the combined effects of ion and electron anisotropies as well as relative drifts between the plasma species are considered. Less than $1\%$ of the observations exceed this threshold, and, for these intervals, the
proton, electron, and $\alpha$-particle components all significantly
contribute to the system's unstable growth.

According to an analytical model of the coupling between
the effects of temperature anisotropy and drifts \citep{ibscher2014},  the
combined effects of these free-energy sources yield a threshold in
the region of parameter space with $\beta_{\parallel \mathrm p}<1$ and $T_{\perp\mathrm
  p}<T_{\parallel\mathrm p}$. This is consistent with the lack of solar-wind observations in this region of parameter space (see Fig.~\ref{fig:brazil}). However,
\citet{bale2009} do not find enhanced fluctuations or other
indications of unstable-mode generation in this region, and \citet{vafin2019} explain the lack of data in this region through collisional effects.  The coupling
of temperature anisotropy and beams has been incorporated into an
improved threshold model for limiting proton-temperature-anisotropy
observations \citep{vafin2018}, which may be tested in future in-situ
observations of low-$\beta_{\parallel \mathrm p}$ systems such as the
near-Sun solar wind.  \citet{verscharen2013b}
provide testable limits on temperature anisotropy and
$\alpha$-particle drifts, which \citet{bourouaine2013} find to largely agree with
solar-wind observations. Numerical
simulations \citep[e.g., by][]{maneva2018} are also used to study the
simultaneous impact of drifts and temperature anisotropies.
The coupling between electrons and ions
modifies the solar-wind expansion, preventing a
uniform progression of the bulk thermodynamic properties toward the
firehose threshold \citep{yoon2017c}.  This effect occurs in addition to
the effects of collisions on drawing the solar wind toward
isotropy (see Sect.~\ref{sec:col:relax}), which is found to be important but insufficient for a complete description of the solar wind's observed state \citep{yoon2016}.

\begin{figure}
  \centerline{\includegraphics[width = \textwidth]
{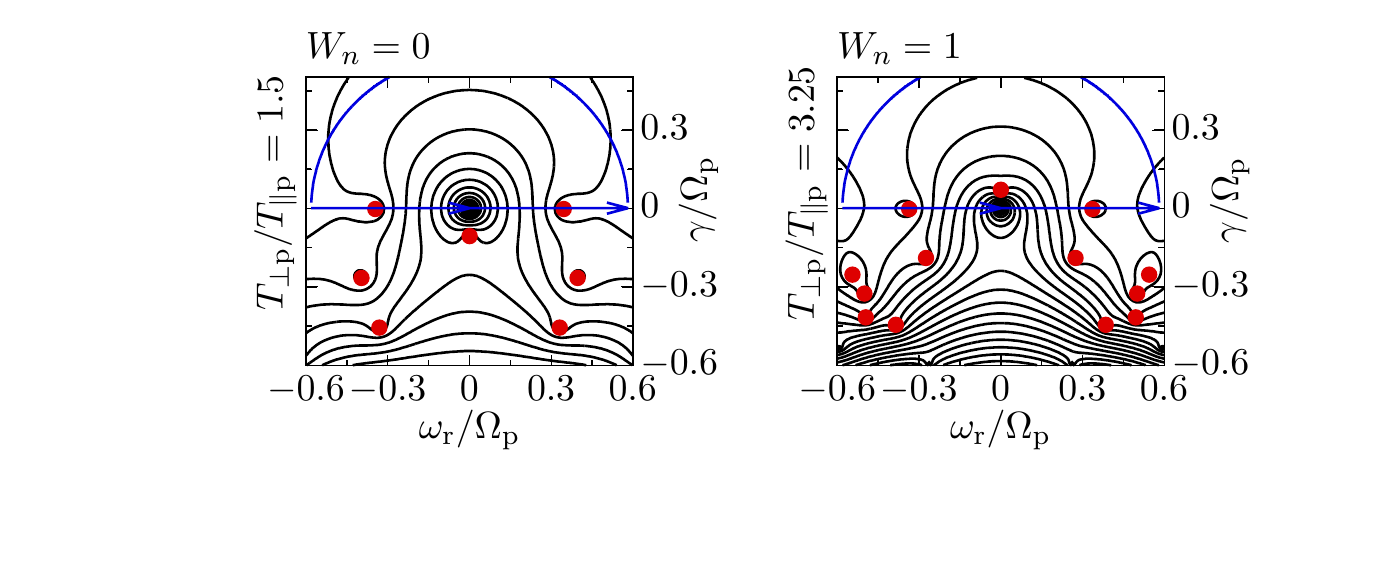}}%
  \caption{Illustration of the \emph{Nyquist instability criterion}. Black lines indicate isocontours of $\mathrm{det}\,\left[\vec {\mathcal D}(\vec k,\omega_{\textrm r}+i\gamma)\right]$ for a
    stable (left) and unstable (right) system, with the normal-mode solutions
    indicated with red dots. The contour integral is performed over
    the entire upper half plane, symbolized by the blue curve (which would formally extend out to $\omega_{\mathrm r}\rightarrow \pm \infty$). 
    Applying the residue theorem yields a non-negative integer $W_n$ equal
    to the number of unstable modes supported by the system.}
\label{fig:nyquist}
\end{figure}

Instead of relying solely on analytical threshold models,
which are formally valid for low-dimensional sub-spaces (e.g., $\beta_{\parallel \mathrm p}$ and $T_{\perp \mathrm p}/T_{\parallel \mathrm p}$ only) of the full parameter space that characterizes the solar wind, the \emph{Nyquist instability criterion} accounts for the simultaneous effects of all
wave--particle free-energy sources \citep{nyquist1932}.  This
method determines whether a system supports any growing modes at a
particular given wavevector $\vec{k}$ by performing a complex contour
integration, which is illustrated in Fig.~\ref{fig:nyquist}. The normal modes of
a system are the solutions to
$\mathrm{det}\,\left[\vec{\mathcal D}(\vec{k},\omega)\right]=0$ according to Equation~(\ref{dispersion_relation}), where $\vec{\mathcal D}$ is the
system's dispersion tensor. As described in Sect.~\ref{sec:waves:lin}, the form of $\vec{\mathcal D}$ depends on the set of system
parameters such as temperature, density, and drift of each
plasma component. The number of modes satisfying $\gamma>0$ can be
ascertained by applying the residue theorem to the integral
\begin{equation}\label{nyqint}
W_n = \frac{1}{2 \pi i} \oint \frac{\mathrm d \omega}{\mathrm{det}\,\left[\vec{\mathcal D}(\vec{k},\omega)\right]},
\end{equation}
where the contour is taken over the upper half plane of complex
frequency space $\omega=\omega_{\mathrm r}+i\gamma$. The integration in Equation~(\ref{nyqint}) is much easier to compute than the determination of the dispersion relation for all individual potentially unstable modes. This method has more than half a century of productive use in the study of plasma stability  \citep{jackson1958,buneman1959,penrose1960,gardner1963}. 

\citet{klein2017c} present a modern automatic
implementation of the Nyquist instability criterion for the case of an
arbitrary number of drifting bi-Maxwellian
components.  The application of this criterion to a
statistically random set of solar-wind observations modeled as a
collection of proton core, proton beam, and $\alpha$-particle
components (each with distinct anisotropies, densities, and drifts)
finds that a majority of intervals are unstable \citep{klein2018}. Most
of the unstable modes are resonant instabilities at ion-kinetic
scales and with growth rates less than the instrument integration time
and convected kinetic scales. About $10\%$ of the intervals have
instabilities with growth rates of order the nonlinear turbulent
cascade rate $1/\tau_{\mathrm{nl}}$ at proton-kinetic scales, which indicates that they 
may grow quickly enough to compete with the background turbulence.

\subsection{Wave--wave instabilities}\label{wwinst}

Wave--wave instabilities, in contrast to wave--particle instabilities,
depend sensitively on the amplitudes of the plasma fluctuations.  The
finite amplitudes of fluctuating waves lead to violations of the
linearization used to derive the wave--particle instabilities discussed
in Sect.~\ref{wpinst}. Instead, nonlinear effects allow for
wave--wave coupling to lead to unstable wave growth, which places limits on
the amplitudes of magnetic and velocity fluctuations.


\subsubsection{Parametric-decay instability}

The \emph{parametric-decay instability} (PDI) is a classic wave--wave
instability first described by \citet{galeev1963} and
\citet{sagdeev1969} for a three-wave interaction. It belongs to a broader class of parametric instabilities that also includes beat and modulational instabilities \citep{hollweg1994}.
In the
low-$\beta_{\mathrm p}$ limit, the PDI causes a
finite-amplitude forward-propagating Alfv\'en wave, known as the
\emph{pump mode}, to decay into a backward-propagating Alfv\'en wave
and a forward-propagating acoustic wave.  \citet{goldstein1978}
provides a generalization of this instability for circularly-polarized Alfv\'en
waves in 
finite-$\beta_{\mathrm p}$ plasmas.  The dynamics of such instabilities are important for the
evolution of the solar wind. As described in Sect.~\ref{slow_waves}, the compressive acoustic mode can efficiently dissipate and thus heat the plasma \citep{barnes1966}. Furthermore, the generation of counter-propagating Alfv\'en waves is essential
for driving the turbulent cascade (see Sect.~\ref{sec:waveturb}). \citet{malara1996} show that, even in the large-amplitude limit and when the pump mode is non-monochromatic, the PDI continues to operate without a significant reduction in its growth rate. 
Theoretical work suggests that the PDI may develop an inverse cascade near the Sun and, therefore, be essential in driving solar-wind
turbulence \citep{chandran2018}.

A number of numerical simulations investigate the presence and
effects of decay instabilities under conditions approximating the solar wind 
\citep{matteini2010a,verscharen2012a,tenerani2013,tenerani2017,shoda2018a,shoda2018b}.
 A recent analysis of solar-wind
observations at 1~au \citep{bowen2018} 
indicates a strong correlation between observed compressive fluctuations and higher estimated PDI growth rates, which is consistent with the parametric decay of Alfv\'en modes. Parametric instabilities are also observed in laboratory plasma experiments \citep{dorfman2016}. 

\subsubsection{Limits on large-amplitude magnetic fluctuations }

In addition to decay instabilities, finite-amplitude waves are capable of \emph{self-destabilization}. Linearly polarized, large-amplitude Alfv\'en waves drive compressions in the plasma, which reduce the amplitude of the Alfv\'enic fluctuations if $\delta |\vec B|\neq 0$  \citep[see also Sect.~\ref{sec:alfven} of this review;][]{hollweg1971}. This effect may lead to the observed preference for Alfv\'enic fluctuations with $B=\mathrm{constant}$. A related example of such behavior occurs if the 
amplitude $\delta B_\perp/B_0$ of the perpendicular magnetic fluctuations exceeds the threshold $\sim \beta_{\mathrm p}^{-1/2}$  \citep{squire2016}. Beyond this limit, the pressure anisotropy associated with
the wave fluctuations exceeds the parallel-firehose limit and destroys the
restoring force associated with the magnetic tension, which destabilizes
the wave. Numerical simulations confirm signatures of this instability, which are currently also being sought in solar-wind observations under high-$\beta_{\mathrm p}$ conditions \citep{squire2017b,squire2017a,tenerani2018}



\subsection{The fluctuating-anisotropy effect}

 Large-scale compressive fluctuations with finite
amplitudes and $\omega_{\mathrm r}\ll \Omega_{\mathrm p}$ modify the plasma  moments, including $\beta_j$ and  $T_{\perp j}/T_{\parallel j}$ according to Equations~(\ref{cgl1}) and~(\ref{cgl2}).  These and potentially other plasma moments (like the relative drifts between species) fluctuate with the large-scale compressive fluctuations \citep{squire2017b,squire2017a,tenerani2018}. If the amplitude of these fluctuations is sufficiently large, these modifications can
move the system from a stable to an unstable configuration with respect to anisotropy-driven kinetic microinstabilities \citep{verscharen2016}. The instability then acts to modify the velocity distribution, e.g., by pitch-angle scattering particles. It suppresses further growth of the anisotropy, which leads to a reduction in the amplitude of the large-scale compressive fluctuations and an isotropization of the particles. Whether this process occurs depends on the polarization and
amplitude of the large-scale compressive mode. Compressive ion-acoustic
 modes (see Sect.~\ref{slow_waves}) with reasonable magnetic fluctuation amplitudes ($\delta |\vec B|/B_0 \gtrsim 0.04$) can trigger this effect with temperature-anisotropy-driven instabilities under typical solar-wind conditions at 1~au.  This \emph{fluctuating-anisotropy effect} can be generalized to a \emph{fluctuating-moment effect}, which includes, for instance, variations in relative drift speeds that may trigger additional instabilities.

\section{Conclusions}\label{sec:conclusions}

We briefly summarize our discussion of the multi-scale nature of the solar wind, give an outlook on future developments in the field, and outline the broader impact of this research topic.

\subsection{Summary}

\begin{figure}
  \includegraphics[width=\textwidth]{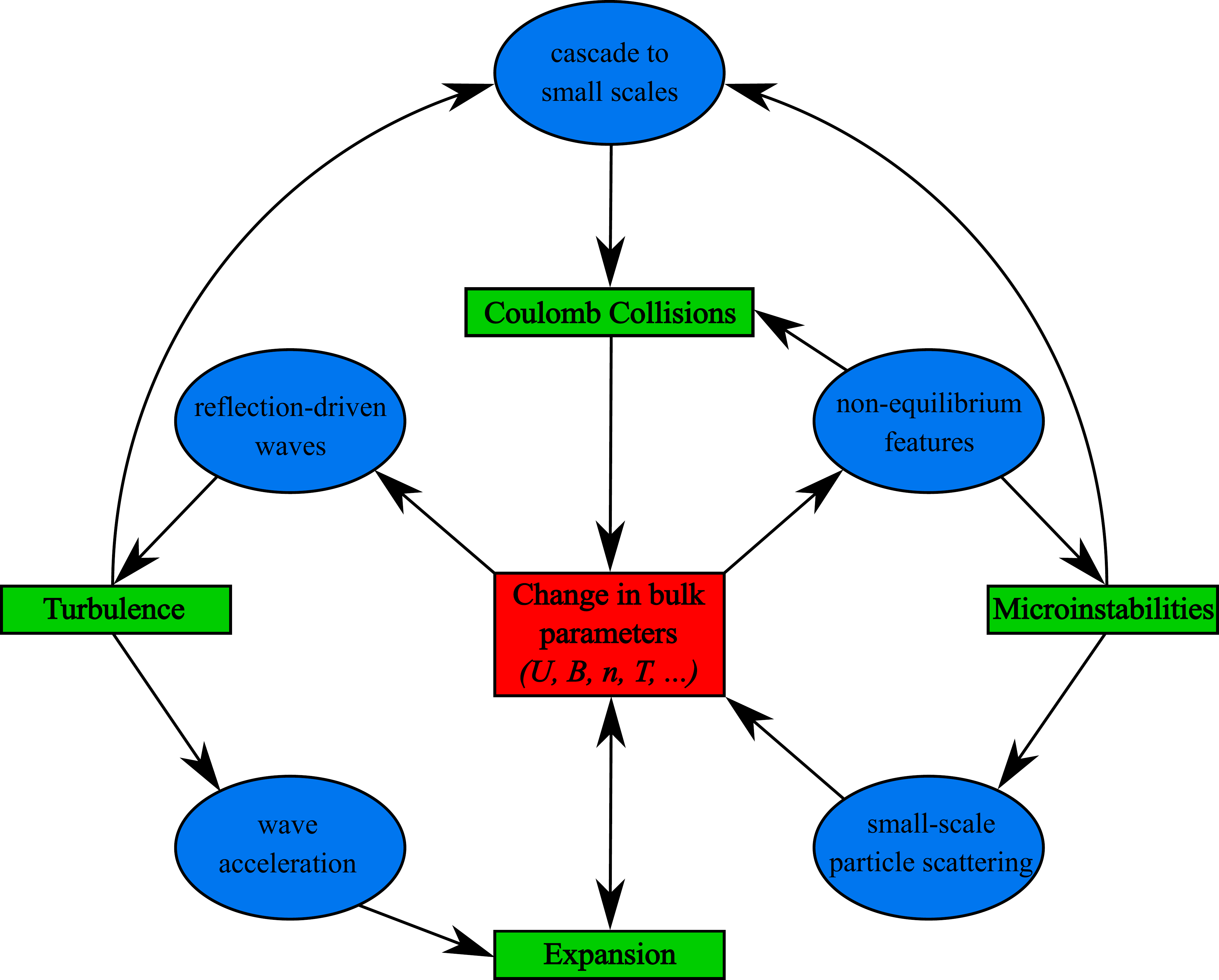}
\caption{Summary of the multi-scale couplings in the solar wind. We describe the effects of collisions in Sect.~\ref{sec:col}, the effects of waves in Sect.~\ref{sec:waves}, the effects of turbulence in Sect.~\ref{sec:turbulence}, and the effects of microinstabilities in Sect.~\ref{sec:inst}. The arrows illustrate the connections and interactions discussed in this review article. }
\label{fig:summary_diagram}       
\end{figure}

As we summarize in Fig.~\ref{fig:summary_diagram}, the dynamics and thermodynamics of the solar wind result from an intricate multi-scale coupling between global expansion effects and local kinetic processes. The global expansion shapes particle distribution functions slowly compared to most of the collective plasma timescales and creates the ubiquitous non-equilibrium features of solar-wind particles. It also generates gradients in the plasma bulk parameters that drive Sunward-propagating waves, which subsequently interact with anti-Sunward-propagating waves to generate turbulence. By creating microphysical features and turbulence, the expansion couples to small scales  and sets the stage for collisional relaxation, the dissipation of waves and turbulence, and kinetic microinstabilities to act locally. On the other hand, these local processes couple to the global scales and modify the large-scale plasma flow by, for example, accelerating the plasma, changing the plasma temperatures, introducing temperature anisotropies, regulating heat flux, or generating electromagnetic structures for particles to scatter on. These effects then modify the expansion.  Figure~\ref{fig:summary_diagram} includes some processes (e.g., reflection-driven waves) which we will discuss in the next major update of this Living Review.

We derive our understanding of the solar wind's multi-scale evolution from detailed measurements of its particles and fields.  In-situ observations provide perspective on small-scale processes, while remote observations provide perspective on large-scale processes. Therefore, we rely on the combination of in-situ and remote observations, in concert with theoretical modeling efforts and numerical simulations to elucidate the multi-scale evolution of the solar wind. This review describes the current state of the art of the field based on a combination of observational discoveries and fundamental plasma physics.

\subsection{Future outlook}

 Major new space missions such as Parker Solar Probe \citep[PSP;][]{fox2016} and Solar Orbiter \citep[SO;][]{mueller2013} are dedicated to the study of the processes at the heart of this review. 
 
 PSP, which launched in August 2018 and achieved its first perihelion in November 2018, is beginning to measure in-situ plasma properties with unprecedented energy and temporal resolution and at unexplored heliocentric distances (see Fig.~\ref{fig:missions}). New findings derived from PSP will transform our understanding of plasma processes near the Sun. PSP is expected to provide our first in-situ observations of the corona, which are anticipated to draw together the heliospheric and solar communities and to enable novel combinations of in-situ and remote observations. 
 
 SO will measure the solar-wind properties through both in-situ measurements of the local plasma conditions and remote observations of the Sun's surface. A major goal for SO is \emph{linkage science}: connecting processes in and near the Sun with the   behavior of solar-wind plasma across all relevant scales. SO's inclined orbit will carry it out of the ecliptic plane and enable it to sample solar wind from polar coronal holes with its more extensive instrumentation package compared to PSP. Both PSP and SO will drive research into the multi-scale nature of the solar wind for  decades.

Other heliospheric missions that are currently being developed and proposed will directly address the topics of this review. These include mission concepts to investigate the nature of waves and turbulence through \emph{multi-point} and \emph{multi-scale measurements} as well as mission concepts to resolve the smallest natural plasma scales in the solar wind \citep[e.g.,][]{nap2016,klein2019,matthaeus2019,tenbarge2019,verscharen2019b}. These efforts demonstrate that the heliophysics community understands the need to investigate the multi-scale couplings of plasma processes and their impact on the dynamics and thermodynamics of the solar wind.

We also anticipate major advances in modeling in the near future. Previously,  numerical simulations of processes that connect over large scale separations required computational resources too great for them to be practical. Therefore, most models either focused on global expansion dynamics (e.g., global MHD simulations) or on local plasma processes (e.g., homogeneous-box particle-in-cell simulations).\footnote{Notable exceptions to this dichotomy in global and local scales include expanding-box models and ad-hoc inclusions of kinetic processes through effective transport coefficients in global models.} However, our increasing numerical capabilities will allow us to simulate self-consistently the coupling across scales of global and local processes in the near future. Even though a full particle-in-cell model of the heliosphere with realistic properties may still lie decades in the future, the ongoing improvement in our modeling capabilities will advance our understanding of the multi-scale nature of the solar wind.

\subsection{Broader impact}

All magnetized plasmas exhibit a broad range of characteristic length scales and timescales. These span from the largest scales of the system to its microscopic scales: those of plasma oscillations, particle gyration, and electrostatic and electromagnetic shielding. The vast system sizes of space and astrophysical plasmas lead to especially large separations among these characteristic plasma scales. The solar wind exemplifies such a multi-scale astrophysical plasma, and the combination of solar-wind observations with fundamental plasma physics has improved our understanding of astrophysical plasma throughout the Universe. The solar wind's expansion through the heliosphere introduces additional global scales that couple to the small-scale plasma processes.
We anticipate that, in the coming years, the connection of small-scale kinetic processes with the large-scale thermodynamics of astrophysical plasmas will be a major research focus not only in heliophysics but throughout the astrophysics community.  

The solar wind is the ideal place to study the multi-scale nature of astrophysical plasmas. The conditions of space and astrophysical plasmas cannot be reproduced and sampled with comparable accuracy in laboratories. With the notable exception of the very local interstellar medium, the only astrophysical plasmas that have been observed in situ are in the heliosphere.

Research into this topic serves a broader impact beyond the purely academic understanding of space and astrophysical plasmas. The study of the solar wind's multi-scale nature enables a better understanding of its dynamics and thermodynamics based on first principles. This knowledge will be invaluable to the design of physics-based models for space weather and to guiding our efforts toward the successful prediction of space hazards for our increasingly technological and spacefaring society.


\begin{acknowledgements}
This work was supported by the STFC Ernest Rutherford Fellowship ST/P003826/1, the STFC Consolidated Grant ST/S000240/1, as well as NASA grants NNX16AM23G and NNX16AG81G. 

We acknowledge the use of data from Wind's SWE instrument (PIs K.~W.~Ogilvie and A.~F.~Vi\~{n}as), from Wind's MFI instrument (PI A.~Szabo), and from Ulysses' SWOOPS instrument (PI D.~J.~McComas). Lynn Wilson is Wind's Project Scientist. We extend our gratitude to Chadi Salem for providing processed Wind/3DP (PI S.~D.~Bale) data to create Fig.~\ref{fig:sh}. We also acknowledge the use of data from ESA's Cluster Science Archive \citep{laakso2010} from the EFW instrument (PI M.~Andr\'e), the FGM instrument (PI C.~Carr), the STAFF instrument (PI P.~Canu), the CIS instrument (PI I.~Dandouras), and the PEACE instrument (PI A.~Fazakerley). We acknowledge vigorous feedback on Fig.~\ref{fig:missions} from D.~J.~McComas at the 2019 SHINE Meeting.

Data access was provided by the National Space Science Data Center (NSSDC) Space Physics Data Facility (SPDF) and NASA/GSFC's Space Physics Data Facility's CDAWeb and COHOWeb services. This review has made use of the SAO/NASA Astrophysics Data System (ADS). 
\end{acknowledgements}

\bibliographystyle{spbasic}      
\bibliography{multiscale_solarwind}   

\end{document}